\documentclass[znote,zbstdefault]{./zeus_paper}
\usepackage[english]{babel}
\chardef\usc=95
\chardef\til=126
\catcode`\@=11 
\DeclareRobustCommand\xdotspace{\futurelet\@let@token\@xdotspace}
\def\@xdotspace{%
  \ifx\@let@token.\else
  \ifx\@let@token\bgroup.\else
  \ifx\@let@token\egroup.\else
  \ifx\@let@token\/.\else
  \ifx\@let@token\ .\else
  \ifx\@let@token~.\else
  \ifx\@let@token!.\else
  \ifx\@let@token,.\else
  \ifx\@let@token:.\else
  \ifx\@let@token;.\else
  \ifx\@let@token?.\else
  \ifx\@let@token/.\else
  \ifx\@let@token'.\else
  \ifx\@let@token).\else
  \ifx\@let@token-.\else
  \ifx\@let@token\@xobeysp.\else
  \ifx\@let@token\space.\else
  \ifx\@let@token\@sptoken.\else
   .\space
   \fi\fi\fi\fi\fi\fi\fi\fi\fi\fi\fi\fi\fi\fi\fi\fi\fi\fi}
\catcode`\@=12 

\newcommand{\stru}[2]{%
   \relax\ifmmode\hbox{\vrule height#1 depth#2 width0pt}%
   \else\vrule height#1 depth#2 width0pt\fi}

\newcommand{\Ronum}[1]{\uppercase\expandafter{\romannumeral#1}}
\newcommand{\ronum}[1]{\expandafter{\romannumeral#1}}
\DeclareRobustCommand{\LaTeXZ}{%
  \LaTeX\kern-.05em4\kern-.1em
  {\raisebox{-0.2ex}{$\scriptstyle\text{ZEUS}$}}\xspace}

\DeclareMathAlphabet{\mathbf}{OT1}{cmr}{bx}{sl}
\newcommand{\eVdist}{\kern-0.06667em}

\newcommand{\slashfrac}[2]{%
  \raisebox{0.5ex}{\ensuremath #1}\kern-0.12em/\kern-0.08em
  \raisebox{-.8ex}{\ensuremath #2}}

\newcommand{\sqr}[3]{%
    {\vcenter{\hrule height.#3ex\hbox{\vrule width.#2ex height#1ex
     \kern#1ex\vrule width.#3ex}\hrule height.#2ex}}}

\catcode`\@=11 
\newcommand{\parenbar}{\mathpalette\p@renb@r}
\def\p@renb@r#1#2{\vbox{%
  \ifx#1\scriptscriptstyle \dimen@.7em\dimen@ii.2em\else
  \ifx#1\scriptstyle \dimen@.8em\dimen@ii.25em\else
  \dimen@1em\dimen@ii.4em\fi\fi \offinterlineskip
  \ialign{\hfill##\hfill\cr
    \vbox{\hrule width\dimen@ii}\cr
    \noalign{\vskip-.3ex}%
    \hbox to\dimen@{$\mathchar300\hfil\mathchar301$}\cr
    \noalign{\vskip-.3ex}%
    $#1#2$\cr}}}
\catcode`\@=12 

\newcommand{\IP}{{\rm I$\kern-0.01667em$P}\xspace}

\newcommand{\tot}{{\rm tot}}

\mathchardef\qsm=63
\mathchardef\pls=43
\mathchardef\mns=512
\mathchardef\plm=518
\mathchardef\eql=61
\mathchardef\smallleft=300
\mathchardef\smallright=301
\mathchardef\les=316
\mathchardef\gre=318
\mathchardef\leq=532
\mathchardef\grq=533

\catcode`\@=11 
\newcounter{pict@width}
\newcounter{pict@height}
\newlength{\pict@scale}
\newcommand{\psfigadd}[4]{%
\setcounter{pict@width}{1*\ratio{#2+\pict@scale/2}{\pict@scale}}
\setcounter{pict@height}{1*\ratio{#3+\pict@scale/2}{\pict@scale}}
\setlength{\unitlength}{\pict@scale}
\hbox to #2{\hspace{-\fill}\begin{picture}(\thepict@width,\thepict@height)
\put(0,0){\psfig{figure=#1,width=#2,height=#3,clip=}}
\SetScale{0.283466457}
\SetWidth{1.763889}
{#4}
\end{picture}}
}
\newcounter{pict@widthfst}
\newcounter{pict@widthscd}
\newcounter{pict@widthtot}
\newcommand{\psfigaddtwo}[7]{%
\setcounter{pict@widthfst}{1*\ratio{#2+\pict@scale/2}{\pict@scale}}
\setcounter{pict@widthscd}{1*\ratio{#2+#4+\pict@scale/2}{\pict@scale}}
\setcounter{pict@widthtot}{1*\ratio{#2+#4+#6+\pict@scale/2}{\pict@scale}}
\setcounter{pict@height}{1*\ratio{#3+\pict@scale/2}{\pict@scale}}
\setlength{\unitlength}{\pict@scale}
\hbox{\hspace{-\fill}\begin{picture}(\thepict@widthtot,\thepict@height)
\put(0,0){\psfig{figure=#1,width=#2,height=#3,clip=}}
\put(\thepict@widthscd,0){\psfig{figure=#5,width=#6,height=#3,clip=}}
\SetScale{0.283466457}
\SetWidth{1.763889}
{#7}
\end{picture}}
}
\newcommand{\psfigror}[4]{%
\setcounter{pict@width}{1*\ratio{#2+\pict@scale/2}{\pict@scale}}
\setcounter{pict@height}{1*\ratio{#3+\pict@scale/2}{\pict@scale}}
\setlength{\unitlength}{\pict@scale}
\hbox{\begin{picture}(\thepict@width,\thepict@height)
\put(0,\thepict@height){\psfig{figure=#1,width=#3,height=#2,clip=,angle=270}}
\SetScale{0.283466457}
\SetWidth{1.763889}
{#4}
\end{picture}}
}
\newcommand{\psfigrol}[4]{%
\setcounter{pict@width}{1*\ratio{#2+\pict@scale/2}{\pict@scale}}
\setcounter{pict@height}{1*\ratio{#3+\pict@scale/2}{\pict@scale}}
\setlength{\unitlength}{\pict@scale}
\hbox{\begin{picture}(\thepict@width,\thepict@height)
\put(0,0){\psfig{figure=#1,width=#3,height=#2,clip=,angle=90}}
\SetScale{0.283466457}
\SetWidth{1.763889}
{#4}
\end{picture}}
}
\catcode`\@=12 
\newlength\listtextwidth

\catcode`\@=11 
\newlength{\@tabfninsert}
\newlength{\@tabfnwidth}
\newcommand{\tabfootnote}[2]{%
  \setlength{\@tabfninsert}{0.8em}
  \setlength{\@tabfnwidth}{\textwidth}
  \addtolength{\@tabfnwidth}{-\@tabfninsert}
  \addtolength{\@tabfnwidth}{-0.4em}
  \noindent\makebox[\@tabfninsert][r]{\footnotesize$^{#1}$\hfil}\hfill%
  \parbox[t]{\@tabfnwidth}{\footnotesize #2\hfill}}
\catcode`\@=12 

\begin{document}
\prepnum{DESY--08--000}
\newcommand{\pythia}{{\sc Pythia}}
\newcommand{\herwig}{{\sc Herwig}}
\newcommand{\jimmy}{{\sc Jimmy}}
\newcommand{\etcut}{\mbox{$E_T^{\rm CUT}$}}
\newcommand{\etgap}{\mbox{$E_T^{\rm GAP}$}}
\newcommand{\xg}{\mbox{$x_\gamma^{\rm OBS}$}}
\newcommand{\xp}{\mbox{$x_p^{\rm OBS}$}}
\newcommand{\GeV}{\mbox{\rm ~GeV}}
\newcommand{\pom} {I\hspace{-0.2em}P}
\newcommand{\xpom}{\mbox{$x_{_{\pom}}$}}
\newcommand{\reg} {I\hspace{-0.2em}R}
\def\citeCTD{{\cite{%
nim:a279:290,*npps:b32:181,*nim:a338:254%
}}\xspace}
\def\citeCAL{{\cite{%
nim:a309:77,*nim:a309:101,*nim:a321:356,*nim:a336:23%
}}\xspace}

\prepnum{DESY--08--011}

\title{
Deep inelastic inclusive and diffractive scattering at $Q^2$ values
from 25 to 320 GeV$^2$ with the ZEUS forward plug calorimeter }
                    
\author{ZEUS Collaboration}
\date{28 January 2008}

\abstract{
Deep inelastic scattering and its diffractive component, $ep \to
e^{\prime}\gamma^* p \to e^{\prime}XN$, have been studied at HERA with
the ZEUS detector using an integrated luminosity of 52.4
pb$^{-1}$. The $M_X$ method has been used to extract the diffractive
contribution. A wide range in the centre-of-mass energy $W$ (37 -- 245
GeV), photon virtuality $Q^2$ (20 -- 450 GeV$^2$) and mass $M_X$ (0.28
-- 35 GeV) is covered. The diffractive cross section for $2 < M_X <
15$ GeV rises strongly with $W$, the rise becoming steeper as $Q^2$
increases. The data are also presented in terms of the diffractive
structure function, $F^{\rm D(3)}_2$, of the proton. For fixed $Q^2$
and fixed $M_X$, $\xpom F^{\rm D(3)}_2$ shows a strong rise as $\xpom
\to 0$, where $\xpom$ is the fraction of the proton momentum carried
by the Pomeron.  For Bjorken-$x < 1 \cdot 10^{-3}$, $\xpom F^{\rm
D(3)}_2$ shows positive $\log Q^2$ scaling violations, while for $x
\ge 5 \cdot 10^{-3}$ negative scaling violations are observed. The
diffractive structure function is compatible with being leading
twist. The data show that Regge factorisation is broken.  }

\makezeustitle

\begin{center}                                                                                     
{                      \Large  The ZEUS Collaboration              }                               
\end{center}                                                                                       
  S.~Chekanov$^{   1}$,                                                                            
  M.~Derrick,                                                                                      
  S.~Magill,                                                                                       
  B.~Musgrave,                                                                                     
  D.~Nicholass$^{   2}$,                                                                           
  \mbox{J.~Repond},                                                                                
  R.~Yoshida\\                                                                                     
 {\it Argonne National Laboratory, Argonne, Illinois 60439-4815, USA}~$^{n}$                       
\par \filbreak                                                                                     
  M.C.K.~Mattingly \\                                                                              
 {\it Andrews University, Berrien Springs, Michigan 49104-0380, USA}                               
\par \filbreak                                                                                     
  M.~Jechow,                                                                                       
  N.~Pavel~$^{\dagger}$\\                                                                          
  {\it Institut f\"ur Physik der Humboldt-Universit\"at zu Berlin,                                 
           Berlin, Germany}~$^{b}$
% ...
\newpage                                                                                           
$^{\    1}$ supported by DESY, Germany \\                                                          
$^{\    2}$ also affiliated with University College London, UK \\                                  
% ...
$^{\dagger}$ deceased \\                                                                           
%
\newpage   
                                                           %
                                                           %
\begin{tabular}[h]{rp{14cm}}                                                                       
$^{a}$ &  supported by the Natural Sciences and Engineering Research Council of Canada (NSERC) \\  
$^{b}$ &  supported by the German Federal Ministry for Education and Research (BMBF), under        
          contract numbers 05 HZ6PDA, 05 HZ6GUA, 05 HZ6VFA and 05 HZ4KHA\\                         
$^{c}$ &  supported in part by the MINERVA Gesellschaft f\"ur Forschung GmbH, the Israel Science   
          Foundation (grant no. 293/02-11.2) and the U.S.-Israel Binational Science Foundation \\  
$^{d}$ &  supported by the German-Israeli Foundation and the Israel Science Foundation\\           
$^{e}$ &  supported by the Italian National Institute for Nuclear Physics (INFN) \\                
$^{f}$ &  supported by the Japanese Ministry of Education, Culture, Sports, Science and Technology 
          (MEXT) and its grants for Scientific Research\\                                          
$^{g}$ &  supported by the Korean Ministry of Education and Korea Science and Engineering          
          Foundation\\                                                                             
$^{h}$ &  supported by the Netherlands Foundation for Research on Matter (FOM)\\                   
$^{i}$ &  supported by the Polish State Committee for Scientific Research, grant no.               
          620/E-77/SPB/DESY/P-03/DZ 117/2003-2005 and grant no. 1P03B07427/2004-2006\\             
$^{j}$ &  partially supported by the German Federal Ministry for Education and Research (BMBF)\\   
$^{k}$ &  supported by RF Presidential grant N 8122.2006.2 for the leading                         
          scientific schools and by the Russian Ministry of Education and Science through its      
          grant for Scientific Research on High Energy Physics\\                                   
$^{l}$ &  supported by the Spanish Ministry of Education and Science through funds provided by     
          CICYT\\                                                                                  
$^{m}$ &  supported by the Science and Technology Facilities Council, UK\\                         
$^{n}$ &  supported by the US Department of Energy\\                                               
$^{o}$ &  supported by the US National Science Foundation. Any opinion,                            
findings and conclusions or recommendations expressed in this material                             
are those of the authors and do not necessarily reflect the views of the                           
National Science Foundation.\\                                                                     
$^{p}$ &  supported by the Polish Ministry of Science and Higher Education                         
as a scientific project (2006-2008)\\                                                              
$^{q}$ &  supported by FNRS and its associated funds (IISN and FRIA) and by an Inter-University    
          Attraction Poles Programme subsidised by the Belgian Federal Science Policy Office\\     
$^{r}$ &  supported by the Malaysian Ministry of Science, Technology and                           
Innovation/Akademi Sains Malaysia grant SAGA 66-02-03-0048\\                                       
\end{tabular}                                                                                      
                                                           %
                                                           %

\pagenumbering{arabic}\pagestyle{plain}
\section{Introduction}
\label{sec-int}
The observation of events with a large rapidity gap in deep inelastic
electron (positron) proton scattering (DIS) at HERA by the ZEUS
experiment~\cite{pl:b315:481} has paved the way for a systematic study
of diffraction at large centre-of-mass energies with a variable hard
scale provided by the mass squared, $-Q^2$, of the virtual
photon. Diffraction is defined by the property that the cross section
does not decrease as a power of the centre-of-mass energy. This can be
interpreted as the exchange of a colourless system, the Pomeron, which
leads to the presence of a large rapidity gap between the proton and
the rest of the final state, which is not exponentially suppressed.

Before HERA came into operation, Ingelman and
Schlein~\cite{pl:b152:256}, based on data from
UA8~\cite{pl:b211:239,pl:b297:417}, had suggested that the Pomeron may
have a partonic structure.  Since then, the H1 and ZEUS experiments at
HERA have presented results on diffractive scattering in
photoproduction and deep inelastic electron-proton scattering for many
different final states.  In parallel, a number of theoretical ideas
and models have been developed in order to understand the data within
the framework of Quantum Chromodynamics (QCD)~\cite{hep-ph-0611275}.

Several methods have been employed by H1 and ZEUS for isolating
diffractive contributions experimentally. In the case of exclusive
vector-meson production, resonance signals in the decay mass spectrum
combined with the absence of other substantial activity in the
detector have been used~\cite{np:b695:3,np:b718:3,pmc:a1:6}. The
contribution from inclusive diffraction has been extracted using the
presence of a large rapidity gap ($\eta_{\rm max}$
method~\cite{epj:c48:715}), the detection of the leading
proton~\cite{epj:c38:43,epj:c48:749} or the hadronic mass spectrum
observed in the central detector ($M_X$
method~\cite{zfp:c70:391,*epj:c6:43,np:b713:3}).  The selections based
on $\eta_{\rm max}$ or on a leading proton may include additional
contributions from Reggeon exchange. Such contributions are
exponentially suppressed when using the $M_X$ method.

In this paper, inclusive processes (Fig.~\ref{f:nondifdiag}),
\begin{eqnarray}
\gamma^* p \to \rm anything ,
\end{eqnarray}
and diffractive processes (Fig.~\ref{f:difdiag}), 
\begin{eqnarray}
\gamma^* p \to X N,
\end{eqnarray}
where $N$ is a proton or a low-mass nucleonic state and $X$ is the
hadronic system without $N$, are studied.  The contribution from
diffractive scattering is extracted with the $M_X$ method.  Results on
the proton structure function $F_2$ and on the diffractive cross
section and structure function are presented for a wide range of
centre-of-mass energies, photon virtualities $-Q^2$ and of mass $M_X$
of the diffractively produced hadronic system, using the data from the
ZEUS experiment collected in 1999 and 2000. The results, which will be
referred to as FPC II, are based on integrated luminosities of 11.0
pb$^{-1}$ for $Q^2 = 20 - 40$ GeV$^2$ and 52.4 pb$^{-1}$ for $Q^2 = 40
- 450$ GeV$^2$.

In a previous study, which will be referred to as
FPC~I~\cite{np:b713:3}, results on inclusive and diffractive
scattering were presented for the $Q^2$ values between 2.7 and 55
GeV$^2$ using an integrated luminosity of 4.2 pb$^{-1}$. The combined
data from the FPC~I and FPC~II analyses provide a measurement of the
$Q^2$ dependence of diffraction over a range of two orders of
magnitude.

\section{Experimental set-up and data set}
\label{sec-exp}
The data used for this measurement were taken with the ZEUS detector
in 1999-2000 when positrons of 27.5 GeV collided with protons of 920
GeV. The detector as well as the analysis methods are identical to
those used for the FPC~I study~\cite{np:b713:3}. A detailed
description of the ZEUS detector can be found
elsewhere~\cite{bluebook,pl:b293:465}. A brief outline of the
components that are most relevant for this analysis is given below.

Charged particles were tracked in the central tracking detector
(CTD)~\cite{nim:a279:290,*npps:b32:181,*nim:a338:254}, which operated
in a magnetic field of 1.43 T provided by a thin superconducting
solenoid. The CTD consisted of 72 cylindrical drift chamber layers,
organised in 9 superlayers covering the polar-angle region $15^{\circ}
< \theta < 164^{\circ}$. The transverse momentum resolution for
full-length tracks was $\sigma(p_T)/p_T = 0.0058 p_T \oplus 0.0065
\oplus 0.0014/p_T$, with $p_T$ in GeV.

The high-resolution uranium-scintillator calorimeter (CAL~\citeCAL)
consisted of three parts: the forward (FCAL), the barrel (BCAL) and
the rear (RCAL) calorimeters. Each part was subdivided transversely
into towers and longitudinally into one electromagnetic section (EMC)
and either one (in RCAL) or two (in BCAL and FCAL) hadronic sections
(HAC). The smallest division of the calorimeter was called a cell. The
CAL energy resolutions, as measured under test beam conditions, were
$\sigma(E)/E = 0.18/\sqrt(E)$ for electrons and $\sigma(E)/E =
0.35/\sqrt(E)$ for hadrons, with $E$ in GeV.

The position of electrons scattered at small angles to the
electron-beam direction was determined including the information from
the SRTD~\cite{nim:a401:63,epj:c21:443} which was attached to the
front face of the RCAL and consisted of two planes of scintillator
strips.  The rear hadron-electron separator (RHES~\cite{nim:a277:176})
was inserted in the RCAL.

In 1998, the forward-plug calorimeter (FPC)~\cite{nim:a450:235} was
installed in the $20 \times 20$ cm$^2$ beam hole of the FCAL. The FPC
was used to measure the energy of particles in the
pseudorapidity~\footnote{The ZEUS coordinate system is a right-handed
Cartesian system, with the $Z$ axis pointing in the proton direction,
referred to as the ``forward direction'', and the $X$ axis pointing
left towards the centre of HERA. The coordinate origin is at the
nominal interaction point.} range $\eta \approx 4.0 - 5.0$. The FPC
was a lead-scintillator sandwich calorimeter read out by
wavelength-shifter (WLS) fibres and photomultipliers (PMT). A hole of
3.15 cm radius was provided for the passage of the beams. In the FPC,
15 mm thick lead plates alternated with 2.6 mm thick scintillator
layers. The scintillator layers consisted of tiles forming towers that
were read out individually. The tower cross sections were $24 \times
24$ mm$^2$ in the electromagnetic and 48$ \times 48$ mm$^2$ in the
hadronic section. The measured energy resolution for positrons was
$\sigma_E/E = (0.41 \pm 0.02)/\sqrt{E} \oplus 0.062 \pm 0.002$, with
$E$ in GeV. When installed in the FCAL, the energy resolution for
pions was $\sigma_E/E = (0.65 \pm 0.02)/\sqrt{E} \oplus 0.06 \pm
0.01$, with $E$ in GeV, and the $e/h$ ratio was close to unity.

The luminosity was measured from the rate of the bremsstrahlung
process $ep \to e \gamma p$. The resulting small-angle energetic
photons were measured by the luminosity monitor~\cite{zfp:c63:391}, a
lead-scintillator calorimeter placed in the HERA tunnel at $Z = -107$
m.

A three-level trigger system was used to select events
online~\cite{bluebook,pl:b293:465,uprocn:chep:1992:222}. The first-
and second-level trigger selections were based on the identification
of a scattered positron with impact point on the RCAL surface outside
an area of 36$\times$36 cm$^2$ centred on the beam axis (``set 1'',
integrated luminosity 11.0 pb$^{-1}$), or outside a radius of 30 cm
centred on the beam axis (``set 2'', integrated luminosity 41.4
pb$^{-1}$). In the offline analysis the reconstructed impact point had
to lie outside an area of 40$\times$40 cm$^2$ (set 1) or outside a
radius of 32 cm (set 2).

\section{Reconstruction of kinematics and event selection}
\label{sec-reconkinem}

The methods for extracting the inclusive DIS and diffractive data
samples are identical to those applied in the FPC~I
study~\cite{np:b713:3} and will be described only briefly.

The reaction
\begin{eqnarray}
e(k) \; p(P) \to e(k^{\prime}) + \rm anything, \nonumber
\end{eqnarray}
see Fig.~\ref{f:nondifdiag}, at fixed squared centre-of-mass energy,
$s = (k+P)^2$, is described in terms of $Q^2 \equiv -q^2 = -(k
-k^{\prime})^2$, Bjorken-$x = Q^2/(2 P \cdot q)$ and $s \approx 4 E_e
E_p$, where $E_e$ and $E_p$ denote the positron and proton beam
energies, respectively. For these data, $\sqrt{s} = 318$ GeV. The
fractional energy transferred to the proton in its rest system is $y
\approx Q^2/(sx)$. The centre-of-mass energy of the hadronic final
state, $W$, is given by $W^2 = [P+q]^2 = m^2_p+Q^2(1/x-1)\approx Q^2/x
= ys$, where $m_p$ is the mass of the proton.

In diffraction, proceeding via
\begin{eqnarray} 
\gamma^* \; p(P) \to X + N(N),
\end{eqnarray} 
see Fig.~\ref{f:difdiag}, the incoming proton undergoes a small
perturbation and emerges either intact ($N=p$), or as a low-mass
nucleonic state $N$, in both cases carrying a large fraction, $x_L$,
of the incoming proton momentum. The virtual photon dissociates into a
hadronic system $X$.

Diffraction is parametrised in terms of the mass $M_X$ of the system
$X$, and the mass $M_N$ of the system $N$. Since $t$, the
four-momentum transfer squared between the incoming proton and the
outgoing system $N$, $t = (P-N)^2$, was not measured, the results
presented are integrated over $t$. The measurements performed by ZEUS
with the leading proton spectrometer~\cite{epj:c38:43} show that the
diffractive contribution has a steeply falling $t$ distribution with
typical $|t|$ values well below 0.5 GeV$^2$.

Diffraction was also analysed in terms of the momentum fraction
$\xpom$ of the proton carried by the Pomeron exchanged in the
$t$-channel, $\xpom = [(P - N) \cdot q]/(P \cdot q ) \approx
(M^2_X+Q^2)/(W^2 + Q^2)$, and the fraction of the Pomeron momentum
carried by the struck quark, $\beta = Q^2/[2(P - N) \cdot q ]\approx
Q^2 / (M^2_X+Q^2)$.  The variables $\xpom$ and $\beta$ are related to
the Bjorken scaling variable, $x$, via $x = \beta \xpom$.

The events studied are of the type
\begin{eqnarray}
e p \to e^{\prime} X \; + \; {\rm rest},
\label{eq:xobs}
\end{eqnarray}
where $X$ denotes the hadronic system observed in the detector and
`rest' the particle system escaping detection through the forward
and/or rear beam holes.

The coordinates of the event vertex were determined with tracks
reconstructed in the CTD. Scattered positrons were identified with an
algorithm based on a neural network~\cite{nim:a365:508}. The direction
and energy of the scattered positron were determined from the combined
information given by CAL, SRTD, RHES and CTD.  Fiducial cuts on the
impact point of the reconstructed scattered positron on the CAL
surface were imposed to ensure a reliable measurement of the positron
energy.

The hadronic system was reconstructed from energy-flow objects
(EFO)~\cite{goebel:2001,gennady} which combine the information from
CAL and FPC clusters and from CTD tracks, and which were not assigned
to the scattered positron.
  
If a scattered-positron candidate was found, the following criteria
were imposed to select the DIS events:
\begin{itemize}
\item the scattered-positron energy, $E_e^{\prime}$, be at least 10 GeV;
\item the total measured energy of the hadronic system be at least 400 MeV;
\item $y^{\rm FB}_{\rm JB} > 0.006$, where $y^{\rm FB}_{\rm JB} = \sum_h (E_h - P_{Z,h})/(2E_e)$ summed over all hadronic EFOs in FCAL plus BCAL; or at least 400 MeV be deposited in the BCAL or in the RCAL outside of the ring of towers closest to the beamline;
\item $-54 < Z_{\rm vtx} < 50$ cm, where $Z_{\rm vtx}$ is the $Z$-coordinate of the event vertex;
\item $43 < \sum_{i=e,h} (E_i - P_{Z,i}) < 64$ GeV, where the sum runs over both the scattered positron and all hadronic EFOs. This cut reduces the background  from photoproduction and beam-gas scattering and removes events with large initial-state QED radiation;
\item candidates for QED-Compton (QEDC) events, consisting of a scattered-positron candidate and a photon candidate, with mass $M_{e \gamma}$ less than 0.25 GeV and total transverse momentum less than $1.5$ GeV, were removed. A Monte Carlo (MC) study showed that the number of remaining QEDC events was negligible. 
\end{itemize}

The value of $Q^2$ was reconstructed from the measured energy
$E^{\prime}_e$ and scattering angle $\theta_e$ of the positron, $Q^2 =
2 E_e E^{\prime}_e (1 + \cos \theta_e)$.

In the FPC~I analysis, which covered lower $Q^2$ values, the value of
$W$ was determined as the weighted average of the values given by the
positron and hadron measurement. Here, the value of $W$ was
reconstructed with the double-angle algorithm
(DA)~\cite{proc:HERA:1991:23,*add:1991:43} which relies only on the
measurement of the angles of the scattered positron and of the
hadronic system.

The mass of the system $X$ was determined by summing over all hadronic
EFOs,
\begin{eqnarray}
M^2_X = \left(\sum E_h \right)^2 - \left(\sum p_{X,h} \right)^2 - \left(\sum p_{Y,h} \right)^2 - \left (\sum p_{Z,h} \right)^2 ,                       \nonumber
\label{eq:mxcalc}
\end{eqnarray}
where $P_h = (p_{X,h},p_{Y,h},p_{Z,h},E_{h})$ is the four-momentum vector of a hadronic EFO. All kinematic variables used to describe inclusive and diffractive scattering were derived from $M_X$, $W$ and $Q^2$.

A total of 60 events were found without a vertex, which were due
either to cosmic radiation (45) or to an overlay of cosmic radiation
with DIS (15); these events were discarded.

About 630k events for data set~1 and 1.4M events for data set~2 passed
the selection cuts. The kinematic range for inclusive and diffractive
events was chosen taking into account detector resolution and
statistics. About 930k events were retained which satisfied $37 < W <
245$ GeV and $20 < Q^2 < 450$ GeV$^2$.

The resolutions of the reconstructed kinematic variables were
estimated using MC simulation of diffractive events of the type
$\gamma^*p \to {X} N$ (see Section~\ref{sec-models}). For the $M_{X}$,
$W$ and $Q^2$ bins considered in this analysis, the resolutions are
approximately the same as for the FPC~I analysis: $\frac{\sigma
(W)}{W} = \frac{1}{W^{1/2}}$, $\frac{\sigma(Q^2)}{Q^2} =
\frac{0.25}{(Q^2)^{1/3}}$ and $\frac{\sigma (M_X)}{M_X} =
\frac{c}{M_X^{1/3}}$, where $c = 0.6$ GeV$^{1/3}$ for $M_X <$ 1 GeV
and $c = 0.4$ GeV$^{1/3}$ for $M_X \ge 1$ GeV, with $M_X, W$ in units
of GeV and $Q^2$ in GeV$^2$.

Results are presented for seven bins in $W$, nine bins in $Q^2$ and
six bins in $M_X$, as shown in Table~\ref{t:binning}.

\section{Monte Carlo simulations}
\label{sec-models}
The data were corrected for detector acceptance and resolution, and
for radiative effects, with suitable combinations of several MC
models, following the same procedure and using the same MC models as
in the FPC~I~\cite{np:b713:3} analysis.

Events from inclusive DIS, including radiative effects, were simulated
using the HERACLES 4.6.1~\cite{cpc:69:155}
program with the DJANGOH 1.1~\cite{cpc:81:381} interface to ARIADNE
4~\cite{cpc:71:15} and using the CTEQ4D next-to-leading-order parton
distribution functions~\cite{pr:d55:1280}. In HERACLES, $O(\alpha)$
electroweak corrections are included. The colour-dipole model of
ARIADNE, including boson-gluon fusion, was used to simulate the
$O(\alpha_S)$ plus leading-logarithmic corrections to the quark-parton
model. The Lund string model as implemented in JETSET
7.4~\cite{cpc:82:74} was used by ARIADNE for hadronisation.

Diffractive DIS in which the proton does not dissociate, $ep \to eXp$
(including the production of $\omega$ and $\phi$ mesons via $ep \to e
V p$, $V = \omega, \phi$ but excluding $\rho^0$ production), were
simulated with SATRAP,~which is based on a saturation
model~\cite{pr:d59:014017,*pr:d60:114023} and is interfaced to the
RAPGAP 2.08 framework~\cite{cpc:86:147}. SATRAP was reweighted as a
function of $Q^2/(Q^2+M_X^2)$ and $W$. The production of $\rho^0$
mesons, $ep \to e \rho^0 p$, was simulated with
ZEUSVM~\cite{zeusvm:1996}, which uses a parametrisation of the
measured $\rho^0$ cross sections as well as of the production and
decay angular
distributions~\cite{epj:c6:603,epj:c12:393,pmc:a1:6}.  The QED
radiative effects were simulated with HERACLES. The QCD parton showers
were simulated with LEPTO 6.5~\cite{cpc:101:108}.

Diffractive DIS in which the proton dissociates, $ep \to eXN$, was
simulated with SATRAP interfaced to the model called
SANG~\cite{helim:2002},
which also includes the production of $\rho^0$ mesons. Following the
previous experience (FPC~I), the mass spectrum of the system $N$ was
generated according to $d\sigma /dM^2_N \propto (1/M^2_N) \times
0.89\sqrt{M_N/4}$ for $M_N \le 4$ GeV, and $d\sigma /dM^2_N \propto
(1/M^2_N) \times (2.5/M_N)^{0.25}$ for $M_N > 4$ GeV. This
parametrisation was found to fit the data in the FPC I analysis. The
fragmentation of the system $N$ was simulated using JETSET 7.4.
 
The parameters of SANG, in particular those determining the shape of
the $M_N$ spectrum and the overall normalisation, were checked with a
subset of the data. Events in this subset were required to have a
minimum rapidity gap $\Delta \eta > \eta_{\rm min}$ between at least
one EFO and its nearest neighbours, all with energies greater than 400
MeV. Good sensitivity for double dissociation was obtained with four
event samples for the kinematic regions $\eta_{\rm min}$ = 3.0 , $W =
55 - 135$ GeV, and $\eta_{\rm min}$ = 4.0, $W = 135 - 245$ GeV, for
$Q^2 = 40 - 80$ GeV$^2$ and $80 - 450$ GeV$^2$. The mass of the
hadronic system reconstructed from the energy deposits in FPC+FCAL,
$M_{\rm FFCAL}$, depends approximately linearly on the mass $M^{\rm
gen}_{N}$ of the generated system $N$. Thus, the $M_{\rm FFCAL}$
distribution is sensitive to those proton dissociative events in which
considerable energy of the system $N$ is deposited in FPC and
FCAL. The study showed that this is the case, broadly speaking, when
the mass of $N$ taken at the generator level is $M_{N} > 2.3$ GeV. At
large $M_{\rm FFCAL}$, the distribution is dominated by double
dissociation. Figure~\ref{f:mffcalh} presents the $M_{\rm FFCAL}$
distributions in four representative ($Q^2,W$) regions for the data
compared to the Monte Carlo expectations for $Xp$, $\rho^0p$, $XN$ and
non-diffractive processes. The contribution expected from $XN$ as
predicted by SANG is dominant. Good agreement between the number of
events measured and those predicted is obtained. Since the
contribution from diffraction with $M_{N} > 2.3$ GeV can affect the
determination of the slope $b$ for the non-diffractive contribution
(see Section~\ref{sec-lnmxmethod}) it was subtracted statistically
from the data as a function of $M_X$, $W$ and $Q^2$.

Background from photoproduction, estimated with PYTHIA
5.7~\cite{cpc:82:74}, was negligible and was neglected.

The ZEUS detector response was simulated using a program based on
GEANT 3.13~\cite{tech:cern-dd-ee-84-1}. The generated events were
passed through the detector and trigger simulations and processed by
the same reconstruction and analysis programs as the data.

The measured hadronic energies for data and MC were increased by a
factor of 1.065 in order to achieve an average transverse momentum
balance between the scattered positron and the hadronic system. The
mass $M_X$ reconstructed from the energy-corrected EFOs, in the $M_X$
region analysed, required an additional correction factor of 1.10 as
determined from MC simulation~\footnote{The hadrons produced in
diffractive events, on average, have lower momenta than those for
hadrons from non-peripheral events, so that their fractional energy
loss in the material in front of the calorimeter is larger.}.

Good agreement between data and simulated event distributions was
obtained for both the inclusive and the diffractive samples.

\section{Determination of the diffractive contribution}
\label{sec-lnmxmethod}
The diffractive contribution was extracted from the data using the
$M_X$ method, which has been described
elsewhere~\cite{zfp:c70:391,*epj:c6:43} and which has also been used
for the FPC~I analysis~\cite{np:b713:3}.

In the QCD picture of non-peripheral DIS, $\gamma^*p \to X \; + \;
{\rm rest}$, the hadronic system $X$ measured in the detector is
related to the struck quark and `rest' to the proton remnant, both of
which are coloured states.  The final-state particles are expected to
be uniformly emitted in rapidity along the $\gamma^*p$ collision axis
and to uniformly populate the rapidity gap between the struck quark
and the proton remnant, as described
elsewhere~\cite{feynman:1972:photon}. In this case, the mass $M_X$ is
distributed as
\begin{eqnarray}
\frac{d{\cal N}^{\rm non-diff}}{d \ln M^2_X} = c \cdot \exp(b \cdot \ln M^2_X),
\label{eq:lnmxsquare}
\end{eqnarray}
where $b$ and $c$ are constants~\footnote{Throughout, whenever a
logarithm of a quantity with dimensions of energy is used, a
normalisation in units of GeV is implied. For example, $\ln M^2_X$ is
defined as $\ln (M^2_X/M^2_0)$, where $M_0 = 1$ GeV.}. DJANGOH
predicts, for non-peripheral DIS, $b \approx 1.9$.

The diffractive reaction, $\gamma^*p \to X N$, on the other hand, has
different characteristics. Diffractive scattering shows up as a peak
near $x_L = 1$, the mass of the system $X$ being limited by kinematics
to $M^2_X/W^2 \stackrel{<}{\sim} 1 - x_L$. Moreover, the distance in
rapidity between the outgoing nucleon system $N$ and the system $X$ is
$\Delta \eta \approx \ln(1/(1-x_L))$,
becoming large when $x_L$ is close to one. Combined with the limited
values of $M_X$ and the peaking of the diffractive cross section near
$x_L = 1$, this leads to a large separation in rapidity between $N$
and any other hadronic activity in the event. For the vast majority of
diffractive events, the decay particles from the system $N$ leave
undetected through the forward beam hole. For a wide range of $M_X$
values, the particles of the system $X$ are emitted entirely within
the acceptance of the detector. Monte Carlo studies show that $X$ can
be reliably reconstructed over the full $M_X$ range of this analysis.

Regge phenomenology predicts the shape of the $M_X$ distribution for
peripheral processes. Diffractive production by Pomeron exchange in
the $t$-channel, which dominates $x_L$ values close to unity, leads to
an approximately constant $\ln M^2_X$ distribution ($b \approx
0$). Figure~\ref{f:lnmxsel} shows distributions of $\ln M^2_X$ for the
data (from which the contribution from double dissociation with $M_N >
2.3$ GeV, as predicted by SANG, has been subtracted) for low- and
high-$W$ bins at low and high $Q^2$ together with the expectations
from MC simulations for non-peripheral DIS (DJANGOH) and for
diffractive processes (SATRAP + ZEUSVM and SANG for $M_N < 2.3$
GeV). The observed distributions agree well with the expectation for a
non-diffractive component giving rise to an exponentially growing $\ln
M^2_X$ distribution, and for a diffractive component producing an
almost constant distribution in a substantial part of the $\ln M^2_X$
range. At high $W$ there is a clear signal for a contribution from
diffraction. At low $W$ the diffractive contribution is seen to be
small.

The $\ln M^2_X$ spectra for all ($W,Q^2$) bins studied in this
analysis are displayed in Fig.~\ref{f:lnmxall}. They are compared with
the MC predictions for the contributions from non-peripheral and
diffractive production. The MC simulations are in good agreement with
the data. It can be seen that the events at low and medium values of
$\ln M^2_X$ originate predominantly from diffractive production.

The assumption of an exponential rise of the $\ln M^2_X$ distribution
for non-diffractive processes permits the subtraction of this
component and, therefore, the extraction of the diffractive
contribution without assumptions about its exact $M_X$ dependence. The
distribution is of the form:
\begin{eqnarray}
\frac{dN}{d\ln M^2_X} = D + c \cdot \exp(b \; \ln M^2_X)\; \; \; {\rm for}  \;  \ln {M^2_X < \ln W^2 - \eta_0}, 
\label{eq:lnmxdist}
\end{eqnarray}
with $M_X$ in GeV, $D$ is the diffractive contribution and the second
term represents the non-diffractive contribution. The quantity $(\ln
W^2 - \eta_0)$ specifies the maximum value of $\ln M^2_X$ up to which
the exponential behaviour of the non-diffractive contribution holds. A
value of $\eta_0 = 2.2$ was found from the data. Equation
(\ref{eq:lnmxdist}) was fitted to the data in the limited range $\ln
W^2 - 4.4 < \ln M^2_X < \ln W^2 - \eta_0$ in order to determine the
parameters $b$ and $c$. The parameter $D$ was assumed to be constant
over the fit range, which is suggested by Figs.~\ref{f:lnmxsel}
and~\ref{f:lnmxall} where at high $W$ and high $Q^2$, $dN/\ln M^2_X$
is a slowly varying function when $M_X^2 >
Q^2$~\cite{jetp:81:625,np:b303:634}. However, the diffractive
contribution was not taken from the fit but was obtained from the
observed number of events after subtracting the non-diffractive
contribution determined using the fitted values of $b$ and $c$.

The fit range chosen is smaller than that used for the FPC~I analysis
(viz. for FPC~I: $\ln W^2 - 5.6 < \ln M^2_X < \ln W^2 - 2.2 $). This
change takes account of the observation that at high $Q^2$ and low
values of $M_X$ diffraction is suppressed, as seen in
Fig.~\ref{f:lnmxall}.

The non-diffractive contribution in the ($W, Q^2$) bins was determined
by fitting for every ($W, Q^2$) interval the $\ln M^2_X$ distribution
of the data from which the contribution of $\gamma^*p \to XN$ with
$M_N > 2.3$ GeV as given by SANG, has been subtracted (see
Appendix~\ref{asec-protondissoc} and Tables~\ref{t:fracnondif1}
and~\ref{t:fracnondif2}). A fit of the form of Eq.~(\ref{eq:lnmxdist})
treating $b,c$ and $D$ as fit variables, was used. Note that this is a
difference compared to the FPC I analysis, where for each $(W,Q^2)$
interval, the same value of $b$, obtained as an average over all $W$,
$Q^2$ values, was used. Good fits with $\chi^2$ per degree of freedom
of about unity were obtained. The value of the slope $b$ varied
typically between 1.4 and 1.9. The statistical error of the
diffractive contribution includes the uncertainties on $b$ and $c$.

Only bins of $M_X, W, Q^2$, for which the non-diffractive background
was less than 50\%, were kept for further analysis.

The $M_X$ method used for extracting the diffractive contribution was
tested by performing a ``Monte Carlo experiment'' in which a sample of
simulated non-peripheral DIS events (DJANGOH) and diffractive events
with (SATRAP+ZEUSVM+SANG) and without proton dissociation (SATRAP +
ZEUSVM) was analysed as if it were the data. The resulting diffractive
structure function (as defined in Section~\ref{sec-difff2d} below) is
shown in Fig.~\ref{f:xpfd3vsxpgw} as a function of $\xpom$ for the
$\beta$ and $Q^2$ values used in the analysis. Only the statistical
uncertainties are shown. The extracted structure function agrees with
the diffractive structure function used for generating the events
which validates the self consistency of the analysis procedure.

The extraction of the diffractive contribution was also studied for
the case of a possible contribution from Reggeon exchange interfering
with the contribution from diffraction. The amount of Reggeon-Pomeron
interference allowed by the data~\cite{epj:c38:43} was found to be
smaller than the combined statistical and systematic uncertainties in
the present measurement, see Appendix~\ref{asec-reggeontest}.

\section{Evaluation of cross sections and systematic uncertainties}
\label{sec-evalsig}
The total and diffractive cross sections for $ep$ scattering in a
given ($W,Q^2$) bin were determined from the integrated luminosity,
the number of observed events corrected for background, acceptance and
smearing, and corrected to the QED Born level.

The cross sections and structure functions are presented at chosen
reference values $M_{X {\rm ref}}$, $W_{\rm ref}$ and $Q^2_{\rm
ref}$. This was achieved as follows: first, the cross sections and
structure functions were determined at the weighted average of each
($M_X$, $W$, $Q^2$) bin. They were then transported to the reference
position using the ZEUS NLO QCD fit~\cite{pr:d67:012007} in the case
of the proton structure function $F_2$, and the result of the
BEKW(mod) fit (see Section~\ref{sec-bekw}) for the diffractive cross
sections and structure functions. The resulting changes to the cross
section and structure function values from the average to those at the
reference positions were at the 5 -- 15\% level.

\subsection{Systematic uncertainties} 
\label{subsec-sys}
A study of the main sources contributing to the systematic
uncertainties of the measurements was performed. The systematic
uncertainties were calculated by varying the cuts or modifying the
analysis procedure and repeating the full analysis for every
variation. The size of the variations of cuts and the changes of the
energy scales were chosen commensurate with the resolutions or the
uncertainties of the relevant variables:
\begin{itemize}
\item the acceptance depends on the position measurement of the scattered positron. For set 1 the cut was increased from $40\times 40{\;\rm cm^2}$ to $41\times 41{\;\rm cm^2}$ (systematic uncertainty 1a) and decreased to $39 \times 39{\;\rm cm^2}$ (systematic uncertainty 1b). For set 2, the radius cut was increased from $32{\;\rm cm}$ to $33{\;\rm cm}$ (systematic uncertainty 1a) and decreased to $31{\;\rm cm}$ (systematic uncertainty 1b). This affected the low-$Q^2$ region. Changes below 1\% were observed; 

\item the measured energy of the scattered positron was increased (decreased) by 2\% in the data, but not in the MC (systematic uncertainties 2a,b). In most cases the changes were smaller than 1\%. For a few bins changes up to 3\% were observed. For one bin at high $Q^2$ and high $W$, a change of 7\% was found;

\item the lower cut for the energy of the scattered positron was lowered to 8 GeV (raised to 12 GeV) (systematic uncertainties 3a,b). In most cases the changes were smaller than 1\%. For a few bins changes up to 3\% were found. For one bin at high $Q^2$ and high $W$, a change of 7\% was found;

\item to estimate the systematic uncertainties due to the uncertainty in the hadronic energy, the analysis was repeated after increasing (decreasing) the hadronic energy measured by the CAL by 2\%~\cite{epj:c21:443} in the data but not in MC (systematic uncertainties 4a,b). The changes were below 3\%;

\item the energies measured by the FPC were increased (decreased) by 10\% in the data but not in MC (systematic uncertainties 5a,b). The changes were below 1\%; 
\item to estimate the uncertainties when the hadronic system $h$ is in one of the transition regions: beam/(FPC+FCAL) (polar angle of the hadronic system $\theta_h < 8^{\circ}$);  FCAL/BCAL ($27^{\circ} < \theta_h < 40^{\circ}$) or BCAL/RCAL ($128^{\circ} < \theta_h < 138^{\circ}$), the energy of $h$ was increased in the data by $10\%$ but not in MC (systematic uncertainty 6). This led to changes below 1\%;

\item the minimum hadronic energy cut of 400 MeV as well as the cut $y_{\rm JB} > 0.006$ were increased by 50\% (systematic uncertainty 7). In most cases the changes were below 1\%. For a few bins at $Q^2 \le 35$ GeV$^2$, changes up to 3\% were found;

\item in order to check the simulation of the hadronic final state, the selection on $\sum_{i=e,h}(E_i - P_{Z,i})$ was changed from 43 -- 64 GeV to 35 -- 64 GeV (systematic uncertainty 8), leading for $Q^2 = 25, 35$ GeV$^2$ to maximum changes at the level of 4\%, and to changes up to 6\% for $Q^2 = 320$ GeV$^2$.
\end{itemize}

The above systematic tests apply to the total as well as to the
diffractive cross sections. The following systematic tests apply to
the diffractive cross section only:
\begin{itemize} 
 \item the reconstructed mass $M_X$ of the system ${\cal }X$ was increased (decreased) by 5\% in the data but not in the MC (systematic uncertainties 9a,b). Changes below 1\% were observed except for $Q^2 = 25, 35$ GeV$^2$, where decreasing $M_X$ led to changes up to 5\% at high $y$;

\item the contribution from double dissociation with $M_N > 2.3$ GeV was determined with the reweighted SANG simulation and was subtracted from the data. The diffractive cross section was redetermined by increasing (decreasing) the predicted contribution from SANG by 30\% (systematic uncertainties 10a,b). The resulting changes in the diffractive cross section were well below the statistical uncertainty;

\item the slope $b$ describing the $\ln M^2_X$ dependence of the non-diffractive contribution (see Eq.~(\ref{eq:lnmxdist})) was increased (decreased) by 0.2 units (systematic uncertainties 11a,b); this led to an increase (decrease) of the number of diffractive events for the highest $M_X$ value  at a given $W, Q^2$ by 1 (1.5) times the size of the statistical uncertainty. For the lower $M_X$ values the changes were smaller.
\end{itemize}

The uncertainty in the luminosity measurement was $2\%$ and was neglected. The major sources of systematic uncertainties for the diffractive cross section, $d\sigma^{\rm diff}/dM_X$, were found to be the uncertainties 4a,b; 8; 9a,b, 10a,b; and 11a,b for the largest $M_X$ value at a given value of $W$. The total systematic uncertainty for each bin was determined by adding the individual contributions in quadrature.

\section{Proton structure function $F_2$ and the total $\gamma^*p$ cross section}
\label{sec-f2sigtot}
The differential cross section for inclusive $ep$ scattering mediated by virtual photon exchange is given in terms of the structure functions $F_i$ of the proton by
\begin{eqnarray}
\frac{d^2 \sigma^{e^+ p}}{dx dQ^2} = \frac{2 \pi \alpha^2}{x Q^4} [Y F_2(x,Q^2) - y^2 F_L(x,Q^2)](1 + \delta_r(x,Q^2)),
\end{eqnarray}
where $Y = 1 + (1-y)^2$, $F_2$ is the main component of the cross
section which in the DIS factorisation scheme corresponds to the sum
of the momentum densities of the quarks and antiquarks weighted by the
squares of their charges, $F_L$ is the longitudinal structure function
and $\delta_r$ is a term accounting for radiative corrections.

In the $Q^2$ range considered in this analysis, $Q^2 \le 450$ GeV$^2$,
the contributions from $Z^0$ exchange and $Z^0$ - $\gamma$
interference are at most of the order of 0.4\%
and were ignored. The contribution of $F_L$ to the cross section
relative to that from $F_2$ is given by $(y^2/Y) \cdot (F_L/F_2)$. For
the determination of $F_2$, the $F_L$ contribution was taken from the
ZEUS NLO QCD fit~\cite{pr:d67:012007}. The contribution of $F_L$ to
the cross section in the highest $y$ (= lowest $x$) bin of this
analysis was 3.2\%, decreasing to 1.3\% for the next highest $y$
bin. For the other bins, the $F_L$ contribution is below 1\%. The
resulting uncertainties on $F_2$ are below 1\%.
 
The measured $F_2$ values are listed in Table~\ref{t:f2tab}, and are
shown in Fig.~\ref{f:fpcf2lh} together with those from the FPC~I
analysis. Here, the $F_2$ values of FPC~I measured at $Q^2 = 27$
GeV$^2$ were transported to $Q^2 = 25$ GeV$^2$. Good agreement is
observed between the measurements done at the same values of $Q^2$,
namely 25 and 55 GeV$^2$. The data are compared to the predictions of
the ZEUS NLO QCD fit~\cite{pr:d67:012007} obtained from previous ZEUS
$F_2$ measurements~\cite{epj:c21:443}. The fit describes the data
well.

The proton structure function, $F_2$, rises rapidly as $x \to 0$ for
all values of $Q^2$, the slope increasing as $Q^2$ increases.  The
form
\begin{eqnarray}
F_2 = c \cdot x^{-\lambda}
\label{eq:f2fitlambda} 
\end{eqnarray}
was fitted for every $Q^2$ bin to the $F_2$ data, requiring $x < 0.01$
to exclude the region where valence quarks may dominate. Since, for
fixed $Q^2$, the $x$ dependence of $F_2$ is related to the $W$
dependence of the total $\gamma^* p$ cross section, the power
$\lambda$ can be related to the intercept of the Pomeron trajectory,
$\lambda = \alpha_{\pom}(0) -1$ (see
Section~\ref{subsec-wdepdiff}). For later comparison with the
diffractive results, these $\alpha_{\pom}$ values will be referred to
as $\alpha^{\rm tot}_{\pom}$.  The resulting values for $c$ and
$\alpha^{\rm tot}_{\pom}(0)$ are listed in Table~\ref{t:f2fit}.
Figure~\ref{f:apomcomb} presents the results from this study together
with those from the FPC~I analysis. The parameter $\alpha^{\rm
tot}_{\pom}(0)$ rises approximately linearly with $\ln Q^2$ from
$\alpha^{\rm tot}_{\pom}(0) = 1.155 \pm 0.011({\rm
stat.})^{+0.007}_{-0.011}({\rm syst.})$ at $Q^2 = 2.7$ GeV$^2$, to
$1.322 \pm 0.017$ (statistical and systematic uncertainties added in
quadrature) at $Q^2 = 70$ GeV$^2$, substantially above the `soft
Pomeron' value of $1.096^{+0.012}_{-0.009}$ deduced from hadron-hadron
scattering data~\cite{np:b244:322,*pl:b296:227,*pl:b395:311}. This is
in agreement with previous
observations~\cite{np:b713:3,epj:c7:609,pl:b520:183}. Since the
Pomeron intercept is changing with $Q^2$, the assumption of single
Pomeron exchange cannot be sustained.
 
The total cross section for virtual photon-proton scattering,
$\sigma^{\rm tot} _{\gamma^{\ast} p} \equiv \sigma_T(x,Q^2) +
\sigma_L(x,Q^2)$, where $T(L)$ stands for transverse (longitudinal)
photons, was extracted from the measurement of $F_2$ using the
relation
\begin{eqnarray}
\sigma^{\rm tot} _{\gamma^{\ast} p} = \frac{4 \pi^2 \alpha}{Q^2(1-x)} F_2(x,Q^2),
\end{eqnarray}
which is valid for $4m^2_p x^2 \ll Q^2$~\cite{pr:129:1834}.
The total cross section values are listed in Table~\ref{t:sigtot} for fixed $Q^2$ as a function of $W$. 

The total cross section multiplied by $Q^2$ is shown in
Fig.~\ref{f:sigtotlh} together with the results from the FPC~I
analysis. For fixed value of $Q^2$, $Q^2 \sigma^{\rm tot}
_{\gamma^{\ast} p}$ rises rapidly with $W$. For $Q^2 \le 14$ GeV$^2$,
the rise becomes steeper with increasing $Q^2$, while for $Q^2 \ge 70$
GeV$^2$ the rise becomes less steep as $Q^2$ increases. The $W$
behaviour of $\sigma^{\rm tot}_{\gamma^*p}$ reflects the $x$
dependence of $F_2$ as $x \to 0$, viz. $\sigma^{\rm tot}_{\gamma^*p}
\propto W^{2(\alpha^{\rm tot}_{\pom}(0)-1)}$.

\section{Diffractive cross section}
\label{sec-sigdiff}
The cross section for diffractive scattering via $ep \to eXN$ can be
expressed in terms of the transverse ({\em T}) and longitudinal ({\em
L}) cross sections, $\sigma^{\rm diff}_T$ and $\sigma^{\rm diff}_L$,
for $\gamma^* p \to XN$ as\\
\begin{eqnarray}
\frac{d\sigma^{\rm diff}_{\gamma^* p \to XN}(M_X,W, Q^2)}{dM_X} \equiv \frac{d(\sigma^{\rm diff}_T + \sigma^{\rm diff}_L)}{dM_X} \approx \frac{2 \pi}{\alpha}\frac{Q^2}{(1-y)^2+1}\frac{d\sigma^{\rm diff}_{ep \to eXN}(M_X,W,Q^2)}{dM_X d\ln W^2 dQ^2} .
\end{eqnarray}
Here, a term $(1 - y^2 /[1+(1-y)^2])\sigma^{\rm diff}_L/(\sigma^{\rm diff}_T + \sigma^{\rm diff}_L)$
multiplying $(\sigma^{\rm diff}_T + \sigma^{\rm diff}_L)$ has been 
neglected~\cite{np:b713:3,pr:129:1834,anphy:28:18,pr:167:1365}.
Since $y = W^2/s$, this approximation reduces the diffractive cross
section for $M_X < 2$ GeV by at most 8\% at $W < 200$ GeV, and by at
most 22\% in the highest $W$ bin, 200 -- 245 GeV, under the assumption
that only longitudinal photons contribute. Since the reduction is
always smaller than the total uncertainty of the diffractive cross
section given by the statistical and systematic uncertainties added in
quadrature: no correction was applied.

\subsection{$W$ dependence of the diffractive cross section}
\label{subsec-wdepdiff}
The diffractive cross section $d\sigma^{\rm diff}/dM_X$ for $\gamma^*p
\to XN$, $M_N < 2.3$ GeV, corrected for radiative effects and after
transporting the measured values to the reference values ($M_X, W,
Q^2$) using the BEKW(mod) fit (see Section~\ref{sec-bekw}), is
presented in Tables~\ref{t:dsigdmx1} --~\ref{t:dsigdmx6} and
Figs.~\ref{f:dsigdmxlh2.7.25} and~\ref{f:dsigdmxlh.35.320} as a
function of $W$. The results from the FPC~I and FPC~II analyses are
shown. Where measurements at the same values of $Q^2$ are available,
agreement is observed between the two data sets.

The diffractive cross section $d\sigma^{\rm diff}/dM_X$ varies with
$M_X$, $W$ and $Q^2$. For $M_X = 1.2$~GeV, the diffractive cross
section shows a moderate increase with increasing $W$ and a steep
reduction with $Q^2$, approximately proportional to $1/Q^4$. For
larger $M_X$ values, the diffractive cross section exhibits a
substantial rise with increasing $W$ and a less steep decrease with
$Q^2$ roughly proportional to $1/Q^2$. The diffractive cross section
is significant up to $Q^2 = 320$ GeV$^2$, provided $M_X = 11 - 30$
GeV.

The $W$ dependence was quantified by fitting the form
\begin{eqnarray}
\frac{d\sigma^{\rm diff}_{ \gamma^* p \to XN}}{dM_X} = h \cdot(W/W_0)^{a^{\rm diff}}
\end{eqnarray}
to the data for each $(M_X,Q^2)$ bin with $M_X < 15$ GeV; here $W_0 =
1$ GeV and $h$, $a^{\rm diff}$ are free parameters. The $a^{\rm diff}$
values from the FPC~I and II analyses
are shown in Fig.~\ref{f:difslope12} as a function of $Q^2$ for different $M_X$ intervals. For $M_X > 4$ GeV they range from 0.3 to 0.8 with a trend for $a^{\rm diff}$ to be larger by about 0.2 -- 0.4 units when $Q^2$ is above 20 GeV$^2$.

Under the assumption that the diffractive cross section can be
described by the exchange of a single Pomeron, the parameter $a^{\rm
diff}$ is related to the Pomeron trajectory averaged over $t$:
$\overline{\alpha_{\pom}} = 1+ a^{\rm diff}/4$. In the present
measurement, the diffractive cross section is integrated over $t$,
providing $t$-averaged values of $\alpha_{\pom}$. In the framework of
Regge phenomenology, the cross section for diffractive scattering can
be written as~\cite{Collins:1977:regge},
\begin{eqnarray}
d\sigma/dt = f(t) \cdot e^{2(\alpha_{\pom}(t)-1)\cdot \ln (W/W_0)^2},
\end{eqnarray}
where $f(t)$ characterises the $t$ dependences of the ($\gamma^* \pom
\gamma^*$) and ($p \pom N$) vertices.  Assuming $d\sigma/dt \propto
e^{A \cdot t}$ and $\alpha_{\pom}(t) = \alpha_{\pom}(0) +
\alpha_{\pom}^{\prime} \cdot t$ leads to $\alpha_{\pom}(0) =
\overline{\alpha_{\pom}} + \alpha_{\pom}^{\prime}/A$. Taking $A = 7.9
\pm 0.5({\rm stat.})^{+0.9}_{-0.5}({\rm syst.})$ GeV$^{-2}$, as
measured by ZEUS with the leading proton spectrometer
(LPS)~\cite{epj:c38:43}~\footnote{This value of $A$ has been
determined for $\xpom < 0.01$, where diffraction is dominant in the
ZEUS data. Here it is assumed that $A$ for the diffractive
contribution remains the same in the region $0.01 < \xpom < 0.03$;
$\xpom = 0.03$ is the highest value of $\xpom$ reached in the FPC~I
and FPC~II analyses.}, and $\alpha_{\pom}^{\prime} = 0.25$
GeV$^{-2}$~\cite{np:b244:322,*pl:b296:227}, gives $\alpha_{\pom}(0)
\approx \overline{\alpha_{\pom}} +0.03 = 1.03 + a^{\rm diff}/4$. The
$\alpha_{\pom}(0)$ values deduced from diffractive cross sections are
denoted as $\alpha^{\rm diff}_{\pom}(0)$.

The $\alpha^{\rm diff}_{\pom}(0)$ values for individual $M_X$ bins are
given in Table~\ref{t:fitdifapom}. The combined results from FPC~I and
FPC~II for $2 < M_X < 15$ GeV are given in Table~\ref{t:fitapom215}
and are shown in Fig.~\ref{f:apomcomb} as a function of $Q^2$ for
$\alpha_{\pom}^{\prime} = 0.25$ GeV$^{-2}$ and $\alpha_{\pom}^{\prime}
= 0$. For $Q^2 < 20$ GeV$^2$, $\alpha^{\rm diff}_{\pom}(0)$ is
compatible with the soft-Pomeron value, while a substantial rise with
$Q^2$ above the soft-Pomeron value is observed for $Q^2 > 30$
GeV$^2$. The $\alpha^{\rm diff}_{\pom}(0)$ values lie, however,
consistently below those obtained from $F_2$, with $[{\alpha^{\rm
diff}_{\pom}(0)-1}]/[{\alpha^{\rm tot}_{\pom}(0)-1}] \approx 0.5 -
0.7$. Since the Pomeron intercept is changing with $Q^2$, the Pomeron
observed in deep inelastic scattering does not correspond to a simple
pole in the angular momentum plane.

\subsection{$M_X$ and $Q^2$ dependences of the diffractive cross section at fixed $W$}
Figure~\ref{f:dsdmmxlh} shows the diffractive cross section multiplied
by a factor of $Q^2$ as a function of $M_X$ for $W = 220$ GeV. For
$Q^2$ values up to about $55$ GeV$^2$ masses $M_X$ below 5 GeV are
prevalent. As $Q^2$ increases, the maximum shifts to larger values of
$M_X$.

The $Q^2$ dependence of diffraction was studied in terms of the
diffractive cross section multiplied by the factor $Q^2 \cdot
(Q^2+M_X^2)$ since scaling of the diffractive structure function
implies that the quantity $Q^2 \cdot (Q^2+M_X^2)\frac{d\sigma^{\rm
diff}_{\gamma^*p \to XN}}{dM^2_X}$ (see below) should be independent
of $Q^2$, up to logarithmic terms. Figure~\ref{f:dsdmq2lh} and
Tables~\ref{t:q2q2mx2dsdmx12},~\ref{t:q2q2mx2dsdmx34},~\ref{t:q2q2mx2dsdmx56}
show $Q^2 \cdot (Q^2+M_X^2)\frac{d\sigma^{\rm diff}_{\gamma^*p \to
XN}}{dM^2_X}$ as a function of $Q^2$ separately for $M_X$ = 1.2, 3, 6
GeV and $M_X$ = 11, 20, 30 GeV. In both cases the data lie within a
band of about $\pm 25\%$ width for fixed $Q^2$ for the $M_X$ values
given. For the lower $M_X$ region, $Q^2 \cdot
(Q^2+M_X^2)\frac{d\sigma^{\rm diff}_{\gamma^*p \to XN}}{dM^2_X}$ is
approximately constant up to $Q^2 \approx 30-40$ GeV$^2$, followed by
a decrease proportional to $\log Q^2$. For larger $M_X$ values, the
data show a weak dependence on $\log Q^2$. A similar behaviour is
observed for lower values of $W$. Thus, the scaling behaviour of
$d\sigma^{\rm diff}_{\gamma^*p \to XN}/{dM^2_X}$ is of the form
$1/[Q^2(Q^2+M^2_X)]$.

\subsection{Diffractive contribution to the total cross section}
The relationship between the total and diffractive cross sections can
be derived under certain assumptions. For instance, the imaginary part
of the amplitude for elastic scattering, $A_{\gamma^* p \to \gamma^*
p}(t,W,Q^2)$, at $t = 0$ can be assumed to be linked to the total
cross section by a generalisation of the optical theorem to virtual
photon scattering. Assuming that $\sigma^{\rm tot}_{\gamma^*p} \propto
W^{2 \lambda}$ and that the elastic and inclusive diffractive
amplitudes at $t=0$ are purely imaginary and have the same $W$ and
$Q^2$ dependences, then $A_{\gamma^* p \to \gamma^* p}(t=0,W,Q^2)$ is
proportional to $W^{2 \lambda}$. Neglecting the real part of the
scattering amplitudes, the rise of the diffractive cross section with
$W$ should then be proportional to $W^{4 \lambda}$, so that the ratio
of the diffractive cross section to the total $\gamma^* p$ cross
section,
\begin{eqnarray}
r^{\rm diff}_{\rm tot} \equiv \frac{\sigma^{\rm diff}}{\sigma^{\rm tot}} = \frac{\int^{M_b}_{M_a} dM_X d\sigma^{\rm diff}_{\gamma^* p \to XN, M_N < 2.3 {\rm GeV}}/dM_X}{\sigma^{\rm tot}_{\gamma^* p}},
\end{eqnarray}
should behave as $r^{\rm diff}_{\rm tot}  \propto W^{2 \lambda}$. 

The ratio $r^{\rm diff}_{\rm tot}$ was determined for all $M_a < M_X <
M_b$ intervals, with the $\sigma^{\rm tot}_{\gamma^* p}$ values taken
from this analysis. The ratio $r^{\rm diff}_{\rm tot}$ is listed in
Tables~\ref{t:rdiftoth1} --~\ref{t:rdiftoth6} and is shown in
Fig.~\ref{f:rdiftot} for the FPC~II data, and in Fig.~\ref{f:rdiftotl}
for those from the FPC~I analysis. The relative contribution of
diffraction to the total cross section is approximately independent of
$W$. It is substantial when $M_X^2 > Q^2$. For $Q^2 =$ 25 -- 320
GeV$^2$, diffraction with $M_X < 2$ GeV accounts for about 0.1 to
0.4$\%$ of the total cross section, while the $M_X$ intervals 15 -- 25
GeV and 25 -- 35 GeV together account for 3 -- 4$\%$.

The ratio $r = \sigma^{\rm diff}(0.28 < M_X < 35 {\rm \; GeV}, M_N <
2.3 {\rm \; GeV})/\sigma^{\rm tot}$ was evaluated as a function of
$Q^2$ for the highest $W$ bin ($200 < W < 245$ GeV) which provides the
best coverage in $M_X$.  Both FPC~I and FPC~II data are listed in
Table~\ref{t:rdiftotalmx} and shown in Fig.~\ref{f:rdiftot220}. The
ratio $r$ is $15.8^{+1.1}_{-1.0}\%$ at $Q^2 = 4$ GeV$^2$, decreasing
to 5.0$^{+0.9} _{-0.9}$ \% at $Q^2 =190$ GeV$^2$. The data are well
described by the form $r = a - b \cdot \ln (1+Q^2)$. Considering both
statistical and systematic uncertainties, the fit yielded $a = 0.2069
\pm 0.0075$ and $b=0.0320\pm 0.0020$, which is shown by the line in
Fig.~\ref{f:rdiftot220}. The figure shows that the ratio $r$ of the
diffractive to the total cross section is decreasing logarithmically
with $Q^2$.

\section{Diffractive structure function of the proton}
\label{sec-difff2d}
The diffractive structure function of the proton, $F^{\rm
D(3)}_2(\beta,\xpom,Q^2)$, is related to the diffractive cross section
for $W^2 \gg Q^2$ as follows:
\begin{eqnarray}
\frac{1}{2M_X} \frac{d\sigma^{\rm diff}_{\gamma^*p \to XN}(M_X,W,Q^2)}{dM_X} = \frac{4 \pi^2 \alpha}{Q^2(Q^2+M^2_X)} \xpom F^{\rm D(3)}_2(\beta,\xpom,Q^2).
\end{eqnarray}
With this definition, $F^{\rm D(3)}_2$ will include also contributions
from longitudinal photons. If $F^{\rm D(3)}_2$ is interpreted in terms
of quark densities, it specifies the probability to find, in a proton
undergoing a diffractive reaction, a quark carrying a fraction $x =
\beta \xpom$ of the proton momentum.

\subsection{$\xpom F^{\rm D(3)}_2$ as a function of $\xpom$}
Figure~\ref{f:f2d3vsxph} shows $\xpom F^{\rm D(3)}_2$ for the FPC~II
data set as a function of $\xpom$ for fixed $Q^2$ and fixed $M_X$, or,
equivalently fixed $\beta$: $\xpom F^{\rm D(3)}_2$ rises approximately
proportional to $\ln 1/\xpom$ as $\xpom \to 0$. This rise reflects the
increase of the diffractive cross section $d\sigma^{\rm diff}/dM_X$
with $W$. Figures~\ref{f:f2d3vsxplh1} and~\ref{f:f2d3vsxplh2} show
that the combined FPC~I and FPC~II data exhibit this rise for most
$Q^2$ values from 2.7 to 320 GeV$^2$. The data are also provided in
Tables~\ref{t:f2d301} --~\ref{t:f2d305}.

\subsection{$\xpom F^{\rm D(3)}_2$ as a function of $Q^2$}
The $Q^2$ dependence of $\xpom F^{\rm D(3)}_2$ for fixed $\beta$ and
$\xpom$ is provided in Tables~\ref{t:f2d3xpbetvsq2a}
--~\ref{t:f2d3xpbetvsq2h} and is presented in Fig.~\ref{f:f2d3vsq2lh}
for the FPC~I and FPC~II data. Fits of the form
\begin{eqnarray} 
\xpom F^{\rm D(3)}_2 & = & c + a \cdot \ln(1+Q^2)
\end{eqnarray} 
yielded the values of $c$ and $a$ given in Table~\ref{t:fitxpfd} for
selected values of $\xpom$, $\beta$ with six or more data
points. Figure~\ref{f:f2d3vsq2lh} and the fit results show that with
increasing $\beta$ the slope $a$ changes from positive values,
corresponding to positive logarithmic scaling violations, to constancy
or negative logarithmic scaling violations. The data are dominated by
positive scaling violations in the region characterised roughly by
$\xpom \beta = x < 1 \cdot 10^{-3}$, by negative scaling violations
for $x \ge 5 \cdot 10^{-3}$, and by constancy in between.

The data contradict the assumption of Regge
factorisation~\cite{pl:b152:256}, that the diffractive structure
function $\xpom F^{\rm D(3)}_2(\beta,\xpom,Q^2)$ factorises into a
term that depends only on $\xpom$ and a second term that depends only
on $\beta$ and $Q^2$. This can be seen in Table~\ref{t:fitxpfd} which
gives the fit results for fixed $\beta = 0.4$ and $\beta = 0.7$, where
the term $a$ shows a strong dependence on $\xpom$.

The $Q^2$ dependence of $\xpom F^{\rm D(3)}_2$ was also studied for
selected values of $\xpom = 0.0001, 0.0003$, $0.001, 0.003, 0.01$ and
of $\beta$. These choices of $\xpom$ and $\beta$ values were made for
the purpose of comparison with the results from
H1~\cite{epj:c48:749}. The values of the diffractive structure
function at these values of $\xpom$ and $\beta$ were obtained from
those at the measured $\xpom$, $\beta$ values by using the BEKW(mod)
fit to the combined FPC I and FPC II data with a total of 427 measured
points (see below). Only points for which the ratio of the transported
to the measured value of $\xpom F^{\rm D(3)}_2$ was within 0.75 --
1.33 were retained, corresponding to about half of the data
sample. Since the $\xpom F^{\rm D(3)}_2$ data from H1 had been
determined for $M_N < 1.6$ GeV while those from this measurement are
presented for $M_N < 2.3$ GeV, the H1 data may have to be increased by
a factor of 1.1 to 1.2 for an absolute comparison; no correction has
been applied.

The measurements of $\xpom F^{\rm D(3)}_2$ by ZEUS and by H1 are
compared in Figs.~\ref{f:f2d3vsq2bxp00031zh}
--~\ref{f:f2d3vsq2bxp03zh}
as a function of $Q^2$ for fixed values of $\xpom$ and $\beta$.  For
$\xpom = 0.0003$ the H1 points at $\beta = 0.27$ and 0.43 are lower
than those from ZEUS by 10 -- 40$\%$ while at $\beta = 0.67$ they are
in agreement. For $\xpom = 0.001$ and $\beta = 0.08 - 0.5$ the H1
points are lower by about 10 - 30 $\%$ while at $\beta = 0.8$ and $Q^2
\le 7$ GeV$^2$ they are higher by about 40$\%$. For $\xpom = 0.003$
the H1 points at $\beta = 0.027 - 0.43$ are lower by about 10 -
30$\%$; at $\beta = 0.67$ the H1 results agree within about
15$\%$. For $\xpom = 0.01$ there is good agreement between the two
measurements for most values of $\beta$. For $\xpom = 0.03$ and $\beta
\le 0.27$ the H1 points agree with those of ZEUS within the errors,
while for $\beta \ge 0.43$ the H1 points are always higher. These
differences are not understood.

\subsection{$\xpom F^{\rm D(3)}_2$ as a function of $\beta$}
The $\beta$ dependence of $\xpom F^{\rm D(3)}_2$ for the FPC~I and
FPC~II data is shown in
Figs.~\ref{f:f2d3vsbetah12} -- ~\ref{f:f2d3vsbetah5} for fixed $\xpom$
and $Q^2$. The values of $\xpom F^{\rm D(3)}_2$ at the chosen $\xpom$
values were obtained from those at the measured $\xpom$ values using
the BEKW(mod) fit (see below). The diffractive structure function
exhibits a fall towards $\beta =1$ and a broad maximum around $\beta =
0.5$. The broad maximum is approximately of the form $\beta (1-\beta)$
as expected when the virtual photon turns into a $q \bar{q}$
system. For $\xpom \ge 0.005$, $\xpom F^{\rm D(3)}_2$ rises as $\beta
\to 0$ which is suggestive for the formation of $q \bar{q}g$ states
via gluon radiation. For $\xpom = 0.0025$ and $0.005$ there is some
excess at high $\beta \ge 0.95$. Since here the $q \bar{q}$
contribution from transverse photons is expected to be small, the
excess suggests diffractive contributions from longitudinal photons.
    
\subsection{Comparison with the BEKW parametrisation}
\label{sec-bekw}
Further insight into the $\xpom F^{\rm D(3)}_2$ data can be gained
with the help of the BEKW parametrisation~\cite{epj:c7:443} which
considers the contributions from the transitions: transverse photon
$\to q\overline{q}$, longitudinal photon $\to q\overline{q}$ and
transverse photon $\to q\overline{q}g$. In the BEKW parametrisation,
the incoming virtual photon fluctuates into a $q\overline{q}$ or
$q\overline{q}g$ dipole which interacts with the target proton via
two-gluon exchange. The $\beta$ spectrum and the scaling behaviour in
$Q^2$ are derived from the wave functions of the incoming transverse
($T$) or longitudinal ($L$) photon on the light cone in the
non-perturbative limit. The $\xpom$ dependence of the cross section is
not predicted by BEKW but is to be determined by
experiment. Specifically
\begin{eqnarray}
\xpom F^{D(3)}_2(\beta,\xpom,Q^2) & = & c_T \cdot F^T_{q\overline{q}} + c_L \cdot F^L_{q\overline{q}} + c_g \cdot F^T_{q\overline{q}g},
\label{eq:bekw}
\end{eqnarray}
where 
\begin{eqnarray}
F^T_{q\overline{q}} & = & \left (\frac{x_0}{\xpom} \right )^{n_T(Q^2)}\cdot \beta(1 - \beta), \\
\label{eq:bekwqqT}
F^L_{q\overline{q}} & = & \left (\frac{x_0}{\xpom} \right )^{n_L(Q^2)} \cdot \frac{Q^2_0}{Q^2+Q^2_0} \cdot \left [\ln \left(\frac{7}{4} + \frac{Q^2}{4 \beta Q^2_0} \right) \right ]^2 \cdot \beta^3 (1 - 2\beta)^2, \\
\label{eq:bekwqqL}
F^T_{q\overline{q}g} & = & \left (\frac{x_0}{\xpom} \right )^{n_g(Q^2)} \cdot \ln \left(1+\frac{Q^2}{Q^2_0}\right)\cdot (1-\beta)^{\gamma}.
\label{eq:bekwqqg}
\end{eqnarray}
The contribution from longitudinal photons coupling to $q
\overline{q}$ is limited to $\beta$ values close to unity. The $q
\overline{q}$ contribution from transverse photons is expected to have
a broad maximum around $\beta = 0.5$, while the $q \overline{q} g$
contribution becomes important at small $\beta$, provided the power
$\gamma$ is large. The original BEKW parametrisation also includes a
higher-twist term for $q \overline{q}$ produced by transverse
photons. The present data are insensitive to this term, and it has,
therefore, been neglected.

For $F^{L}_{q \overline{q}}$, the term $(\frac{Q^2_0}{Q^2})$ provided
by BEKW was replaced by the factor $(\frac{Q^2_0}{Q^2+Q^2_0})$ to
avoid problems as $Q^2 \to 0$. The powers $n_{T,L,g}(Q^2)$ were
assumed by BEKW to be of the form $n(Q^2) = n_0 + n_1 \cdot \ln [1 +
\ln(\frac{Q^2}{Q_0^2})]$. The rise of $\alpha_{\pom}(0)$ with $\ln
Q^2$ observed in the present data suggested using the form $n(Q^2) =
n_0 + n_1\ln(1+\frac{Q^2}{Q_0^2})$. This modified BEKW form will be
referred to as BEKW(mod). Taking $x_0 = 0.01$ and $Q^2_0 = 0.4$
GeV$^2$, the BEKW(mod) form gives a good description of the
data. According to the fit, the coefficients $n_0$ can be set to zero,
and the coefficient $n_1$ can be assumed to be the same for $T$, $L$
and $g$.

The fits of BEKW(mod) to the data from this analysis (FPC~II), to the
data from the FPC~I analysis and to the combined FPC~I and FPC~II data
led to the results shown in Table~\ref{t:fitbekw}.

Figures~\ref{f:f2d3vsxplh1} and~\ref{f:f2d3vsxplh2} compare the
$\xpom$ dependence of the $\xpom F^{\rm D(3)}_2(\beta,\xpom,Q^2)$ data
from the FPC~I and FPC~II analyses with the BEKW(mod) fit. The fit
gives a good description of the total of 427 data points.

The measured $Q^2$ and $\beta$ dependences of the diffractive
structure function are also well reproduced by the BEKW(mod) fit, see
Figs.~\ref{f:f2d3vsq2lh},~\ref{f:f2d3vsbetah12} --
\ref{f:f2d3vsbetah5}. Based on the BEKW(mod) fit, the data show that
the $(q\overline{q})_T$ contribution from transverse photons dominates
the diffractive structure function for $0.2 < \beta < 0.9$. In the
region $\beta > 0.95$, the contribution from longitudinal photons,
$(q\overline{q})_L$, is dominant. This reflects, at least in part, the
increase of the contribution from longitudinal compared to transverse
photons in the production of $\rho^0$ mesons~\cite{pmc:a1:6}. For
$\beta \le 0.15$, the largest contribution is due to gluon emission as
described by the term $(q\overline{q}g)_T$. These conclusions hold for
all $Q^2$ values studied.

\section{Summary and conclusions}
Inclusive and diffractive scattering has been measured with data taken
in 1999-2000 with the ZEUS detector augmented by the forward-plug
calorimeter (FPC), for $Q^2$ between 25 and 320 GeV$^2$ using an
integrated luminosity of 52.4 pb$^{-1}$. Where appropriate, the
results from a previous study (FPC~I) using 4.2 pb$^{-1}$ and covering
the region $Q^2 = 2.7$ -- $55$ GeV$^2$, were included.

The proton structure function, $F_2(x,Q^2)$, shows a rapid rise as $x
\to 0$ at all $Q^2$ values. The rise for the region $x < 0.01$ has
been parametrised in terms of the Pomeron trajectory $\alpha^{\rm
tot}_{\pom}(0)$, showing a rapid increase of $\alpha^{\rm
tot}_{\pom}(0)\propto \ln Q^2$ for $Q^2$ values between 2.7 and 70
GeV$^2$.

The total cross section for virtual-photon proton scattering
multiplied by $Q^2$, $Q^2\sigma^{\rm tot}_{\gamma^{\ast} p}$, shows a
rapid rise with increasing $W$, reflecting the rise of $F_2$ as $x \to
0$; at lower $Q^2$ values (2.7 -- 55 GeV$^2$), this rise becomes
steeper as $Q^2$ increases. At higher $Q^2$ values, the trend is
reversed.

The diffractive cross section, $d\sigma^{\rm diff}_{\gamma^*p \to
XN}/dM_X$, $M_N < 2.3$ GeV, was studied as a function of the hadronic
centre-of-mass energy $W$, of the mass $M_X$ of the diffractively
produced system $X$ and for different $Q^2$ values. For $M_X = 1.2$
GeV, the cross section decreases rapidly with increasing $Q^2$. For
larger $M_X$ values a strong rise with $W$ is observed up to $M_X$
values of 11~GeV. The intercept of the Pomeron trajectory deduced from
the data rises with increasing $Q^2$ but its size is not as large as
observed for $F_2(x,Q^2)$, $[{\alpha^{\rm
diff}_{\pom}(0)-1}]/[{\alpha^{\rm tot}_{\pom}(0)-1}] \approx 0.5 -
0.7$. For fixed $Q^2$, the ratio of the diffractive cross section for
$0.28 < M_X < 35$~GeV to the total cross section is independent of
$W$. For $W = 200 - 245$ GeV this ratio decreases $\propto \ln
(1+Q^2)$ from $15.8 \pm 0.7({\rm stat.})^{+0.9}_{-0.7}({\rm syst.})\%$
at $Q^2 = 4$ GeV$^2$ to $5.0 \pm 0.4({\rm stat.})^{+0.8}_{-0.8}({\rm
syst.})\%$ at $Q^2 = 190$ GeV$^2$.

Diffraction has also been studied in terms of the diffractive
structure function of the proton, $F^{\rm
D(3)}_2(\beta,\xpom,Q^2)$. For fixed $M_X$, $\xpom F^{\rm D(3)}_2 $
shows a strong rise as $\xpom \to 0$ for all $Q^2$ between 2.7 and 320
GeV$^2$. The $\xpom$ dependence of $\xpom F^{D(3)}_2 $ varies only
modestly with $Q^2$. The data show positive scaling violations
proportional to $\ln Q^2$ in the region $\xpom \beta = x < 2 \cdot
10^{-3}$, and constancy with $Q^2$ or negative scaling violations
proportional to $\ln Q^2$ for $x \ge 2 \cdot 10^{-3}$. Therefore, in
the $Q^2$ region studied, the diffractive structure function is
consistent with being of leading twist.

The data contradict Regge factorisation: the diffractive structure
function $F^{\rm D(3)}_2(\beta,\xpom,Q^2)$ does not factorise into a
term which depends only on $\xpom$ and a second term which depends
only on $\beta$ and $Q^2$.
                                         
A good description of $\xpom F^{\rm D(3)}_2$ as a function of $\xpom$,
$\beta$ and $Q^2$ has been obtained by fitting the data with the
BEKW(mod) parametrisation.  This fit implies that the region $0.25 <
\beta < 0.9$ is dominated by the $\gamma^* \to (q\overline{q})_T$
contribution, the region $\beta > 0.95$ is dominated by the $\gamma^*
\to (q\overline{q})_L$ term, while the rise of $\xpom F^{D(3)}_2$ as
$\beta \to 0$ results from gluon emission described by the $\gamma^*
\to (q\overline{q}g)_T$ term.
\\
\\
\\
\\

{\Large \bf Acknowledgements}

We thank the DESY directorate for their strong support and
encouragement. The effort of the HERA machine group is gratefully
acknowledged. We thank the DESY computing and network services for
their support. The construction, testing and installation of the ZEUS
detector has been made possible by the effort of many people not
listed as authors. We would like to thank, in particular,
J. Hauschildt and K. L\"{o}ffler (DESY), R. Feller, E. M\"{o}ller and
H. Prause (I. Inst. for Exp. Phys., Univ.  Hamburg), A. Maniatis
(II. Inst. for Exp. Phys., Univ. Hamburg), N. Wilfert and the members
of the mechanical workshop of the Faculty of Physics from
Univ. Freiburg for their work on the FPC.





\newpage
\appendix
{
\vspace*{-1cm}
{\Large\bf Appendix}
\pagestyle{plain}
\section{Subtraction of the contribution from proton dissociation with $M_N > 2.3$ GeV}
\label{asec-protondissoc}
The contribution from proton dissociation with $M_N > 2.3$ GeV to the
diffractive data sample was determined with SANG and subtracted from
the data sample. Tables~\ref{t:fracnondif1} and~\ref{t:fracnondif2}
give for every $Q^2$, $W$, $M_X$ bin, for which diffractive cross
sections are quoted in Tables~\ref{t:dsigdmx1} -- ~\ref{t:dsigdmx6},
the fraction of events from $M_N > 2.3$ GeV:
\begin{eqnarray}
\frac{{\cal N}^{\rm SANG(M_{\cal N} > {2.3 GeV})}}{{\cal N}^{\rm event}-{\cal N}^{\rm non-diff}-{\cal N}^{\rm SANG(M_{\cal N} > \rm{2.3 GeV})}}.
\end{eqnarray}
For 84\% of the bins, the fraction of events for proton dissociation
with $M_N > 2.3$ GeV that are subtracted, is less than or equal to
20\%.

\section{Extracting the diffractive contribution in the\\ presence of Reggeon exchange}
\label{asec-reggeontest}
For this analysis the effect of Reggeon exchange interfering with the
diffractive component was studied. A positive interference between
Pomeron (${I\hspace{-0.2em}P}$) and Reggeon exchange (${I
\hspace{-0.2em}R}$), which reproduces the rise observed in the LPS
data~\cite{epj:c38:43} for $x_{\pom} F^{\rm D(3)}_2$ as $\xpom >
0.03$, can be achieved by the exchange of the $f$-meson
trajectory. The LPS data were fit to the form \begin{eqnarray}
x_{\pom} F^{\rm D(3)}_2(\beta,x_{\pom},Q^2) = \left[d_1 \cdot
\sqrt{x_{\pom} F^{\rm D(3)\rm BEKW}_2} + d_2 \cdot
\sqrt{x_{\pom}/0.01}\hspace*{0.2cm} \right]^2
\end{eqnarray} 
where $x_{\pom} F^{\rm D(3)\rm BEKW}_2$ is taken from the fit to the
FPC~I and FPC~II data, see Section 9.5.1, and the second term
represents the Reggeon contribution.  The fit to the LPS data yielded
$d_1 = 0.768 \pm 0.020$ and $d_2 = 0.0177\pm 0.0019$, with $\chi^2 =
135$ for 78 degrees of freedom.

In order to determine the possible contribution from Reggeon exchange
and Reggeon-Pomeron interference (${I\hspace{-0.2em}R}^2 + 2 \cdot
{I\hspace{-0.2em}P} \cdot {I\hspace{-0.2em}R}$) to the diffractive
data, Monte Carlo (MC) events were generated according to
\begin{eqnarray} 
  x_{\pom} F^{D(3)(2{I\hspace{-0.2em}R+I\hspace{-0.2em}P)}}_2(\beta,x_{\pom},Q^2) = 2d_1 \cdot d_2\cdot \sqrt{x_{\pom} F^{\rm D(3)\rm BEKW}_2 \cdot x_{\pom}/0.01 } + d_2^2 \cdot \frac{x_{\pom}}{0.01} 
\end{eqnarray} 
These MC events were subjected to the same analysis procedure as the
data. The Reggeon plus Reggeon-Pomeron interference contribution
(${I\hspace{-0.2em}R}^2 + 2 \cdot {I\hspace{-0.2em}P} \cdot
{I\hspace{-0.2em}R}$) to the diffractive cross section $d\sigma^{\rm
diff}/dM_X$ was found to be smaller than the combined statistical and
systematic uncertainty for all but 3 of the 166 data points. No
correction was applied to the data.

\vfill\eject

\providecommand{\etal}{et al.\xspace}
\providecommand{\coll}{Coll.\xspace}
\catcode`\@=11
\def\@bibitem#1{
\ifmc@bstsupport
  \mc@iftail{#1}%
    {;\newline\ignorespaces}%
    {\ifmc@first\else.\fi\orig@bibitem{#1}}
  \mc@firstfalse
\else
  \mc@iftail{#1}%
    {\ignorespaces}%
    {\orig@bibitem{#1}}%
\fi}%
\catcode`\@=12



\begin{table}[p]
\begin{center}
\scriptsize
\begin{tabular}{||c|c|c|c|c|c|c|c|c|c||}
\hline
$Q^2$ (GeV$^2$) & 20 - 30 & 30 - 40 & 40 - 50 & 50 - 60 & 60 - 80 & 80 - 100 & 100 - 150 & 150 - 250 & 250 - 450 \\
$Q^2_{\rm ref}$ (GeV$^2$) & 25 & 35 & 45 & 55 & 70 & 90 & 120 & 190 & 320 \\
\hline
$W$ (GeV) & 37 - 55 & 55 - 74 & 74 - 99 & 99 - 134 & 134 - 164 & 164 - 200 & 200 - 245 & & \\
$W_{\rm ref}$ (GeV) & 45 & 65 & 85 & 115 & 150 & 180 & 220 & & \\
\hline
$M_X$ (GeV) & 0.28 - 2 & 2 - 4 & 4 - 8 & 8 - 15 & 15 - 25 & 25 - 35 & & & \\         $M_{X {\rm ref}}$ (GeV) & 1.2 & 3 & 6 & 11 & 20 & 30 & & & \\
\hline
\end{tabular}
\normalsize
\caption{Binning and reference values for $Q^2$, $W$ and $M_X$.}
\label{t:binning}
\end{center}
\end{table}

\begin{table}[p]
\vspace*{-1cm}
\begin{center}
\begin{tabular}{||cc|cccccc||}
\hline
$Q^2$ & $W$ & $M_X = 1.2$ & $M_X=3$ & $M_X=6$ & $M_X=11$ & $M_X=20$ & $M_X=30$ \\
(GeV$^2$) & (GeV) &   (GeV) & (GeV) & (GeV) & (GeV) & (GeV) & (GeV) \\
 & & & & & & & \\
\hline
\hline
   25&  45& 0.06 & 0.09 &      &      &      &      \\
   25&  65& 0.08 & 0.06 & 0.11 &      &      &      \\
   25&  85& 0.06 & 0.06 & 0.08 & 0.20 &      &      \\
   25& 115& 0.08 & 0.06 & 0.08 & 0.20 &      &      \\
   25& 150& 0.09 & 0.05 & 0.07 & 0.17 &      &      \\
   25& 180& 0.05 & 0.05 & 0.05 & 0.11 & 0.23 &      \\
   25& 220& 0.07 & 0.04 & 0.06 & 0.10 & 0.20 & 0.47 \\
\hline
   35&  45& 0.04 & 0.05 &      &      &      &      \\
   35&  65& 0.02 & 0.07 & 0.12 &      &      &      \\
   35&  85& 0.08 & 0.09 & 0.09 &      &      &      \\
   35& 115& 0.05 & 0.07 & 0.10 & 0.15 &      &      \\
   35& 150& 0.13 & 0.06 & 0.07 & 0.12 & 0.25 &      \\
   35& 180& 0.06 & 0.07 & 0.06 & 0.13 & 0.29 &      \\
   35& 220& 0.05 & 0.07 & 0.06 & 0.11 & 0.23 & 0.36 \\
\hline
   45&  45& 0.20 &      &      &      &      &      \\
   45&  65& 0.11 & 0.08 & 0.13 &      &      &      \\
   45&  85& 0.03 & 0.06 & 0.09 &      &      &      \\
   45& 115& 0.06 & 0.06 & 0.10 & 0.17 &      &      \\
   45& 150& 0.07 & 0.08 & 0.08 & 0.13 & 0.33 &      \\
   45& 180& 0.07 & 0.03 & 0.08 & 0.12 & 0.24 &      \\ 
   45& 220& 0.07 & 0.05 & 0.05 & 0.09 & 0.15 & 0.30 \\  
\hline
   55&  45& 0.09 & 0.07 &      &      &      &      \\
   55&  65& 0.20 & 0.09 &      &      &      &      \\
   55&  85& 0.08 & 0.12 & 0.08 &      &      &      \\
   55& 115& 0.03 & 0.05 & 0.09 & 0.18 &      &      \\
   55& 150& 0.10 & 0.07 & 0.07 & 0.11 & 0.26 &      \\
   55& 180& 0.11 & 0.06 & 0.09 & 0.12 & 0.26 &      \\
   55& 220& 0.07 & 0.09 & 0.06 & 0.08 & 0.22 & 0.29 \\
\hline
\end{tabular}
\caption{Fraction of events from proton dissociation with $M_N > 2.3$ GeV in the diffractive data sample, as determined with SANG in bins of $Q^2, W, M_X$, for $Q^2 = 25 - 55$ GeV$^2$.}
\label{t:fracnondif1}
\end{center}
\end{table}

\begin{table}[p]
\vspace*{-1cm}
\begin{center}
\begin{tabular}{||cc|cccccc||}
\hline
$Q^2$ & $W$ & $M_X = 1.2$ & $M_X=3$ & $M_X=6$ & $M_X=11$ & $M_X=20$ & $M_X=30$ \\
(GeV$^2$) & (GeV) &   (GeV) & (GeV) & (GeV) & (GeV) & (GeV) & (GeV) \\
 & & & & & & & \\
\hline
   70&  65& 0.00 & 0.06 &      &      &      &      \\
   70&  85& 0.06 & 0.08 & 0.10 &      &      &      \\
   70& 115& 0.00 & 0.04 & 0.08 & 0.16 &      &      \\
   70& 150& 0.13 & 0.04 & 0.09 & 0.11 &      &      \\
   70& 180& 0.24 & 0.03 & 0.09 & 0.12 & 0.22 &      \\
\hline
\end{tabular}
\caption{Fraction of events from proton dissociation with $M_N > 2.3$ GeV in the diffractive data sample, as determined with SANG in bins of $Q^2, W, M_X$, for $Q^2 = 70 - 320$ GeV$^2$.}
\label{t:fracnondif2}
\end{center}
\end{table}

\begin{table}[p]
\vspace*{-1cm}
\begin{center}
\begin{tabular}{||ccc|ccc||}
\hline
$Q^2$ (GeV$^2$) & $x$ &  $F_2\pm \rm stat. \pm syst.$  & $Q^2$ (GeV$^2$) & $x$ &  $F_2 \pm \rm stat. \pm syst.$ \\
\hline\hline
 25 & 0.012200& 0.609$\pm 0.005^{+ 0.026}_{- 0.013}$ &90 & 0.042570& 0.514$\pm 0.006^{+ 0.011}_{- 0.012}$ \\
 25 & 0.005884& 0.768$\pm 0.007^{+ 0.017}_{- 0.016}$ &90 & 0.020860& 0.639$\pm 0.008^{+ 0.013}_{- 0.013}$ \\
 25 & 0.003449& 0.895$\pm 0.007^{+ 0.018}_{- 0.019}$ &90 & 0.012300& 0.756$\pm 0.009^{+ 0.015}_{- 0.016}$ \\
 25 & 0.001887& 1.054$\pm 0.008^{+ 0.022}_{- 0.022}$ &90 & 0.006760& 0.919$\pm 0.010^{+ 0.019}_{- 0.019}$ \\
 25 & 0.001110& 1.259$\pm 0.012^{+ 0.026}_{- 0.027}$ &90 & 0.003984& 1.104$\pm 0.014^{+ 0.023}_{- 0.023}$ \\
 25 & 0.000771& 1.360$\pm 0.014^{+ 0.032}_{- 0.030}$ &90 & 0.002770& 1.229$\pm 0.018^{+ 0.025}_{- 0.026}$ \\
 25 & 0.000516& 1.464$\pm 0.017^{+ 0.056}_{- 0.053}$ &90 & 0.001856& 1.359$\pm 0.022^{+ 0.043}_{- 0.037}$ \\
\end{tabular}
\caption{Proton structure function $F_2$.}
\label{t:f2tab}
\end{center}
\end{table}

\begin{table}[p]
\begin{center}
\begin{tabular}{||c|cc||}
\hline
$Q^2$ (GeV$^2$)  & $c$   & $\alpha^{\rm tot}_{\pom}(0)$\\
\hline\hline
 25 & $0.184 \pm 0.015$  & $1.279 \pm 0.013$ \\
 35 & $0.199 \pm 0.016$  & $1.276 \pm 0.014$ \\
 45 & $0.181 \pm 0.019$  & $1.298 \pm 0.018$ \\
 55 & $0.167 \pm 0.017$  & $1.322 \pm 0.017$ \\
 70 & $0.175 \pm 0.016$  & $1.322 \pm 0.017$ \\
 90 & $0.196 \pm 0.031$  & $1.311 \pm 0.029$ \\
120 & $0.214 \pm 0.032$  & $1.301 \pm 0.028$ \\
\hline
\end{tabular}
\caption{The results of the fits of $F_2$ data for $x < 0.01$ in bins of $Q^2$ to $F_2(x,Q^2) = c \cdot x^{-\lambda}$, where $\alpha^{\rm tot}_{\pom}(0) = 1+\lambda$. The errors give the statistical and systematic uncertainties added in quadrature.}
\label{t:f2fit}
\end{center}
\end{table}

\begin{table}[p]
\vspace*{-1.3cm}
\begin{center}
\begin{tabular}{||ccc|ccc||}
\hline
$Q^2$ & $W$ &  $\sigma^{\rm tot}_{\gamma^*p} \pm {\rm stat. \pm syst.}$ & $Q^2$ & $W$ &  $\sigma^{\rm tot}_{\gamma^*p} \pm {\rm stat. \pm syst.}$\\
(GeV$^2$) & (GeV) & ($\rm{\mu}$b) & (GeV$^2$) & (GeV) & ($\rm{\mu}$b) \\ 
\hline\hline
 25 & 45  &  2.733 $\pm  0.022^{+ 0.118}_{-0.060}$ & 90& 45&  0.641$\pm  0.008^{+ 0.014}_{-0.015}$\\
 25 & 65  &  3.448 $\pm  0.030^{+ 0.076}_{-0.072}$ & 90& 65&  0.797$\pm  0.010^{+ 0.016}_{-0.016}$\\   
 25 & 85  &  4.016 $\pm  0.033^{+ 0.081}_{-0.048}$ & 90& 85&  0.942$\pm  0.011^{+ 0.019}_{-0.020}$\\  
 25 & 115 &  4.731 $\pm  0.038^{+ 0.097}_{-0.097}$ & 90&115&  1.146$\pm  0.013^{+ 0.023}_{-0.023}$\\ 
 25 & 150 &  5.651 $\pm  0.055^{+ 0.118}_{-0.122}$ & 90&150&  1.376$\pm  0.017^{+ 0.028}_{-0.028}$\\ 
 25 & 180 &  6.104 $\pm  0.061^{+ 0.145}_{-0.135}$ & 90&180&  1.532$\pm  0.022^{+ 0.032}_{-0.032}$\\
\end{tabular}
\caption{Total $\gamma^*p$ cross section $\sigma^{\tot}_{\gamma^* p}$.}
\label{t:sigtot}
\end{center}
\end{table}

\begin{table}[p]
\small
\begin{center}
\begin{tabular}{||llll|llll||}
\hline
$M_X$ & $Q^2$ & $W$ & $\frac{d\sigma^{\rm diff}_{\gamma^* p \to XN}}{dM_X}$ & $M_X$ & $Q^2$ & $W$ & $\frac{d\sigma^{\rm diff}_{\gamma^* p \to XN}}{dM_X}$ \\  
 & & & $\pm$ stat. $\pm$ syst. & & & &  $\pm$ stat. $\pm$ syst. \\
 (GeV) & (GeV$^2$) & (GeV) &  (nb/GeV) & (GeV) & (GeV$^2$) & (GeV) &  (nb/GeV)  \\
\hline\hline
  1.2& 25& 45&     5.57$\pm  0.91^{+  1.21}_{-  2.39}$&  1.2& 90& 45&     0.23$\pm  0.10^{+  0.03}_{-  0.08}$ \\
  1.2& 25& 65&     6.39$\pm  0.88^{+  0.54}_{-  1.13}$&  1.2& 90& 65&     0.14$\pm  0.11^{+  0.07}_{-  0.11}$ \\
  1.2& 25& 85&     8.44$\pm  0.98^{+  1.09}_{-  0.97}$&  1.2& 90& 85&     0.11$\pm  0.06^{+  0.06}_{-  0.04}$ \\
  1.2& 25&115&     8.29$\pm  1.02^{+  1.36}_{-  1.27}$&  1.2& 90&115&     0.64$\pm  0.20^{+  0.04}_{-  0.13}$ \\
  1.2& 25&150&     7.02$\pm  1.15^{+  1.07}_{-  1.12}$&  1.2& 90&150&     0.82$\pm  0.34^{+  0.21}_{-  0.18}$ \\
\end{tabular}
\caption{Cross section for diffractive scattering, $\gamma^*p \to XN$, $M_N < 2.3$ {\rm GeV}, for $M_X = 1.2$ {\rm GeV} in bins of $W$ and $Q^2$.}
\label{t:dsigdmx1}
\end{center}
\normalsize
\end{table} 
\clearpage

\begin{table}[p]
\small
\begin{center}
\begin{tabular}{||llll|llll||}
\hline
$M_X$ & $Q^2$ & $W$ & $\frac{d\sigma^{\rm diff}_{\gamma^* p \to XN}}{dM_X}$ & $M_X$ & $Q^2$ & $W$ & $\frac{d\sigma^{\rm diff}_{\gamma^* p \to XN}}{dM_X}$ \\  
 & & & $\pm$ stat. $\pm$ syst. & & & &  $\pm$ stat. $\pm$ syst. \\
 (GeV) & (GeV$^2$) & (GeV) &  (nb/GeV) & (GeV) & (GeV$^2$) & (GeV) &  (nb/GeV)  \\
\hline\hline
  3& 25& 45&    11.68$\pm  2.24^{+ 4.77}_{- 7.18}$ & & & & \\
  3& 25& 65&    27.26$\pm  2.25^{+ 2.16}_{- 4.40}$&  3& 90& 65&     0.96$\pm  0.29^{+  0.22}_{-  0.55}$ \\
  3& 25& 85&    27.78$\pm  1.91^{+ 2.12}_{- 3.20}$&  3& 90& 85&     1.24$\pm  0.27^{+  0.09}_{-  0.21}$ \\
  3& 25&115&    31.88$\pm  2.20^{+ 2.66}_{- 1.38}$&  3& 90&115&     1.26$\pm  0.24^{+  0.07}_{-  0.13}$ \\
  3& 25&150&    42.69$\pm 3.18^{+  3.26}_{- 6.22}$&  3& 90&150&     1.18$\pm  0.32^{+  0.36}_{-  0.08}$ \\
\end{tabular}
\caption{Cross section for diffractive scattering, $\gamma^*p \to XN$, $M_N < 2.3$ {\rm GeV}, for $M_X = 3$ {\rm GeV} in bins of $W$ and $Q^2$.}
\label{t:dsigdmx2}
\end{center}
\normalsize
\end{table}

\begin{table}[p]
\small
\begin{center}
\begin{tabular}{||llll|llll||}
\hline
$M_X$ & $Q^2$ & $W$ & $\frac{d\sigma^{\rm diff}_{\gamma^* p \to XN}}{dM_X}$ & $M_X$ & $Q^2$ & $W$ & $\frac{d\sigma^{\rm diff}_{\gamma^* p \to XN}}{dM_X}$ \\  
 & & & $\pm$ stat. $\pm$ syst. & & & &  $\pm$ stat. $\pm$ syst. \\
 (GeV) & (GeV$^2$) & (GeV) &  (nb/GeV) & (GeV) & (GeV$^2$) & (GeV) &  (nb/GeV)  \\
\hline
\hline
  6& 25& 65&    22.6$\pm  4.3^{+  5.9}_{-  9.7}$  & & & & \\
  6& 25& 85&    28.4$\pm  1.7^{+  1.8}_{-  2.3}$  & 6& 90& 85&     2.9$\pm  0.4^{+  0.4}_{-  0.5}$ \\
  6& 25&115&    35.8$\pm  1.9^{+  2.7}_{-  2.6}$  & 6& 90&115&     3.2$\pm  0.4^{+  0.6}_{-  0.5}$ \\
  6& 25&150&    36.1$\pm  2.4^{+  1.5}_{-  2.7}$  & 6& 90&150&     4.0$\pm  0.5^{+  0.2}_{-  0.3}$ \\
  6& 25&180&    44.3$\pm  2.6^{+  1.3}_{-  1.9}$  & 6& 90&180&     4.2$\pm  0.5^{+  0.6}_{-  0.4}$ \\
\end{tabular}
\caption{Cross section for diffractive scattering, $\gamma^*p \to XN$, $M_N < 2.3$ {\rm GeV}, for $M_X = 6$ {\rm GeV} in bins of $W$ and $Q^2$.}
\label{t:dsigdmx3}
\end{center}
\normalsize
\end{table}

\begin{table}[p]
\small
\begin{center}
\begin{tabular}{||llll|llll||}
\hline
$M_X$ & $Q^2$ & $W$ & $\frac{d\sigma^{\rm diff}_{\gamma^* p \to XN}}{dM_X}$ & $M_X$ & $Q^2$ & $W$ & $\frac{d\sigma^{\rm diff}_{\gamma^* p \to XN}}{dM_X}$ \\  
 & & & $\pm$ stat. $\pm$ syst. & & & &  $\pm$ stat. $\pm$ syst. \\
 (GeV) & (GeV$^2$) & (GeV) &  (nb/GeV) & (GeV) & (GeV$^2$) & (GeV) &  (nb/GeV)  \\
\hline\hline
 11& 25& 85&    16.6$\pm  4.4^{+  4.2}_{-  4.8}$ & & & & \\
 11& 25&115&    16.5$\pm  1.2^{+  2.6}_{-  4.0}$& 11& 90&115&     3.5$\pm  0.6^{+  0.6}_{-  0.7}$ \\
 11& 25&150&    16.8$\pm  1.5^{+  2.7}_{-  4.6}$& 11& 90&150&     4.1$\pm  0.4^{+  0.4}_{-  0.4}$ \\
 11& 25&180&    22.9$\pm  1.5^{+  1.4}_{-  1.8}$& 11& 90&180&     4.7$\pm  0.4^{+  0.3}_{-  0.2}$ \\
 11& 25&220&    24.6$\pm  1.6^{+  1.5}_{-  1.6}$& 11& 90&220&     5.5$\pm  0.5^{+  0.4}_{- 0.3}$ \\
\end{tabular}
\caption{Cross section for diffractive scattering, $\gamma^*p \to XN$, $M_N < 2.3$ {\rm GeV}, for $M_X = 11$ {\rm GeV} in bins of $W$ and $Q^2$.}
\label{t:dsigdmx4}
\end{center}
\normalsize
\end{table}

\begin{table}[p]
\small
\begin{center}
\begin{tabular}{||llll|llll||}
\hline
$M_X$ & $Q^2$ & $W$ & $\frac{d\sigma^{\rm diff}_{\gamma^* p \to XN}}{dM_X}$ & $M_X$ & $Q^2$ & $W$ & $\frac{d\sigma^{\rm diff}_{\gamma^* p \to XN}}{dM_X}$ \\  
 & & & $\pm$ stat. $\pm$ syst. & & & &  $\pm$ stat. $\pm$ syst. \\
 (GeV) & (GeV$^2$) & (GeV) &  (nb/GeV) & (GeV) & (GeV$^2$) & (GeV) &  (nb/GeV)  \\
\hline\hline
& & & &                                                     20& 90&150&     2.8$\pm  0.3^{+  0.7}_{-  0.8}$ \\
 20& 25&180&    13.8$\pm  1.2^{+ 2.4}_{- 3.2}$& 20& 90&180&      3.0$\pm  0.5^{+  0.5}_{-  0.4}$ \\
 20& 25&220&    14.5$\pm  1.6^{+ 2.1}_{- 2.0}$& 20& 90&220&     3.3$\pm  0.4^{+  0.3}_{-  0.3}$ \\
\hline
 20& 35&150&     9.2$\pm  2.4^{+ 1.3}_{-  1.4}$ & & & & \\
 20& 35&180&     8.1$\pm  2.5^{+ 2.2}_{-  2.1}$& 20&120&180&     2.1$\pm  0.3^{+  0.3}_{-  0.3}$ \\
 20& 35&220&     9.9$\pm  1.1^{+ 1.3}_{-  1.8}$& 20&120&220&     2.0$\pm  0.2^{+  0.2}_{-  0.4}$ \\
\hline
 20& 45&150&     5.3$\pm  0.5^{+ 1.5}_{-  1.9}$& 20&190&150&     0.9$\pm  0.1^{+  0.4}_{-  0.3}$ \\
 20& 45&180&     7.2$\pm  0.7^{+ 0.8}_{-  1.0}$& 20&190&180&     1.1$\pm  0.2^{+  0.1}_{-  0.2}$ \\
 20& 45&220&     9.1$\pm  0.6^{+ 0.5}_{-  0.7}$& 20&190&220&     1.1$\pm  0.1^{+  0.1}_{-  0.1}$ \\
\hline
 20& 55&150&     5.2$\pm  1.2^{+ 1.0}_{-  1.1}$& 20&320&150&     0.37$\pm  0.08^{+  0.11}_{-  0.11}$ \\
 20& 55&180&     4.5$\pm  0.5^{+ 1.0}_{-  1.4}$& 20&320&180&     0.23$\pm  0.08^{+  0.08}_{-  0.12}$ \\
 20& 55&220&     5.1$\pm  0.4^{+  0.7}_{- 0.8}$& 20&320&220&     0.49$\pm  0.11^{+  0.08}_{-  0.11}$ \\
\hline
  20& 70&180&     3.8$\pm  0.3^{+  0.5}_{-  0.6}$ \\
  20& 70&220&     4.6$\pm  0.4^{+  0.3}_{-  0.8}$ \\ 
\cline{1-4}
\end{tabular}
\caption{Cross section for diffractive scattering, $\gamma^*p \to XN$, $M_N < 2.3$ {\rm GeV}, for $M_X = 20$ {\rm GeV} in bins of $W$ and $Q^2$.}
\label{t:dsigdmx5}
\end{center}
\normalsize
\end{table}

\begin{table}[p]
\small
\begin{center}
\begin{tabular}{||llll|llll||}
\hline
$M_X$ & $Q^2$ & $W$ & $\frac{d\sigma^{\rm diff}_{\gamma^* p \to XN}}{dM_X}$ & $M_X$ & $Q^2$ & $W$ & $\frac{d\sigma^{\rm diff}_{\gamma^* p \to XN}}{dM_X}$ \\  
 & & & $\pm$ stat. $\pm$ syst. & & & &  $\pm$ stat. $\pm$ syst. \\
 (GeV) & (GeV$^2$) & (GeV) &  (nb/GeV) & (GeV) & (GeV$^2$) & (GeV) &  (nb/GeV)  \\
\hline\hline
& & & &                                                30& 90&180&     2.4$\pm  1.3^{+ 0.6}_{- 0.9}$ \\
 30& 25&220&    8.2$\pm  3.3^{+ 2.7}_{-  2.7}$& 30& 90&220&     3.4$\pm  0.6^{+ 0.6}_{- 0.5}$ \\
\hline
 30& 35&220&    6.6$\pm  1.1^{+ 1.8}_{-  2.2}$& 30&190&180&     0.91$\pm  0.47^{+  0.19}_{- 0.43}$ \\
\cline{1-4}
 30& 45&180&    5.5$\pm  1.7^{+ 1.2}_{-  1.6}$& 30&190&220&     1.01$\pm  0.15^{+  0.25}_{- 0.25}$ \\ 
\cline{5-8} 
 30& 45&220&    6.3$\pm  1.1^{+ 0.9}_{-  1.2}$ \\
\cline{1-4}
 30& 55&220&    4.6$\pm  0.6^{+ 0.8}_{-  2.1}$ \\
\cline{1-4}
 30& 70&220&    2.9$\pm  0.7^{+ 0.6}_{-  0.7}$ \\
\cline{1-4}
\end{tabular}
\caption{Cross section for diffractive scattering, $\gamma^*p \to XN$, $M_N < 2.3$ {\rm GeV}, for $M_X = 30$ {\rm GeV} in bins of $W$ and $Q^2$.}
\label{t:dsigdmx6}
\end{center}
\normalsize
\end{table}

\begin{table}[p]
\begin{center}
\small
\begin{tabular}{||c|c|c||}
\hline
 $M_X$ (GeV) & $Q^2$ (GeV$^2$) & $\alpha_{\pom}(0)$ \\
\hline
 1.2 & 25 & 1.150 $\pm$ 0.047 \\
 1.2 & 35 & 1.128 $\pm$ 0.091 \\
 1.2 & 45 & 1.248 $\pm$ 0.062 \\
 1.2 & 55 & 1.214 $\pm$ 0.068 \\
\hline
 3& 25 & 1.142 $\pm$ 0.029 \\
 3& 35 & 1.156 $\pm$ 0.038 \\
 3& 45 & 1.132 $\pm$ 0.039 \\
 3& 55 & 1.174 $\pm$ 0.046 \\
 3& 70 & 1.241 $\pm$ 0.059 \\
 3& 90 & 1.126 $\pm$ 0.079 \\
\hline
 6& 25 & 1.167 $\pm$ 0.029 \\
 6& 35 & 1.198 $\pm$ 0.047 \\
 6& 45 & 1.216 $\pm$ 0.037 \\
 6& 55 & 1.198 $\pm$ 0.042 \\
 6& 70 & 1.178 $\pm$ 0.038 \\
 6& 90 & 1.149 $\pm$ 0.062 \\
 6& 120 & 1.189 $\pm$ 0.080 \\
\hline
11& 25 & 1.190 $\pm$ 0.072 \\
11& 45 & 1.234 $\pm$ 0.056 \\
11& 55 & 1.183 $\pm$ 0.065 \\
11& 70 & 1.172 $\pm$ 0.071 \\
11& 90 & 1.219 $\pm$ 0.086 \\
\hline
\end{tabular}
\caption{The value of $\alpha_{\pom}(0)$ deduced from the $W$ dependence of the diffractive cross section, assuming $\alpha_{\pom}^{\prime} = 0.25$ GeV$^{-2}$, for fixed $M_X$ and $Q^2$, see text.}
\label{t:fitdifapom}
\end{center}
\normalsize
\end{table}
\clearpage

\begin{table}[p]
\begin{center}
\begin{tabular}{||c|c||}
\hline
$Q^2$ (GeV$^2$) & $\alpha_{\pom}(0)$\\
\hline\hline
 2.7& $1.112 \pm 0.023$ \\
 4  & $1.127 \pm 0.014$ \\
 6  & $1.137 \pm 0.015$ \\
 8  & $1.109 \pm 0.012$ \\
 14 & $1.130 \pm 0.014$ \\
 25 & $1.166 \pm 0.028$ \\
 27 & $1.195 \pm 0.021$ \\
 45 & $1.194 \pm 0.026$ \\
 55 & $1.201 \pm 0.035$ \\
 55 & $1.185 \pm 0.030$ \\
 70 & $1.197 \pm 0.033$ \\
 90 & $1.165 \pm 0.044$ \\
\hline
\end{tabular}
\caption{The value of $\alpha_{\pom}(0)$ deduced from the $W$ dependence of the diffractive cross section, assuming $\alpha_{\pom}^{\prime} = 0.25$ GeV$^{-2}$, for fixed $2 < M_X < 15$ {\rm GeV} and $Q^2$, from FPC~I and FPC~II, see text.}
\label{t:fitapom215}
\end{center}
\end{table}

\begin{table}[p]
\begin{center}
\small
\begin{tabular}{||c|c|c|c|c|c||}
\hline
 $M_X$ & $Q^2$ & $Q^2 \cdot (Q^2+M_X^2)\frac{d\sigma^{\rm diff}_{\gamma^*p \to XN}}{dM^2_X}  $ & $M_X$ & $Q^2$ & $Q^2 \cdot (Q^2+M_X^2)\frac{d\sigma^{\rm diff}_{\gamma^*p \to XN}}{dM^2_X}  $ \\
 (GeV) & (GeV$^2$) & ($\mu$b GeV$^2$) & (GeV) & (GeV$^2$) & ($\mu$b GeV$^2$) \\   
\hline \hline
 1.2 &  4 & 4.27$\pm 0.44^{+1.36}_{-0.79}$ &  3 &  4 & 4.98$\pm 0.33^{+0.52}_{-0.53}$\\
 1.2 &  6 & 4.73$\pm 0.52^{+1.10}_{-1.03}$ &  3 &  6 & 5.84$\pm 0.46^{+0.53}_{-0.61}$\\
 1.2 &  8 & 4.44$\pm 0.45^{+0.96}_{-0.66}$ &  3 &  8 & 6.01$\pm 0.40^{+0.59}_{-0.54}$\\
 1.2 & 14 & 4.77$\pm 0.45^{+0.70}_{-0.67}$ &  3 & 14 & 6.12$\pm 0.40^{+0.66}_{-0.58}$\\
 1.2 & 25 & 4.61$\pm 0.63^{+0.95}_{-0.94}$ &  3 & 25 & 5.47$\pm 0.42^{+0.60}_{-0.62}$\\
 1.2 & 27 & 3.07$\pm 0.63^{+0.90}_{-0.71}$ &  3 & 27 & 5.58$\pm 0.56^{+0.82}_{-0.85}$\\
 1.2 & 35 & 3.05$\pm 0.63^{+0.87}_{-0.66}$ &  3 & 35 & 4.20$\pm 0.54^{+0.63}_{-0.82}$\\
 1.2 & 45 & 2.66$\pm 0.50^{+0.77}_{-0.57}$ &  3 & 45 & 3.69$\pm 0.36^{+0.70}_{-0.53}$\\
 1.2 & 55 & 1.83$\pm 0.96^{+1.00}_{-1.47}$ &  3 & 55 & 3.88$\pm 0.84^{+1.20}_{-1.03}$\\
 1.2 & 55 & 1.94$\pm 0.51^{+0.63}_{-0.63}$ &  3 & 55 & 2.89$\pm 0.42^{+0.66}_{-0.67}$\\
 1.2 & 70 & 1.50$\pm 0.45^{+0.63}_{-0.48}$ &  3 & 70 & 3.51$\pm 0.47^{+0.75}_{-0.68}$\\
 1.2 & 90 & 0.78$\pm 0.44^{+0.57}_{-0.46}$ &  3 & 90 & 2.59$\pm 0.59^{+0.67}_{-0.65}$\\
 1.2 &120 & 3.93$\pm 1.37^{+1.38}_{-1.37}$ &  3 &120 & 2.18$\pm 0.49^{+0.62}_{-0.73}$\\
 1.2 &190 & 3.11$\pm 2.07^{+2.32}_{-2.07}$ &  3 &190 & 2.19$\pm 0.70^{+0.72}_{-1.49}$\\
     &    &                                &  3 &320 & 1.46$\pm 1.01^{+1.01}_{-1.02}$\\
\hline
\end{tabular}
\caption{The diffractive cross section multiplied by $Q^2(Q^2+M^2_X)$,  \newline $Q^2(Q^2+M^2_X)d\sigma^{\rm diff}_{\gamma^*p \to XN}/dM^2_X$, $M_N < 2.3$\GeV, for $W = 220$\GeV as a function of $Q^2$ for $M_X = 1.2$ and $3.0 $\GeV. The first uncertainties are statistical and the second are the systematic uncertainties.} 
\label{t:q2q2mx2dsdmx12}
\end{center}
\normalsize
\end{table}

\begin{table}[p]
\begin{center}
\small
\begin{tabular}{||c|c|c|c|c|c||}
\hline
 $M_X$ & $Q^2$ & $Q^2 \cdot (Q^2+M_X^2)\frac{d\sigma^{\rm diff}_{\gamma^*p \to XN}}{dM^2_X}  $ & $M_X$ & $Q^2$ & $Q^2 \cdot (Q^2+M_X^2)\frac{d\sigma^{\rm diff}_{\gamma^*p \to XN}}{dM^2_X}  $ \\
 (GeV) & (GeV$^2$) & ($\mu$b GeV$^2$) & (GeV) & (GeV$^2$) & ($\mu$b GeV$^2$) \\   
\hline \hline
   &    &    &                             11 & 2.7& 2.30$\pm 0.25^{+0.39}_{-0.25}$\\
 6 &  4 & 3.53$\pm 0.23^{+0.28}_{-0.39}$ & 11 &  4 & 2.43$\pm 0.18^{+0.21}_{-0.25}$\\
 6 &  6 & 3.99$\pm 0.28^{+0.36}_{-0.30}$ & 11 &  6 & 3.39$\pm 0.27^{+0.29}_{-0.30}$\\
 6 &  8 & 4.14$\pm 0.26^{+0.29}_{-0.34}$ & 11 &  8 & 3.26$\pm 0.23^{+0.34}_{-0.24}$\\
 6 & 14 & 4.31$\pm 0.25^{+0.36}_{-0.30}$ & 11 & 14 & 3.70$\pm 0.23^{+0.27}_{-0.25}$\\
 6 & 25 & 6.11$\pm 0.35^{+0.50}_{-0.45}$ & 11 & 25 & 4.07$\pm 0.26^{+0.37}_{-0.38}$\\
 6 & 27 & 5.78$\pm 0.43^{+0.46}_{-0.58}$ & 11 & 27 & 4.07$\pm 0.32^{+0.33}_{-0.46}$\\
 6 & 35 & 6.52$\pm 0.53^{+0.79}_{-0.88}$ & 11 & 35 & 3.57$\pm 0.33^{+0.42}_{-0.42}$\\
 6 & 45 & 6.16$\pm 0.35^{+0.41}_{-0.60}$ & 11 & 45 & 4.40$\pm 0.25^{+0.36}_{-0.32}$\\
 6 & 55 & 4.35$\pm 0.57^{+0.64}_{-0.69}$ & 11 & 55 & 4.46$\pm 0.51^{+0.54}_{-0.58}$\\
 6 & 55 & 6.30$\pm 0.30^{+0.64}_{-0.57}$ & 11 & 55 & 4.76$\pm 0.31^{+0.37}_{-0.52}$\\
 6 & 70 & 4.71$\pm 0.38^{+0.43}_{-0.60}$ & 11 & 70 & 3.98$\pm 0.26^{+0.42}_{-0.32}$\\
 6 & 90 & 4.22$\pm 0.49^{+0.64}_{-0.77}$ & 11 & 90 & 4.78$\pm 0.40^{+0.50}_{-0.50}$\\
 6 &120 & 3.50$\pm 0.45^{+0.81}_{-0.50}$ & 11 &120 & 4.30$\pm 0.38^{+0.44}_{-0.49}$\\
 6 &190 & 1.50$\pm 0.32^{+0.53}_{-0.42}$ & 11 &190 & 3.62$\pm 0.45^{+0.57}_{-0.60}$\\
   &    &    &                             11 &320 & 3.12$\pm 0.78^{+0.82}_{-1.09}$\\
\hline
\end{tabular}
\caption{The diffractive cross section multiplied by $Q^2(Q^2+M^2_X)$, \newline  $Q^2(Q^2+M^2_X)d\sigma^{\rm diff}_{\gamma^*p \to XN}/dM^2_X$, $M_N < 2.3$\GeV, for $W = 220$\GeV as a function of $Q^2$ for $M_X = 6$ and $11 $\GeV. The first uncertainties are statistical and the second are the systematic uncertainties.} 
\label{t:q2q2mx2dsdmx34}
\end{center}
\normalsize
\end{table}
\clearpage

\begin{table}[p]
\begin{center}
\small
\begin{tabular}{||c|c|c|c|c|c||}
\hline
 $M_X$ & $Q^2$ & $Q^2 \cdot (Q^2+M_X^2)\frac{d\sigma^{\rm diff}_{\gamma^*p \to XN}}{dM^2_X}  $ & $M_X$ & $Q^2$ & $Q^2 \cdot (Q^2+M_X^2)\frac{d\sigma^{\rm diff}_{\gamma^*p \to XN}}{dM^2_X}  $ \\
 (GeV) & (GeV$^2$) & ($\mu$b GeV$^2$) & (GeV) & (GeV$^2$) & ($\mu$b GeV$^2$) \\   
\hline \hline
 20 & 2.7& 2.91$\pm 0.36^{+0.44}_{-0.41}$ & 30 & 2.7& 2.57$\pm 0.75^{+0.84}_{-0.79}$\\
 20 &  4 & 2.70$\pm 0.28^{+0.32}_{-0.30}$ & 30 &  4 & 2.49$\pm 0.71^{+0.72}_{-0.77}$\\
 20 &  6 & 2.92$\pm 0.32^{+0.37}_{-0.41}$ & 30 &  6 & 2.65$\pm 0.79^{+0.89}_{-0.85}$\\
 20 &  8 & 3.23$\pm 0.34^{+0.36}_{-0.51}$ & 30 &  8 & 2.80$\pm 0.89^{+1.05}_{-0.90}$\\
 20 & 14 & 3.46$\pm 0.33^{+0.36}_{-0.36}$ & 30 & 14 & 3.60$\pm 0.96^{+1.02}_{-1.04}$\\
 20 & 25 & 3.85$\pm 0.42^{+0.71}_{-0.69}$ & 30 & 25 & 3.17$\pm 1.26^{+1.62}_{-1.64}$\\
 20 & 27 & 3.08$\pm 0.41^{+0.49}_{-0.43}$ & 30 & 27 & 2.82$\pm 0.98^{+1.03}_{-1.04}$\\
 20 & 35 & 3.75$\pm 0.41^{+0.65}_{-0.80}$ & 30 & 35 & 3.57$\pm 0.61^{+1.17}_{-1.34}$\\
 20 & 45 & 4.53$\pm 0.28^{+0.39}_{-0.44}$ & 30 & 45 & 4.45$\pm 0.75^{+0.98}_{-1.12}$\\
 20 & 55 & 2.93$\pm 0.54^{+0.87}_{-0.58}$ &    &    &    \\
 20 & 55 & 3.20$\pm 0.27^{+0.53}_{-0.57}$ & 30 & 55 & 4.04$\pm 0.51^{+0.88}_{-1.93}$\\
 20 & 70 & 3.80$\pm 0.33^{+0.42}_{-0.70}$ & 30 & 70 & 3.32$\pm 0.84^{+1.07}_{-1.19}$\\
 20 & 90 & 3.66$\pm 0.38^{+0.54}_{-0.53}$ & 30 & 90 & 5.05$\pm 0.96^{+1.30}_{-1.19}$\\
 20 &120 & 3.11$\pm 0.32^{+0.48}_{-0.68}$ &    &    &   \\
 20 &190 & 3.14$\pm 0.38^{+0.48}_{-0.51}$ & 30 &190 & 3.50$\pm 0.51^{+1.00}_{-1.00}$\\
 20 & 320& 2.82$\pm 0.61^{+0.77}_{-0.90}$ &    &    &   \\
\hline
\end{tabular}
\caption{The diffractive cross section multiplied by $Q^2(Q^2+M^2_X)$, \newline  $Q^2(Q^2+M^2_X)d\sigma^{\rm diff}_{\gamma^*p \to XN}/dM^2_X$, $M_N < 2.3$\GeV, for $W = 220$\GeV as a function of $Q^2$ for $M_X = 20$ and $30 $\GeV. The first uncertainties are statistical and the second are the systematic uncertainties.} 
\label{t:q2q2mx2dsdmx56}
\end{center}
\normalsize
\end{table}
\clearpage

\begin{table}[p]
\vspace*{-1cm}
\small
\begin{center}
\begin{tabular}{||lrc|lrc||}
\hline
$Q^2$ & $W$ &$\frac{\int^{M_b}_{M_a} dM_X d\sigma^{\rm diff}_{\gamma^* p \to XN}/dM_X}{\sigma^{\rm tot}}$ & $Q^2$ & $W$ &$\frac{\int^{M_b}_{M_a} dM_X d\sigma^{\rm diff}_{\gamma^* p \to XN}/dM_X}{\sigma^{\rm tot}}$\\  
 & & $\pm$stat.$\pm$syst. &  &  &$\pm$stat.$\pm$syst. \\
(GeV$^2$)&(GeV)& &(GeV$^2$)&(GeV)&    \\
\hline\hline 
 25& 45& 0.00346$\pm 0.00057^{+ 0.00075}_{- 0.00148}$& 90& 45& 0.00059$\pm 0.00026^{+ 0.00008}_{- 0.00022}$ \\
 25& 65& 0.00317$\pm 0.00044^{+ 0.00027}_{- 0.00056}$& 90& 65& 0.00029$\pm 0.00024^{+ 0.00014}_{- 0.00023}$ \\
 25& 85& 0.00360$\pm 0.00042^{+ 0.00047}_{- 0.00041}$& 90& 85& 0.00019$\pm 0.00011^{+ 0.00011}_{- 0.00007}$ \\
 25&115& 0.00301$\pm 0.00037^{+ 0.00049}_{- 0.00046}$& 90&115& 0.00096$\pm 0.00030^{+ 0.00007}_{- 0.00019}$ \\
 25&150& 0.00213$\pm 0.00035^{+ 0.00032}_{- 0.00034}$& 90&150& 0.00102$\pm 0.00042^{+ 0.00026}_{- 0.00022}$ \\
\end{tabular}
\caption{Ratio of the cross section for diffractive scattering, $\gamma^*p \to XN$, $M_N < 2.3$ {\rm GeV}, integrated over $M_X = 0.28 - 2$ {\rm GeV}, to the total cross section.}
\label{t:rdiftoth1}
\end{center}
\normalsize
\end{table}

\begin{table}[p]
\vspace*{-1cm}
\small
\begin{center}
\begin{tabular}{||lrc|lrc||}
\hline
$Q^2$ & $W$ &$\frac{\int^{M_b}_{M_a} dM_X d\sigma^{\rm diff}_{\gamma^* p \to XN}/dM_X}{\sigma^{\rm tot}}$ & $Q^2$ & $W$ &$\frac{\int^{M_b}_{M_a} dM_X d\sigma^{\rm diff}_{\gamma^* p \to XN}/dM_X}{\sigma^{\rm tot}}$\\  
 & & $\pm$stat.$\pm$syst. &  &  &$\pm$stat.$\pm$syst. \\
(GeV$^2$)&(GeV)& &(GeV$^2$)&(GeV)&    \\
\hline\hline
 25& 45& 0.00844$\pm 0.00162^{+ 0.00345}_{- 0.00519}$ & & &\\
 25& 65& 0.01572$\pm 0.00131^{+ 0.00125}_{- 0.00254}$& 90& 65& 0.00235$\pm 0.00071^{+ 0.00054}_{- 0.00135}$ \\
 25& 85& 0.01379$\pm 0.00096^{+ 0.00105}_{- 0.00159}$& 90& 85& 0.00260$\pm 0.00057^{+ 0.00019}_{- 0.00043}$ \\
 25&115& 0.01346$\pm 0.00094^{+ 0.00112}_{- 0.00058}$& 90&115& 0.00219$\pm 0.00041^{+ 0.00013}_{- 0.00022}$ \\
 25&150& 0.01509$\pm 0.00113^{+ 0.00115}_{- 0.00220}$& 90&150& 0.00171$\pm 0.00046^{+ 0.00052}_{- 0.00011}$ \\
\end{tabular}
\caption{Ratio of the cross section for diffractive scattering, $\gamma^*p \to XN$, $M_N < 2.3$ {\rm GeV}, integrated over $M_X = 2 - 4$ {\rm GeV}, to the total cross section.}
\label{t:rdiftoth2}
\end{center}
\normalsize
\end{table}

\begin{table}[p]
\vspace*{-1cm}
\small
\begin{center}
\begin{tabular}{||lrc|lrc||}
\hline
$Q^2$ & $W$ &$\frac{\int^{M_b}_{M_a} dM_X d\sigma^{\rm diff}_{\gamma^* p \to XN}/dM_X}{\sigma^{\rm tot}}$ & $Q^2$ & $W$ &$\frac{\int^{M_b}_{M_a} dM_X d\sigma^{\rm diff}_{\gamma^* p \to XN}/dM_X}{\sigma^{\rm tot}}$\\  
 & & $\pm$stat.$\pm$syst. &  &  &$\pm$stat.$\pm$syst. \\
(GeV$^2$)&(GeV)& &(GeV$^2$)&(GeV)&    \\
\hline\hline
 25& 65& 0.02602$\pm 0.00497^{+ 0.00681}_{- 0.01117}$ & & & \\
 25& 85& 0.02823$\pm 0.00173^{+ 0.00179}_{- 0.00230}$& 90& 85& 0.01227$\pm 0.00168^{+ 0.00175}_{- 0.00222}$ \\
 25&115& 0.03019$\pm 0.00159^{+ 0.00226}_{- 0.00221}$& 90&115& 0.01094$\pm 0.00125^{+ 0.00190}_{- 0.00160}$ \\
 25&150& 0.02549$\pm 0.00173^{+ 0.00109}_{- 0.00193}$& 90&150& 0.01171$\pm 0.00134^{+ 0.00064}_{- 0.00082}$ \\
 25&180& 0.02898$\pm 0.00170^{+ 0.00083}_{- 0.00123}$& 90&180& 0.01090$\pm 0.00131^{+ 0.00149}_{- 0.00113}$ \\
\end{tabular}
\caption{Ratio of the cross section for diffractive scattering, $\gamma^*p \to XN$, $M_N < 2.3$ {\rm GeV}, integrated over $M_X = 4 - 8$ {\rm GeV}, to the total cross section.}
\label{t:rdiftoth3}
\end{center}
\normalsize
\end{table}

\begin{table}[p]
\vspace*{-1cm}
\small
\begin{center}
\begin{tabular}{||lrc|lrc||}
\hline
$Q^2$ & $W$ &$\frac{\int^{M_b}_{M_a} dM_X d\sigma^{\rm diff}_{\gamma^* p \to XN}/dM_X}{\sigma^{\rm tot}}$ & $Q^2$ & $W$ &$\frac{\int^{M_b}_{M_a} dM_X d\sigma^{\rm diff}_{\gamma^* p \to XN}/dM_X}{\sigma^{\rm tot}}$\\  
 & & $\pm$stat.$\pm$syst. &  &  &$\pm$stat.$\pm$syst. \\
(GeV$^2$)&(GeV)& &(GeV$^2$)&(GeV)&    \\
\hline\hline
 25& 85& 0.02888$\pm 0.00764^{+ 0.00727}_{- 0.00826}$ & & & \\
 25&115& 0.02434$\pm 0.00177^{+ 0.00377}_{- 0.00594}$& 90&115& 0.02133$\pm 0.00348^{+ 0.00350}_{- 0.00452}$ \\
 25&150& 0.02078$\pm 0.00184^{+ 0.00337}_{- 0.00564}$& 90&150& 0.02066$\pm 0.00183^{+ 0.00193}_{- 0.00177}$ \\
 25&180& 0.02618$\pm 0.00170^{+ 0.00163}_{- 0.00207}$& 90&180& 0.02131$\pm 0.00196^{+ 0.00128}_{- 0.00106}$ \\
 25&220& 0.02614$\pm 0.00171^{+ 0.00164}_{- 0.00173}$& 90&220& 0.02285$\pm 0.00196^{+ 0.00144}_{- 0.00139}$ \\
\end{tabular}
\caption{Ratio of the cross section for diffractive scattering, $\gamma^*p \to XN$, $M_N < 2.3$ {\rm GeV}, integrated over $M_X = 8 - 15$ {\rm GeV}, to the total cross section.}
\label{t:rdiftoth4}
\end{center}
\normalsize
\end{table}

\begin{table}[p]
\vspace*{-1cm}
\small
\begin{center}
\begin{tabular}{||lrc|lrc||}
\hline
$Q^2$ & $W$ &$\frac{\int^{M_b}_{M_a} dM_X d\sigma^{\rm diff}_{\gamma^* p \to XN}/dM_X}{\sigma^{\rm tot}}$ & $Q^2$ & $W$ &$\frac{\int^{M_b}_{M_a} dM_X d\sigma^{\rm diff}_{\gamma^* p \to XN}/dM_X}{\sigma^{\rm tot}}$\\  
 & & $\pm$stat.$\pm$syst. &  &  &$\pm$stat.$\pm$syst. \\
(GeV$^2$)&(GeV)& &(GeV$^2$)&(GeV)&    \\
\hline\hline
 25&180& 0.02250$\pm 0.00197^{+ 0.00390}_{- 0.00516}$& 90&180& 0.01969$\pm 0.00293^{+ 0.00292}_{- 0.00284}$ \\
 25&220& 0.02207$\pm 0.00243^{+ 0.00324}_{- 0.00310}$& 90&220& 0.01957$\pm 0.00207^{+ 0.00201}_{- 0.00196}$ \\
\hline
 35&150& 0.02464$\pm 0.00632^{+ 0.00357}_{- 0.00371}$ & & & \\
 35&180& 0.01927$\pm 0.00601^{+ 0.00526}_{- 0.00501}$&120&180& 0.01939$\pm 0.00300^{+ 0.00270}_{- 0.00308}$ \\
 35&220& 0.02054$\pm 0.00229^{+ 0.00277}_{- 0.00372}$&120&220& 0.01623$\pm 0.00166^{+ 0.00189}_{- 0.00312}$ \\
\hline
 45&150& 0.01819$\pm 0.00180^{+ 0.00510}_{- 0.00661}$&190&150& 0.01617$\pm 0.00239^{+ 0.00719}_{- 0.00517}$ \\
 45&180& 0.02273$\pm 0.00220^{+ 0.00246}_{- 0.00313}$&190&180& 0.01800$\pm 0.00299^{+ 0.00175}_{- 0.00366}$ \\
 45&220& 0.02445$\pm 0.00153^{+ 0.00144}_{- 0.00185}$&190&220& 0.01687$\pm 0.00207^{+ 0.00158}_{- 0.00181}$ \\
\hline
 55&150& 0.02160$\pm 0.00501^{+ 0.00429}_{- 0.00461}$&320&150& 0.01389$\pm 0.00296^{+ 0.00409}_{- 0.00425}$ \\
 55&180& 0.01668$\pm 0.00179^{+ 0.00378}_{- 0.00525}$&320&180& 0.00704$\pm 0.00254^{+ 0.00260}_{- 0.00377}$ \\
 55&220& 0.01742$\pm 0.00149^{+ 0.00245}_{- 0.00271}$&320&220& 0.01311$\pm 0.00287^{+ 0.00217}_{- 0.00308}$ \\
\hline
 70&180& 0.01904$\pm 0.00149^{+ 0.00243}_{- 0.00290}$ \\
 70&220& 0.02028$\pm 0.00176^{+ 0.00140}_{- 0.00334}$ \\                       
\cline{1-3}
\end{tabular}
\caption{Ratio of the cross section for diffractive scattering, $\gamma^*p \to XN$, $M_N < 2.3$ {\rm GeV}, integrated over $M_X = 15 - 25$ {\rm GeV}, to the total cross section.}
\label{t:rdiftoth5}
\end{center}
\normalsize
\end{table}

\begin{table}[p]
\small
\begin{center}
\begin{tabular}{||lrc|lrc||}
\hline
$Q^2$ & $W$ &$\frac{\int^{M_b}_{M_a} dM_X d\sigma^{\rm diff}_{\gamma^* p \to XN}/dM_X}{\sigma^{\rm tot}}$ & $Q^2$ & $W$ &$\frac{\int^{M_b}_{M_a} dM_X d\sigma^{\rm diff}_{\gamma^* p \to XN}/dM_X}{\sigma^{\rm tot}}$\\  
 & & $\pm$stat.$\pm$syst. &  &  &$\pm$stat.$\pm$syst. \\
(GeV$^2$)&(GeV)& &(GeV$^2$)&(GeV)&    \\
\hline\hline
   & & & 90&180& 0.01581$\pm 0.00827^{+ 0.00364}_{- 0.00571}$ \\
 25&220& 0.01252$\pm 0.00496^{+ 0.00403}_{- 0.00417}$ & 90&220& 0.02010$\pm 0.00381^{+ 0.00348}_{- 0.00283}$ \\
\hline
 35&220& 0.01366$\pm 0.00234^{+ 0.00381}_{- 0.00458}$ & & & \\
\hline
 45&180& 0.01725$\pm 0.00539^{+ 0.00369}_{- 0.00507}$&190&180& 0.01465$\pm 0.00761^{+ 0.00310}_{- 0.00689}$ \\
 45&220& 0.01694$\pm 0.00286^{+ 0.00239}_{- 0.00317}$&190&220& 0.01525$\pm 0.00222^{+ 0.00378}_{- 0.00376}$ \\
\hline
 55&220& 0.01569$\pm 0.00198^{+ 0.00277}_{- 0.00723}$ \\
\cline{1-3}
 70&220& 0.01289$\pm 0.00327^{+ 0.00254}_{- 0.00327}$ \\
\cline{1-3}
\end{tabular}
\caption{Ratio of the cross section for diffractive scattering, $\gamma^*p \to XN$, $M_N < 2.3$ {\rm GeV}, integrated over $M_X = 25 - 35$ {\rm GeV}, to the total cross section.}
\label{t:rdiftoth6}
\end{center}
\normalsize
\end{table}

\begin{table}[p]
\small
\begin{center}
\begin{tabular}{||rr||}
\hline
$Q^2$ & $\sigma^{\rm diff}_{0.28 < M_X < 35 \rm GeV}/{\sigma^{\rm tot}} \pm stat.  \pm syst.$\\
\hline\hline
 4 & $0.158\pm 0.007^{+0.009}_{-0.007}$ \hspace*{1cm}\\
 6 & $0.149\pm 0.007^{+0.005}_{-0.005}$ \hspace*{1cm}\\
 8 & $0.134\pm 0.006^{+0.005}_{-0.004}$ \hspace*{1cm}\\
14 & $0.118\pm 0.005^{+0.003}_{-0.002}$ \hspace*{1cm}\\
25 & $0.106\pm 0.006^{+0.012}_{-0.012}$ \hspace*{1cm}\\ 
27 & $0.096\pm 0.006^{+0.003}_{-0.004}$ \hspace*{1cm}\\
35 & $0.090\pm 0.005^{+0.011}_{-0.014}$ \hspace*{1cm}\\
45 & $0.095\pm 0.004^{+0.007}_{-0.009}$ \hspace*{1cm}\\
55 & $0.084\pm 0.003^{+0.008}_{-0.014}$ \hspace*{1cm} \\
70 & $0.071\pm 0.004^{+0.007}_{-0.009}$ \hspace*{1cm}\\
90 & $0.075\pm 0.005^{+0.008}_{-0.008}$ \hspace*{1cm}\\
120 & $0.044\pm 0.003^{+0.005}_{-0.005}$ \hspace*{1cm}\\
190 & $0.050\pm 0.004^{+0.008}_{-0.008}$ \hspace*{1cm}\\
\hline 
\end{tabular}
\caption{Ratio of the cross section for diffractive scattering, $\gamma^*p \to XN$, $M_N < 2.3$ {\rm GeV}, integrated over $M_X = 0.28 - 35$ {\rm GeV}, to the total cross section, for $W = 220$ GeV.}
\label{t:rdiftotalmx}
\end{center}
\normalsize
\end{table}

\begin{table}[p]
\small
\begin{center}
\begin{tabular}{||llll|llll||}
\hline
$\beta$ & $\xpom$ & $Q^2$ & $\xpom F^{D(3)}_2$ & $\beta$ & $\xpom$& $Q^2$ & $\xpom F^{D(3)}_2$  \\
& & & $\pm$ stat. $\pm$ syst. & & & & $\pm$ stat. $\pm$ syst.   \\
& & \hspace*{-0.4cm}(GeV$^2$) & & & & \hspace*{-0.4cm}(GeV$^2$)  & \\
\hline\hline
0.9455&0.01290& 25& 0.0137$\pm0.0022^{+0.0030}_{-0.0059}$&0.9605&0.01769& 35& 0.0092$\pm0.0033^{+0.0075}_{-0.0070}$ \\
0.9455&0.00622& 25& 0.0157$\pm0.0022^{+0.0013}_{-0.0028}$&0.9605&0.00855& 35& 0.0176$\pm0.0046^{+0.0026}_{-0.0040}$ \\
0.9455&0.00365& 25& 0.0207$\pm0.0024^{+0.0027}_{-0.0024}$&0.9605&0.00502& 35& 0.0125$\pm0.0033^{+0.0005}_{-0.0010}$ \\
0.9455&0.00200& 25& 0.0203$\pm0.0025^{+0.0033}_{-0.0031}$&0.9605&0.00275& 35& 0.0216$\pm0.0046^{+0.0021}_{-0.0081}$ \\
0.9455&0.00117& 25& 0.0172$\pm0.0028^{+0.0026}_{-0.0027}$&0.9605&0.00162& 35& 0.0110$\pm0.0038^{+0.0010}_{-0.0024}$ \\
\hline
\end{tabular}
\caption{The diffractive structure function multiplied by $\xpom$, $\xpom F^{D(3)}_2(\beta,\xpom,Q^2)$, for diffractive scattering, $\gamma^*p \to XN$, $M_N < 2.3$ {\rm GeV}, for $Q^2 = 25$ and $35$\GeV$^2$, in bins of $\beta$ and $\xpom$.} 
\label{t:f2d301}
\end{center}
\normalsize
\end{table}
\clearpage

\begin{table}[p]
\small
\begin{center}
\begin{tabular}{||llll|llll||}
\hline
$\beta$ & $\xpom$ & $Q^2$ & $\xpom F^{D(3)}_2$ & $\beta$ & $\xpom$& $Q^2$ & $\xpom F^{D(3)}_2$  \\
& & & $\pm$ stat. $\pm$ syst. & & & & $\pm$ stat. $\pm$ syst.   \\
& & \hspace*{-0.4cm}(GeV$^2$) & & & & \hspace*{-0.4cm}(GeV$^2$)  & \\
\hline\hline
0.9690&0.02243& 45& 0.0060$\pm0.0022^{+0.0029}_{-0.0050}$&0.9745&0.02713& 55& 0.0071$\pm0.0028^{+0.0011}_{-0.0022}$ \\
0.9690&0.01088& 45& 0.0065$\pm0.0020^{+0.0022}_{-0.0025}$&0.9745&0.01319& 55& 0.0091$\pm0.0029^{+0.0012}_{-0.0037}$ \\
0.9690&0.00639& 45& 0.0108$\pm0.0020^{+0.0021}_{-0.0017}$&0.9745&0.00775& 55& 0.0086$\pm0.0027^{+0.0010}_{-0.0018}$ \\
0.9690&0.00350& 45& 0.0172$\pm0.0030^{+0.0035}_{-0.0007}$&0.9745&0.00425& 55& 0.0167$\pm0.0038^{+0.0014}_{-0.0013}$ \\
0.9690&0.00206& 45& 0.0221$\pm0.0041^{+0.0028}_{-0.0028}$&0.9745&0.00250& 55& 0.0247$\pm0.0069^{+0.0009}_{-0.0014}$ \\
\hline
\end{tabular}
\caption{The diffractive structure function multiplied by $\xpom$, $\xpom F^{D(3)}_2(\beta,\xpom,Q^2)$, for diffractive scattering, $\gamma^*p \to XN$, $M_N < 2.3$ {\rm GeV}, for $Q^2 = 45$ and $55$\GeV$^2$, in bins of $\beta$ and $\xpom$.} 
\label{t:f2d302}
\end{center}
\normalsize
\end{table}
\clearpage

\begin{table}[p]
\small
\begin{center}
\begin{tabular}{||llll|llll||}
\hline
$\beta$ & $\xpom$ & $Q^2$ & $\xpom F^{D(3)}_2$ & $\beta$ & $\xpom$& $Q^2$ & $\xpom F^{D(3)}_2$  \\
& & & $\pm$ stat. $\pm$ syst. & & & & $\pm$ stat. $\pm$ syst.   \\
& & \hspace*{-0.4cm}(GeV$^2$) & & & & \hspace*{-0.4cm}(GeV$^2$)  & \\
\hline\hline
& & & &                                                    0.9843&0.04323& 90& 0.0070$\pm0.0031^{+0.0010}_{-0.0026}$ \\
0.9798&0.01663& 70& 0.0100$\pm0.0034^{+0.0018}_{-0.0048}$&0.9843&0.02119& 90& 0.0042$\pm0.0034^{+0.0020}_{-0.0034}$ \\
0.9798&0.00979& 70& 0.0100$\pm0.0030^{+0.0004}_{-0.0016}$&0.9843&0.01250& 90& 0.0032$\pm0.0018^{+0.0018}_{-0.0012}$ \\
0.9798&0.00537& 70& 0.0138$\pm0.0035^{+0.0022}_{-0.0006}$&0.9843&0.00687& 90& 0.0196$\pm0.0061^{+0.0013}_{-0.0039}$ \\
0.9798&0.00316& 70& 0.0094$\pm0.0033^{+0.0015}_{-0.0014}$&0.9843&0.00405& 90& 0.0249$\pm0.0103^{+0.0065}_{-0.0055}$ \\
\hline
\end{tabular}
\caption{The diffractive structure function multiplied by $\xpom$, $\xpom F^{D(3)}_2(\beta,\xpom,Q^2)$, for diffractive scattering, $\gamma^*p \to XN$, $M_N < 2.3$ {\rm GeV}, for $Q^2 = 70$ and $90$\GeV$^2$, in bins of $\beta$ and $\xpom$.} 
\label{t:f2d303}
\end{center}
\normalsize
\end{table}
\clearpage

\begin{table}[p]
\small
\begin{center}
\begin{tabular}{||llll|llll||}
\hline
$\beta$ & $\xpom$ & $Q^2$ & $\xpom F^{D(3)}_2$ & $\beta$ & $\xpom$& $Q^2$ & $\xpom F^{D(3)}_2$  \\
& & & $\pm$ stat. $\pm$ syst. & & & & $\pm$ stat. $\pm$ syst.   \\
& & \hspace*{-0.4cm}(GeV$^2$) & & & & \hspace*{-0.4cm}(GeV$^2$)  & \\
\hline\hline
& & & &                                                   0.9925&0.08643&190& 0.0064$\pm0.0040^{+0.0002}_{-0.0006}$ \\
& & & &                                                   0.9925&0.04336&190& 0.0035$\pm0.0088^{+0.0038}_{-0.0012}$ \\
0.9881&0.01653&120& 0.0051$\pm0.0022^{+0.0009}_{-0.0013}$&0.9925&0.02582&190& 0.0042$\pm0.0078^{+0.0104}_{-0.2232}$ \\
0.9881&0.00910&120& 0.0126$\pm0.0054^{+0.0018}_{-0.0048}$ & & & &\\
0.9881&0.00537&120& 0.0197$\pm0.0077^{+0.0074}_{-0.0033}$ & & & &\\
\hline
\end{tabular}
\caption{The diffractive structure function multiplied by $\xpom$, $\xpom F^{D(3)}_2(\beta,\xpom,Q^2)$, for diffractive scattering, $\gamma^*p \to XN$, $M_N < 2.3$ {\rm GeV}, for $Q^2 = 120$ and $190$\GeV$^2$, in bins of $\beta$ and $\xpom$.} 
\label{t:f2d304}
\end{center}
\normalsize
\end{table}
\clearpage

\begin{table}[p]
\small
\begin{center}
\begin{tabular}{||llll||}
\hline
$\beta$ & $\xpom$ & $Q^2$ & $\xpom F^{D(3)}_2$ \\
& & & $\pm$ stat. $\pm$ syst.   \\
& & \hspace*{-0.4cm}(GeV$^2$) & \\
\hline\hline
0.9726&0.07239&320& 0.0006$\pm0.0010^{+0.0005}_{-0.0006}$ \\
0.9726&0.04361&320& 0.0011$\pm0.0013^{+0.0002}_{-0.0014}$ \\
0.9726&0.02429&320& 0.0104$\pm0.0074^{+0.0005}_{-0.0070}$ \\
0.9726&0.01442&320& 0.0057$\pm0.0126^{+0.0023}_{-0.0025}$ \\
0.9726&0.01006&320& 0.0182$\pm0.0117^{+0.0041}_{-0.0008}$ \\
0.9726&0.00675&320& 0.0130$\pm0.0090^{+0.0003}_{-0.0008}$ \\
\hline
0.8989&0.02628&320& 0.0141$\pm0.0051^{+0.0028}_{-0.0038}$ \\
0.8989&0.01560&320& 0.0084$\pm0.0044^{+0.0009}_{-0.0022}$ \\
0.8989&0.01088&320& 0.0134$\pm0.0056^{+0.0021}_{-0.0018}$ \\
0.8989&0.00731&320& 0.0074$\pm0.0057^{+0.0050}_{-0.0024}$ \\
\hline
0.7256&0.03256&320& 0.0116$\pm0.0026^{+0.0027}_{-0.0059}$ \\
0.7256&0.01933&320& 0.0190$\pm0.0042^{+0.0026}_{-0.0030}$ \\
0.7256&0.01348&320& 0.0252$\pm0.0051^{+0.0024}_{-0.0060}$ \\
0.7256&0.00905&320& 0.0278$\pm0.0069^{+0.0022}_{-0.0068}$ \\
\hline
0.4444&0.03155&320& 0.0192$\pm0.0041^{+0.0057}_{-0.0059}$ \\
0.4444&0.02200&320& 0.0116$\pm0.0042^{+0.0043}_{-0.0062}$ \\
0.4444&0.01478&320& 0.0251$\pm0.0055^{+0.0042}_{-0.0059}$ \\
\hline                                                   
\end{tabular}
\caption{The diffractive structure function multiplied by $\xpom$, $\xpom F^{D(3)}_2(\beta,\xpom,Q^2)$, for diffractive scattering, $\gamma^*p \to XN$, $M_N < 2.3$ {\rm GeV}, for $Q^2 = 320$\GeV$^2$, in bins of $\beta$ and $\xpom$. 
} 
\label{t:f2d305}
\end{center}
\normalsize
\end{table}
\clearpage

\begin{table}[p]
\small
\begin{center}
\begin{tabular}{||llll||}
\hline
$\xpom$ & $\beta$ & $Q^2$ & $\xpom F^{D(3)}_2$ \\
& & \hspace*{-0.4cm}(GeV$^2$) & \\
\hline\hline
0.00015&0.700& 2.7& $0.0378^{+0.0066}_{-0.0058}$ \\
0.00015&0.700& 4  & $0.0480^{+0.0093}_{-0.0072}$ \\
0.00015&0.900& 6  & $0.0343^{+0.0082}_{-0.0083}$ \\
0.00015&0.900& 8  & $0.0358^{+0.0086}_{-0.0064}$ \\
\hline
0.0003&0.400& 2.7& $0.0477^{+0.0076}_{-0.0074}$ \\
0.0003&0.400& 4  & $0.0476^{+0.0059}_{-0.0059}$ \\
0.0003&0.400& 6  & $0.0523^{+0.0063}_{-0.0068}$ \\
0.0003&0.400& 8  & $0.0539^{+0.0064}_{-0.0060}$ \\
\hline
0.0003&0.700& 2.7& $0.0382^{+0.0057}_{-0.0054}$ \\
0.0003&0.700& 4  & $0.0394^{+0.0066}_{-0.0056}$ \\
0.0003&0.900& 6  & $0.0331^{+0.0054}_{-0.0047}$ \\
0.0003&0.900& 8  & $0.0431^{+0.0056}_{-0.0076}$ \\
0.0003&0.900& 14 & $0.0439^{+0.0076}_{-0.0074}$ \\
\hline
0.0006&0.400& 2.7& $0.0438^{+0.0063}_{-0.0067}$ \\
0.0006&0.400& 4  & $0.0460^{+0.0051}_{-0.0046}$ \\
0.0006&0.400& 6  & $0.0535^{+0.0053}_{-0.0057}$ \\
0.0006&0.400& 8  & $0.0473^{+0.0052}_{-0.0045}$ \\
\hline
0.0006&0.700& 2.7& $0.0376^{+0.0060}_{-0.0059}$ \\
0.0006&0.700& 4  & $0.0437^{+0.0063}_{-0.0060}$ \\
0.0006&0.700& 14 & $0.0468^{+0.0058}_{-0.0052}$ \\
0.0006&0.700& 27 & $0.0575^{+0.0102}_{-0.0105}$ \\
\hline
0.0006&0.900& 6  & $0.0300^{+0.0045}_{-0.0054}$ \\
0.0006&0.900& 8  & $0.0370^{+0.0050}_{-0.0049}$ \\
0.0006&0.900& 14 & $0.0305^{+0.0043}_{-0.0046}$ \\
0.0006&0.900& 27 & $0.0340^{+0.0122}_{-0.0104}$ \\
\hline
\end{tabular}
\caption{The diffractive structure function multiplied by $\xpom$, $\xpom F^{D(3)}_2(\beta,\xpom,Q^2)$, for diffractive scattering, $\gamma^*p \to XN$, $M_N < 2.3$ {\rm GeV}, for fixed $\xpom = 0.00015, 0.0003, 0.0006$ and fixed $\beta$. The errors are the statistical and systematic uncertainties added in quadrature.} 
\label{t:f2d3xpbetvsq2a}
\normalsize
\end{center}
\end{table}
\clearpage

\begin{table}[p]
\small
\begin{center}
\begin{tabular}{||llll||}
\hline
$\xpom$ & $\beta$ & $Q^2$ & $\xpom F^{D(3)}_2$ \\
& & \hspace*{-0.4cm}(GeV$^2$) & \\
\hline\hline
0.0012&0.125& 2.7& $0.0247^{+0.0037}_{-0.0040}$ \\
0.0012&0.125& 4  & $0.0292^{+0.0029}_{-0.0032}$ \\
0.0012&0.125& 6  & $0.0323^{+0.0034}_{-0.0033}$ \\
0.0012&0.125& 8  & $0.0325^{+0.0029}_{-0.0030}$ \\
\hline
0.0012&0.400& 2.7& $0.0306^{+0.0052}_{-0.0048}$ \\
0.0012&0.400& 4  & $0.0335^{+0.0037}_{-0.0040}$ \\
0.0012&0.400& 6  & $0.0366^{+0.0041}_{-0.0040}$ \\
0.0012&0.400& 8  & $0.0409^{+0.0036}_{-0.0038}$ \\
0.0012&0.400&14  & $0.0434^{+0.0035}_{-0.0032}$ \\
0.0012&0.400&25  & $0.0554^{+0.0046}_{-0.0046}$ \\
0.0012&0.400&27  & $0.0510^{+0.0050}_{-0.0059}$ \\
0.0012&0.400&35  & $0.0600^{+0.0087}_{-0.0095}$ \\
0.0012&0.400&45  & $0.0601^{+0.0053}_{-0.0067}$ \\
\hline
0.0012&0.700& 2.7& $0.0352^{+0.0074}_{-0.0066}$ \\
0.0012&0.700& 4  & $0.0358^{+0.0058}_{-0.0063}$ \\
0.0012&0.700& 14 & $0.0401^{+0.0035}_{-0.0035}$ \\
0.0012&0.700& 25 & $0.0540^{+0.0071}_{-0.0091}$ \\
0.0012&0.700& 27 & $0.0456^{+0.0085}_{-0.0078}$ \\
0.0012&0.700& 35 & $0.0524^{+0.0120}_{-0.0088}$ \\
0.0012&0.700& 35 & $0.0416^{+0.0082}_{-0.0097}$ \\
0.0012&0.700& 55 & $0.0401^{+0.0079}_{-0.0082}$ \\
0.0012&0.700& 55 & $0.0575^{+0.0073}_{-0.0068}$ \\
\hline
0.0012&0.900& 6  & $0.0292^{+0.0060}_{-0.0046}$ \\
0.0012&0.900& 8  & $0.0332^{+0.0062}_{-0.0053}$ \\
0.0012&0.900& 14 & $0.0316^{+0.0046}_{-0.0042}$ \\
0.0012&0.900& 25 & $0.0246^{+0.0055}_{-0.0055}$ \\
0.0012&0.900& 25 & $0.0286^{+0.0071}_{-0.0065}$ \\
0.0012&0.900& 27 & $0.0352^{+0.0090}_{-0.0122}$ \\
0.0012&0.900& 45 & $0.0272^{+0.0045}_{-0.0050}$ \\
0.0012&0.900& 55 & $0.0302^{+0.0092}_{-0.0100}$ \\
0.0012&0.900& 55 & $0.0270^{+0.0057}_{-0.0063}$ \\
0.0012&0.900& 70 & $0.0324^{+0.0082}_{-0.0076}$ \\
\hline                                                   
\end{tabular}
\caption{The diffractive structure function multiplied by $\xpom$, $\xpom F^{D(3)}_2(\beta,\xpom,Q^2)$, for diffractive scattering, $\gamma^*p \to XN$, $M_N < 2.3$ {\rm GeV}, for fixed $\xpom = 0.0012$ and fixed $\beta$. The errors are the statistical and systematic uncertainties added in quadrature.}
\label{t:f2d3xpbetvsq2b} 
\end{center}
\normalsize
\end{table}
\clearpage

\begin{table}[p]
\small
\begin{center}
\begin{tabular}{||llll||}
\hline
$\xpom$ & $\beta$ & $Q^2$ & $\xpom F^{D(3)}_2$ \\
& & \hspace*{-0.4cm}(GeV$^2$) & \\
\hline\hline
0.0012&0.970& 35 & $0.0127^{+0.0052}_{-0.0054}$ \\
0.0012&0.970& 45 & $0.0200^{+0.0067}_{-0.0057}$ \\
0.0012&0.970& 55 & $0.0168^{+0.0128}_{-0.0162}$ \\
0.0012&0.970& 55 & $0.0228^{+0.0082}_{-0.0079}$ \\
0.0012&0.970& 70 & $0.0155^{+0.0080}_{-0.0068}$ \\
0.0012&0.970& 120& $0.0093^{+0.0085}_{-0.0075}$ \\
\hline 
0.0025&0.025& 2.7& $0.0205^{+0.0041}_{-0.0032}$ \\
0.0025&0.025& 4  & $0.0221^{+0.0025}_{-0.0028}$ \\
0.0025&0.025& 6  & $0.0324^{+0.0038}_{-0.0039}$ \\
\hline
0.0025&0.125& 2.7& $0.0234^{+0.0044}_{-0.0040}$ \\
0.0025&0.125& 8. & $0.0277^{+0.0035}_{-0.0028}$ \\
0.0025&0.125& 14 & $0.0331^{+0.0032}_{-0.0031}$ \\
\hline
0.0025&0.400& 2.7& $0.0280^{+0.0051}_{-0.0053}$ \\
0.0025&0.400& 6  & $0.0385^{+0.0053}_{-0.0053}$ \\
0.0025&0.400& 8  & $0.0400^{+0.0060}_{-0.0043}$ \\
0.0025&0.400& 14 & $0.0364^{+0.0033}_{-0.0030}$ \\
0.0025&0.400& 25 & $0.0414^{+0.0043}_{-0.0050}$ \\
0.0025&0.400& 27 & $0.0369^{+0.0046}_{-0.0046}$ \\
0.0025&0.400& 35 & $0.0469^{+0.0058}_{-0.0069}$ \\
0.0025&0.400& 45 & $0.0402^{+0.0047}_{-0.0038}$ \\
\hline
0.0025&0.700& 2.7& $0.0301^{+0.0060}_{-0.0052}$ \\
0.0025&0.700& 4  & $0.0327^{+0.0043}_{-0.0053}$ \\
0.0025&0.700& 25 & $0.0429^{+0.0055}_{-0.0046}$ \\
0.0025&0.700& 27 & $0.0373^{+0.0062}_{-0.0061}$ \\
0.0025&0.700& 55 & $0.0427^{+0.0071}_{-0.0086}$ \\
0.0025&0.700& 55 & $0.0347^{+0.0054}_{-0.0045}$ \\
0.0025&0.700& 70 & $0.0380^{+0.0047}_{-0.0058}$ \\
0.0025&0.700& 90 & $0.0390^{+0.0075}_{-0.0085}$ \\
\hline                                                   
\end{tabular}
\caption{The diffractive structure function multiplied by $\xpom$, $\xpom F^{D(3)}_2(\beta,\xpom,Q^2)$, for diffractive scattering, $\gamma^*p \to XN$, $M_N < 2.3$ {\rm GeV}, for fixed $\xpom = 0.0012, 0.0025$ and fixed $\beta$. The errors are the statistical and systematic uncertainties added in quadrature.}
\label{t:f2d3xpbetvsq2c} 
\end{center}
\normalsize
\end{table}
\clearpage

\begin{table}[p]
\small
\begin{center}
\begin{tabular}{||llll||}
\hline
$\xpom$ & $\beta$ & $Q^2$ & $\xpom F^{D(3)}_2$ \\
& & \hspace*{-0.4cm}(GeV$^2$) & \\
\hline\hline
0.0025&0.900& 8  & $0.0319^{+0.0055}_{-0.0052}$ \\
0.0025&0.900& 14 & $0.0320^{+0.0054}_{-0.0063}$ \\
0.0025&0.900& 27 & $0.0148^{+0.0058}_{-0.0057}$ \\
0.0025&0.900& 45 & $0.0251^{+0.0051}_{-0.0049}$ \\
0.0025&0.900& 55 & $0.0190^{+0.0105}_{-0.0084}$ \\
0.0025&0.900& 55 & $0.0244^{+0.0052}_{-0.0069}$ \\
0.0025&0.900& 70 & $0.0279^{+0.0062}_{-0.0067}$ \\
0.0025&0.900& 90 & $0.0226^{+0.0078}_{-0.0077}$ \\
0.0025&0.900& 120& $0.0240^{+0.0087}_{-0.0097}$ \\
\hline
0.0025&0.970& 35 & $0.0212^{+0.0067}_{-0.0102}$ \\
0.0025&0.970& 45 & $0.0207^{+0.0060}_{-0.0060}$ \\
0.0025&0.970& 55 & $0.0255^{+0.0101}_{-0.0102}$ \\
0.0025&0.970& 70 & $0.0074^{+0.0062}_{-0.0057}$ \\
0.0025&0.970& 90 & $0.0353^{+0.0230}_{-0.0240}$ \\
0.0025&0.970&120 & $0.0431^{+0.0213}_{-0.0213}$ \\
\hline
0.0050&0.025& 2.7& $0.0183^{+0.0030}_{-0.0032}$ \\
0.0050&0.025& 4  & $0.0205^{+0.0024}_{-0.0026}$ \\
0.0050&0.025& 6  & $0.0239^{+0.0030}_{-0.0030}$ \\
\hline
0.0050&0.125& 2.7& $0.0218^{+0.0036}_{-0.0035}$ \\
0.0050&0.125& 4  & $0.0214^{+0.0030}_{-0.0028}$ \\
0.0050&0.125& 6  & $0.0229^{+0.0025}_{-0.0028}$ \\
0.0050&0.125& 8  & $0.0257^{+0.0028}_{-0.0025}$ \\
0.0050&0.125& 14 & $0.0260^{+0.0024}_{-0.0025}$ \\
0.0050&0.125& 25 & $0.0327^{+0.0036}_{-0.0037}$ \\
0.0050&0.125& 27 & $0.0290^{+0.0036}_{-0.0037}$ \\
\hline

\end{tabular}
\caption{The diffractive structure function multiplied by $\xpom$, $\xpom F^{D(3)}_2(\beta,\xpom,Q^2)$, for diffractive scattering, $\gamma^*p \to XN$, $M_N < 2.3$ {\rm GeV}, for fixed $\xpom = 0.0025, 0.0050$ and fixed $\beta$. The errors are the statistical and systematic uncertainties added in quadrature.}
\label{t:f2d3xpbetvsq2d} 
\end{center}
\normalsize
\end{table}
\clearpage

\begin{table}[p]
\small
\vspace*{-1cm}
\begin{center}
\begin{tabular}{||llll||}
\hline
$\xpom$ & $\beta$ & $Q^2$ & $\xpom F^{D(3)}_2$ \\
& & \hspace*{-0.4cm}(GeV$^2$) & \\
\hline\hline
0.005&0.400& 2.7& $0.0296^{+0.0053}_{-0.0052}$ \\
0.005&0.400& 4  & $0.0287^{+0.0035}_{-0.0032}$ \\
0.005&0.400& 6  & $0.0313^{+0.0048}_{-0.0038}$ \\
0.005&0.400& 8  & $0.0339^{+0.0038}_{-0.0052}$ \\
0.005&0.400& 14 & $0.0329^{+0.0034}_{-0.0037}$ \\
0.005&0.400& 25 & $0.0395^{+0.0041}_{-0.0041}$ \\
0.005&0.400& 27 & $0.0336^{+0.0041}_{-0.0054}$ \\
0.005&0.400& 35 & $0.0359^{+0.0042}_{-0.0043}$ \\
0.005&0.400& 45 & $0.0354^{+0.0032}_{-0.0035}$ \\
0.005&0.400& 55 & $0.0401^{+0.0041}_{-0.0051}$ \\
0.005&0.400& 70 & $0.0359^{+0.0036}_{-0.0036}$ \\
0.005&0.400& 90 & $0.0396^{+0.0049}_{-0.0048}$ \\
0.005&0.400&120 & $0.0375^{+0.0050}_{-0.0053}$ \\
\hline
0.005&0.700& 14 & $0.0311^{+0.0048}_{-0.0036}$ \\
0.005&0.700& 25 & $0.0365^{+0.0045}_{-0.0055}$ \\
0.005&0.700& 27 & $0.0306^{+0.0067}_{-0.0073}$ \\
0.005&0.700& 35 & $0.0334^{+0.0070}_{-0.0066}$ \\
0.005&0.700& 55 & $0.0267^{+0.0056}_{-0.0053}$ \\
0.005&0.700& 55 & $0.0335^{+0.0040}_{-0.0036}$ \\
0.005&0.700& 70 & $0.0318^{+0.0052}_{-0.0040}$ \\
0.005&0.700& 90 & $0.0348^{+0.0060}_{-0.0058}$ \\
0.005&0.700&120 & $0.0355^{+0.0068}_{-0.0062}$ \\
0.005&0.700&190 & $0.0316^{+0.0063}_{-0.0065}$ \\
\hline
0.005&0.900& 6  & $0.0253^{+0.0041}_{-0.0044}$ \\
0.005&0.900& 8  & $0.0289^{+0.0044}_{-0.0046}$ \\
0.005&0.900& 14 & $0.0236^{+0.0040}_{-0.0043}$ \\
0.005&0.900& 25 & $0.0212^{+0.0040}_{-0.0045}$ \\
0.005&0.900& 27 & $0.0231^{+0.0060}_{-0.0058}$ \\
0.005&0.900& 45 & $0.0239^{+0.0033}_{-0.0045}$ \\
0.005&0.900& 55 & $0.0186^{+0.0071}_{-0.0059}$ \\
0.005&0.900& 55 & $0.0220^{+0.0044}_{-0.0046}$ \\
0.005&0.900& 70 & $0.0215^{+0.0038}_{-0.0044}$ \\
0.005&0.900& 90 & $0.0166^{+0.0059}_{-0.0053}$ \\
0.005&0.900&120 & $0.0136^{+0.0052}_{-0.0049}$ \\
0.005&0.900&190 & $0.0139^{+0.0044}_{-0.0039}$ \\
\hline 
\end{tabular}
\caption{The diffractive structure function multiplied by $\xpom$, $\xpom F^{D(3)}_2(\beta,\xpom,Q^2)$, for diffractive scattering, $\gamma^*p \to XN$, $M_N < 2.3$ {\rm GeV}, for fixed $\xpom = 0.005$ and fixed $\beta$. The errors are the statistical and systematic uncertainties added in quadrature.}
\label{t:f2d3xpbetvsq2e} 
\end{center}
\normalsize
\end{table}
\clearpage

\begin{table}[p]
\small
\vspace*{-1cm}
\begin{center}
\begin{tabular}{||llll||}
\hline
$\xpom$ & $\beta$ & $Q^2$ & $\xpom F^{D(3)}_2$ \\
& & \hspace*{-0.4cm}(GeV$^2$) & \\
\hline\hline
0.005&0.970& 35 & $0.0119^{+0.0045}_{-0.0045}$ \\
0.005&0.970& 45 & $0.0136^{+0.0040}_{-0.0033}$ \\
0.005&0.970& 55 & $0.0136^{+0.0068}_{-0.0072}$ \\
0.005&0.970& 55 & $0.0164^{+0.0055}_{-0.0055}$ \\
0.005&0.970& 70 & $0.0131^{+0.0047}_{-0.0044}$ \\
0.005&0.970& 90 & $0.0255^{+0.0123}_{-0.0127}$ \\
0.005&0.970&120 & $0.0208^{+0.0115}_{-0.0112}$ \\
0.005&0.970&190 & $0.0130^{+0.0055}_{-0.0074}$ \\
\hline
0.010&0.005& 2.7& $0.0230^{+0.0047}_{-0.0043}$ \\
0.010&0.005& 4  & $0.0232^{+0.0040}_{-0.0038}$ \\
\hline
0.010&0.025& 2.7& $0.0189^{+0.0037}_{-0.0036}$ \\
0.010&0.025& 4  & $0.0189^{+0.0034}_{-0.0032}$ \\
0.010&0.025& 6  & $0.0235^{+0.0034}_{-0.0043}$ \\
0.010&0.025& 8  & $0.0286^{+0.0043}_{-0.0053}$ \\
0.010&0.025& 14 & $0.0306^{+0.0046}_{-0.0047}$ \\
\hline
0.010&0.125& 2.7& $0.0180^{+0.0031}_{-0.0031}$ \\
0.010&0.125& 4  & $0.0186^{+0.0030}_{-0.0027}$ \\
0.010&0.125& 6  & $0.0230^{+0.0035}_{-0.0036}$ \\
0.010&0.125& 8  & $0.0220^{+0.0030}_{-0.0029}$ \\
0.010&0.125& 14 & $0.0241^{+0.0031}_{-0.0035}$ \\
0.010&0.125& 25 & $0.0256^{+0.0047}_{-0.0068}$ \\
0.010&0.125& 27 & $0.0218^{+0.0055}_{-0.0046}$ \\
0.010&0.125& 35 & $0.0293^{+0.0087}_{-0.0092}$ \\
0.010&0.125& 45 & $0.0365^{+0.0045}_{-0.0051}$ \\
0.010&0.125& 55 & $0.0245^{+0.0080}_{-0.0070}$ \\
0.010&0.125& 55 & $0.0278^{+0.0060}_{-0.0073}$ \\
0.010&0.125& 70 & $0.0338^{+0.0048}_{-0.0064}$ \\
0.010&0.125& 90 & $0.0355^{+0.0064}_{-0.0063}$ \\
\hline 
\end{tabular}
\caption{The diffractive structure function multiplied by $\xpom$, $\xpom F^{D(3)}_2(\beta,\xpom,Q^2)$, for diffractive scattering, $\gamma^*p \to XN$, $M_N < 2.3$ {\rm GeV}, for fixed $\xpom = 0.005, 0.010$ and fixed $\beta$. The errors are the statistical and systematic uncertainties added in quadrature.}
\label{t:f2d3xpbetvsq2f} 
\end{center}
\normalsize
\end{table}
\clearpage

\begin{table}[p]
\small
\vspace*{-1.5cm}
\begin{center}
\begin{tabular}{||llll||}
\hline
$\xpom$ & $\beta$ & $Q^2$ & $\xpom F^{D(3)}_2$ \\
& & \hspace*{-0.4cm}(GeV$^2$) & \\
\hline\hline
0.010&0.400& 6  & $0.0288^{+0.0039}_{-0.0044}$ \\
0.010&0.400& 8  & $0.0305^{+0.0027}_{-0.0028}$ \\
0.010&0.400& 14 & $0.0273^{+0.0041}_{-0.0040}$ \\
0.010&0.400& 25 & $0.0304^{+0.0061}_{-0.0083}$ \\
0.010&0.400& 27 & $0.0282^{+0.0042}_{-0.0047}$ \\
0.010&0.400& 35 & $0.0326^{+0.0057}_{-0.0072}$ \\
0.010&0.400& 45 & $0.0286^{+0.0040}_{-0.0057}$ \\
0.010&0.400& 55 & $0.0299^{+0.0056}_{-0.0056}$ \\
0.010&0.400& 55 & $0.0344^{+0.0050}_{-0.0057}$ \\
0.010&0.400& 70 & $0.0293^{+0.0044}_{-0.0055}$ \\
0.010&0.400& 90 & $0.0303^{+0.0047}_{-0.0046}$ \\
0.010&0.400& 120& $0.0292^{+0.0046}_{-0.0054}$ \\
0.010&0.400& 190& $0.0313^{+0.0061}_{-0.0064}$ \\
0.010&0.400& 320& $0.0292^{+0.0102}_{-0.0113}$ \\
\hline
0.010&0.700& 14 & $0.0280^{+0.0026}_{-0.0030}$ \\
0.010&0.700& 25 & $0.0344^{+0.0049}_{-0.0068}$ \\
0.010&0.700& 27 & $0.0241^{+0.0073}_{-0.0051}$ \\
0.010&0.700& 35 & $0.0267^{+0.0072}_{-0.0076}$ \\
0.010&0.700& 55 & $0.0213^{+0.0078}_{-0.0057}$ \\
0.010&0.700& 55 & $0.0291^{+0.0041}_{-0.0052}$ \\
0.010&0.700& 70 & $0.0262^{+0.0039}_{-0.0042}$ \\
0.010&0.700& 90 & $0.0267^{+0.0063}_{-0.0058}$ \\
0.010&0.700&120 & $0.0348^{+0.0078}_{-0.0066}$ \\
0.010&0.700&190 & $0.0246^{+0.0044}_{-0.0045}$ \\
0.010&0.700&320 & $0.0292^{+0.0084}_{-0.0106}$ \\
\hline
0.010&0.900& 14 & $0.0246^{+0.0042}_{-0.0038}$ \\
0.010&0.900& 25 & $0.0171^{+0.0054}_{-0.0083}$ \\
0.010&0.900& 27 & $0.0191^{+0.0046}_{-0.0056}$ \\
0.010&0.900& 45 & $0.0180^{+0.0032}_{-0.0046}$ \\
0.010&0.900& 55 & $0.0168^{+0.0052}_{-0.0050}$ \\
0.010&0.900& 55 & $0.0144^{+0.0040}_{-0.0056}$ \\
0.010&0.900& 70 & $0.0138^{+0.0037}_{-0.0042}$ \\
0.010&0.900& 90 & $0.0172^{+0.0044}_{-0.0049}$ \\
0.010&0.900&120 & $0.0196^{+0.0059}_{-0.0062}$ \\
0.010&0.900&190 & $0.0145^{+0.0046}_{-0.0044}$ \\
0.010&0.900&320 & $0.0121^{+0.0076}_{-0.0070}$ \\
\hline 
\end{tabular}
\caption{The diffractive structure function multiplied by $\xpom$, $\xpom F^{D(3)}_2(\beta,\xpom,Q^2)$, for diffractive scattering, $\gamma^*p \to XN$, $M_N < 2.3$ {\rm GeV}, for fixed $\xpom = 0.010$ and fixed $\beta$. The errors are the statistical and systematic uncertainties added in quadrature.}
\label{t:f2d3xpbetvsq2f} 
\end{center}
\normalsize
\end{table}
\clearpage

\begin{table}[p]
\vspace*{-1cm}
\small
\begin{center}
\begin{tabular}{||llll||}
\hline
$\xpom$ & $\beta$ & $Q^2$ & $\xpom F^{D(3)}_2$ \\
& & \hspace*{-0.4cm}(GeV$^2$) & \\
\hline\hline
0.010&0.970& 35 & $0.0161^{+0.0064}_{-0.0069}$ \\
0.010&0.970& 45 & $0.0066^{+0.0037}_{-0.0038}$ \\
0.010&0.970& 55 & $0.0118^{+0.0079}_{-0.0066}$ \\
0.010&0.970& 55 & $0.0092^{+0.0037}_{-0.0046}$ \\
0.010&0.970& 70 & $0.0108^{+0.0045}_{-0.0048}$ \\
0.010&0.970& 90 & $0.0040^{+0.0039}_{-0.0035}$ \\
0.010&0.970&120 & $0.0150^{+0.0094}_{-0.0108}$ \\
0.010&0.970&190 & $0.0083^{+0.0045}_{-0.0042}$ \\
\hline
0.020&0.005& 2.7& $0.0196^{+0.0084}_{-0.0082}$ \\
0.020&0.005& 4  & $0.0172^{+0.0078}_{-0.0079}$ \\
0.020&0.005& 6  & $0.0235^{+0.0106}_{-0.0103}$ \\
0.020&0.005& 8  & $0.0251^{+0.0124}_{-0.0114}$ \\
\hline
0.020&0.025& 2.7& $0.0167^{+0.0060}_{-0.0052}$ \\
0.020&0.025& 4  & $0.0142^{+0.0053}_{-0.0053}$ \\
0.020&0.025& 6  & $0.0188^{+0.0071}_{-0.0061}$ \\
0.020&0.025& 8  & $0.0219^{+0.0083}_{-0.0083}$ \\
0.020&0.025& 14 & $0.0234^{+0.0096}_{-0.0095}$ \\
0.020&0.025& 25 & $0.0282^{+0.0182}_{-0.0182}$ \\
0.020&0.025& 27 & $0.0254^{+0.0128}_{-0.0129}$ \\
0.020&0.025& 35 & $0.0333^{+0.0123}_{-0.0138}$ \\
0.020&0.025& 45 & $0.0435^{+0.0120}_{-0.0132}$ \\
\hline
0.020&0.125& 4  & $0.0183^{+0.0047}_{-0.0044}$ \\
0.020&0.125& 6  & $0.0174^{+0.0052}_{-0.0051}$ \\
0.020&0.125& 8  & $0.0201^{+0.0056}_{-0.0054}$ \\
0.020&0.125& 14 & $0.0225^{+0.0072}_{-0.0069}$ \\
0.020&0.125& 25 & $0.0252^{+0.0114}_{-0.0119}$ \\
0.020&0.125& 27 & $0.0202^{+0.0092}_{-0.0080}$ \\
0.020&0.125& 35 & $0.0282^{+0.0110}_{-0.0111}$ \\
0.020&0.125& 45 & $0.0223^{+0.0070}_{-0.0087}$ \\
0.020&0.125& 55 & $0.0231^{+0.0123}_{-0.0127}$ \\
0.020&0.125& 55 & $0.0287^{+0.0110}_{-0.0112}$ \\
0.020&0.125& 70 & $0.0255^{+0.0104}_{-0.0112}$ \\
0.020&0.125& 90 & $0.0302^{+0.0078}_{-0.0081}$ \\
0.020&0.125& 190& $0.0359^{+0.0115}_{-0.0115}$ \\
\hline 
\end{tabular}
\caption{The diffractive structure function multiplied by $\xpom$, $\xpom F^{D(3)}_2(\beta,\xpom,Q^2)$, for diffractive scattering, $\gamma^*p \to XN$, $M_N < 2.3$ {\rm GeV}, for fixed $\xpom = 0.010, 0.020$ and fixed $\beta$. The errors are the statistical and systematic uncertainties added in quadrature.}
\label{t:f2d3xpbetvsq2g} 
\end{center}
\normalsize
\end{table}
\clearpage

\begin{table}[p]
\small
\vspace*{-1cm}
\begin{center}
\begin{tabular}{||llll||}
\hline
$\xpom$ & $\beta$ & $Q^2$ & $\xpom F^{D(3)}_2$ \\
& & \hspace*{-0.4cm}(GeV$^2$) & \\
\hline\hline
0.020&0.400& 35 & $0.0214^{+0.0103}_{-0.0111}$ \\
0.020&0.400& 45 & $0.0192^{+0.0081}_{-0.0107}$ \\
0.020&0.400& 90 & $0.0247^{+0.0070}_{-0.0077}$ \\
0.020&0.400&120 & $0.0251^{+0.0043}_{-0.0069}$ \\
0.020&0.400&190 & $0.0280^{+0.0071}_{-0.0087}$ \\
0.020&0.400&320 & $0.0119^{+0.0075}_{-0.0088}$ \\
\hline
0.020&0.700& 25 & $0.0149^{+0.0073}_{-0.0100}$ \\
0.020&0.700& 27 & $0.0153^{+0.0037}_{-0.0033}$ \\
0.020&0.700& 35 & $0.0194^{+0.0113}_{-0.0160}$ \\
0.020&0.700& 55 & $0.0144^{+0.0055}_{-0.0052}$ \\
0.020&0.700& 90 & $0.0240^{+0.0058}_{-0.0063}$ \\
0.020&0.700& 120& $0.0155^{+0.0075}_{-0.0097}$ \\
0.020&0.700& 190& $0.0182^{+0.0076}_{-0.0086}$ \\
0.020&0.700& 320& $0.0197^{+0.0068}_{-0.0070}$ \\
\hline
0.020&0.900& 70 & $0.0102^{+0.0041}_{-0.0061}$ \\
0.020&0.900& 90 & $0.0139^{+0.0067}_{-0.0099}$ \\
0.020&0.900& 120& $0.0055^{+0.0047}_{-0.0055}$ \\
0.020&0.900& 190& $0.0102^{+0.0033}_{-0.0034}$ \\
0.020&0.900& 320& $0.0076^{+0.0056}_{-0.0059}$ \\
\hline
0.020&0.970& 35 & $0.0085^{+0.0082}_{-0.0077}$ \\
0.020&0.970& 45 & $0.0062^{+0.0044}_{-0.0061}$ \\
0.020&0.970& 70 & $0.0102^{+0.0052}_{-0.0069}$ \\
0.020&0.970&120 & $0.0058^{+0.0038}_{-0.0039}$ \\
0.020&0.970&320 & $0.0118^{+0.0118}_{-0.0142}$ \\
\hline
0.030&0.025& 45 & $0.0381^{+0.0187}_{-0.0202}$ \\
\hline
0.030&0.125& 90 & $0.0295^{+0.0228}_{-0.0242}$ \\
0.030&0.125&190 & $0.0322^{+0.0246}_{-0.0280}$ \\
\hline
0.030&0.040& 190& $0.0224^{+0.0110}_{-0.0086}$ \\
0.030&0.040& 320& $0.0193^{+0.0081}_{-0.0083}$ \\
\hline
0.030&0.700& 320& $0.0126^{+0.0050}_{-0.0076}$ \\
\hline
0.030&0.900&  55& $0.0089^{+0.0056}_{-0.0069}$ \\
0.030&0.900& 320& $0.0132^{+0.0073}_{-0.0077}$ \\
\hline
0.030&0.970&  55& $0.0120^{+0.0075}_{-0.0079}$ \\
\hline
0.060&0.970&  90& $0.0072^{+0.0046}_{-0.0052}$ \\
0.060&0.970& 190& $0.0072^{+0.0070}_{-0.0070}$ \\
\hline 
\end{tabular}
\caption{The diffractive structure function multiplied by $\xpom$, $\xpom F^{D(3)}_2(\beta,\xpom,Q^2)$, for diffractive scattering, $\gamma^*p \to XN$, $M_N < 2.3$ {\rm GeV}, for fixed $\xpom = 0.020, 0.030$ and fixed $\beta$. The errors are the statistical and systematic uncertainties added in quadrature.}
\label{t:f2d3xpbetvsq2h} 
\end{center}
\normalsize
\end{table}
\clearpage

\begin{table}[p]
\begin{center}
\scriptsize
\begin{tabular}{||c|c|c|c|c||}
\hline
$\xpom$   & $\beta$ = 0.125 & 0.40 & 0.70 & 0.90 \\
\hline
  0.0012 &       & c = 0.0144 $\pm$ 0.0051 & 0.0258 $\pm$ 0.0072 & 0.0347 $\pm$ 0.0082 \\
         &       & a = 0.0117 $\pm$ 0.0020 & 0.0060 $\pm$ 0.0024 & -0.0017 $\pm$ 0.0025 \\
\hline
  0.0025 &       & 0.0278 $\pm$ 0.0062 & 0.0293 $\pm$ 0.0056 &  0.0368 $\pm$ 0.0097 \\
         &       & 0.0037 $\pm$ 0.0022 & 0.0024 $\pm$ 0.0017 & -0.0033 $\pm$ 0.0027 \\
\hline
  0.0050 & 0.0134 $\pm$ 0.0039 & 0.0255 $\pm$ 0.0036 & 0.0317 $\pm$ 0.0095 & 0.0347 $\pm$ 0.0049 \\
         & 0.0051 $\pm$ 0.0017 & 0.0029 $\pm$ 0.0011 & 0.0002 $\pm$ 0.0024 & -0.0037 $\pm$ 0.0013 \\
\hline
  0.0100 & 0.0121 $\pm$ 0.0031 & 0.0288 $\pm$ 0.0042 & 0.0296 $\pm$ 0.0062 & 0.0306 $\pm$ 0.0220 \\
         & 0.0046 $\pm$ 0.0012 & 0.0003 $\pm$ 0.0012 & -0.0005 $\pm$ 0.0016 & -0.0033 $\pm$ 0.0053\\
\hline
\end{tabular}
\normalsize
\caption{Fits of $\xpom F^{D(3)}_2 = c + a \cdot \ln (1+Q^2)$ for the $\xpom,\beta$ - values indicated: shown are the values for $c$ and $a$.}
\label{t:fitxpfd}
\end{center}
\end{table}

\begin{table}[p]
\begin{center}
\scriptsize
\begin{tabular}{||c|c|c|c|c|c|c||}
\hline
Exp't    & $c_T$             & $c_L$             & $c_g$               & $n_1$             & $\gamma$ & $\chi^2/n_D$\\
\hline
FPC II   & 0.120 $\pm$ 0.003 & 0.074 $\pm$ 0.006 & 0.0111 $\pm$ 0.0015 & 0.067 $\pm$ 0.003 & 7.98 $\pm$ 0.92 & 0.75 \\
\hline
FPC I    & 0.115 $\pm$ 0.003 & 0.107 $\pm$ 0.009 & 0.0091 $\pm$ 0.0003 & 0.062 $\pm$ 0.003 & 8.60 $\pm$ 0.56 & 0.62 \\
\hline
FPC I + FPC II & 0.118 $\pm$ 0.002 & 0.087 $\pm$ 0.005 & 0.0090 $\pm$ 0.0003 & 0.062 $\pm$ 0.002 & 8.22 $\pm$ 0.46 & 0.79 \\
\hline
\end{tabular}
\normalsize
\caption{Results from fitting the data from FPC II, from FPC I, and from the combined data from FPC I and FPC II to BEKW(mod). The fit procedure includes the statistical and systematic uncertainties of the data.}
\label{t:fitbekw}
\end{center}
\end{table}
\clearpage       

\begin{figure}[p]
\begin{center}
\vfill
\vspace*{-1.5cm} \epsfig{file=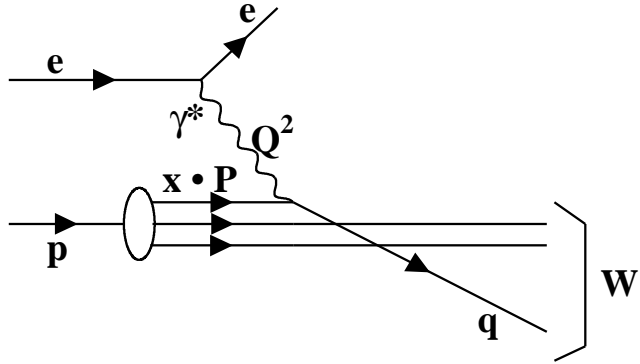,width=11.5cm,clip=}
\end{center}
\vspace*{-1.cm}
\caption{Diagram for non-peripheral deep inelastic scattering.} 
\label{f:nondifdiag}
\vfill
\end{figure}

\begin{figure}[p]
\vfill
\vspace*{-1.cm}
\begin{center}
\epsfig{file=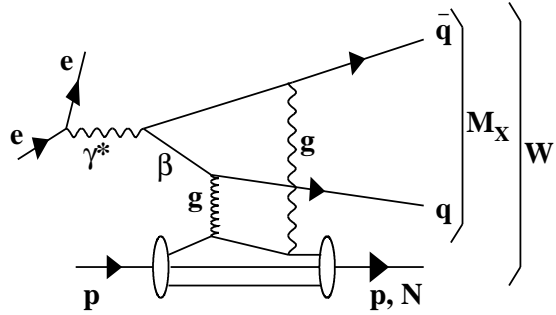,width=9.cm,clip=}
\vspace*{0.5cm}
\epsfig{file=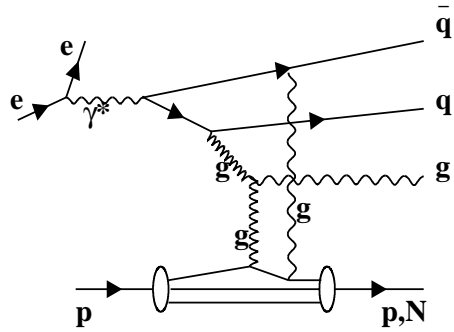,width=9.cm,clip=}
\vfill
\end{center}
\caption{Diagrams of diffractive deep inelastic scattering, $e p \to e X N$, proceeding by the exchange of two gluons.} 
\label{f:difdiag}
\vfill
\end{figure}
\clearpage

\begin{figure}[p]
\vfill
{\epsfig{file=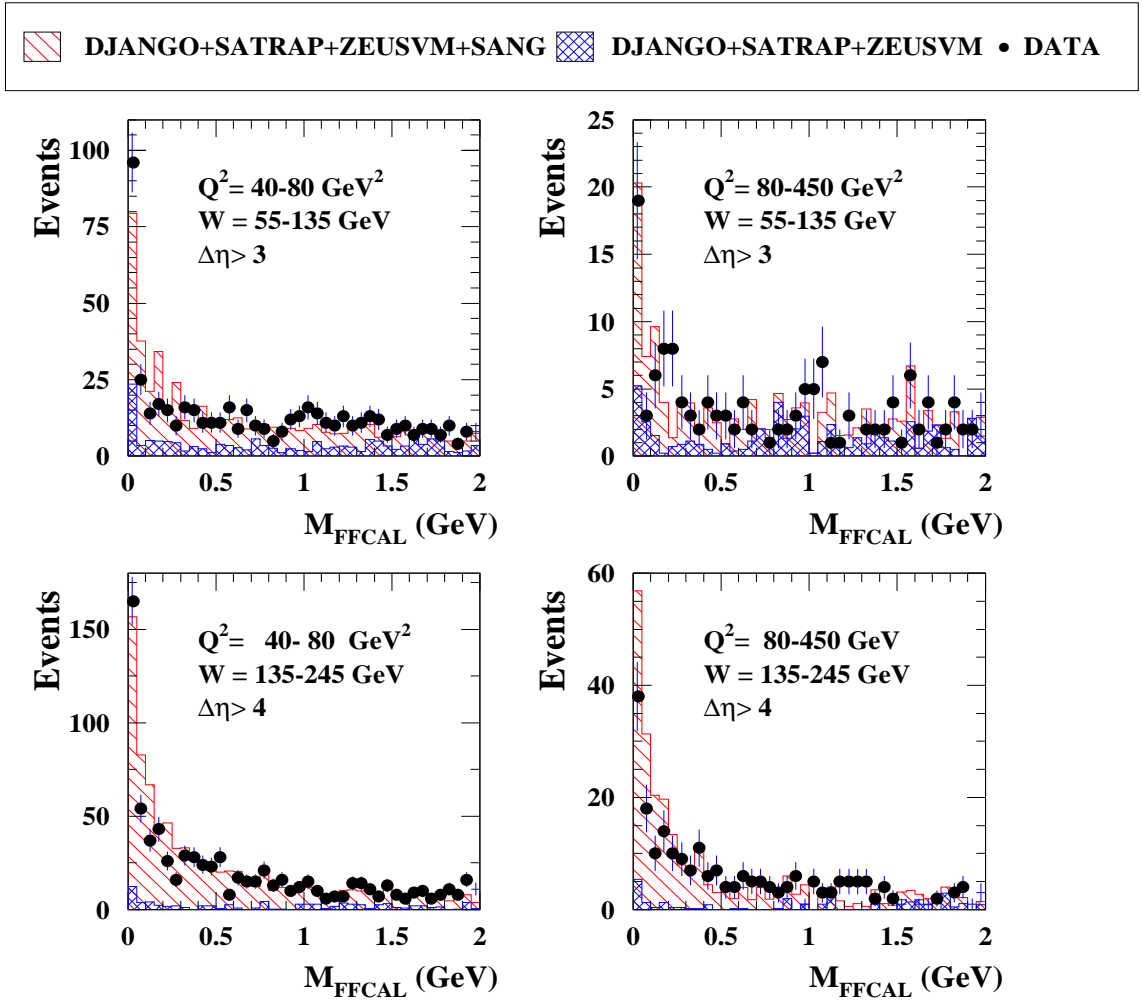,height=15cm,width=16cm}}
\caption
{Distributions of $M_{\rm FFCAL}$ for four different ($Q^2,W,\Delta \eta$) 
regions. The points with error bars show the data. The cross hatched histograms show the MC predictions for the sum of contributions from $Xp$, $\rho^0 p$ and non-peripheral processes; the hatched histograms show the sum of contributions from $Xp$, $\rho^0 p$, diffractive double dissociation ($XN$) and non-peripheral processes. The MC distributions are normalised according to the luminosity of the data.} 

\label{f:mffcalh}
\vfill
\end{figure}
\clearpage

\begin{figure}[p]
\vfill
\hspace*{0.3cm}
{\epsfig{file=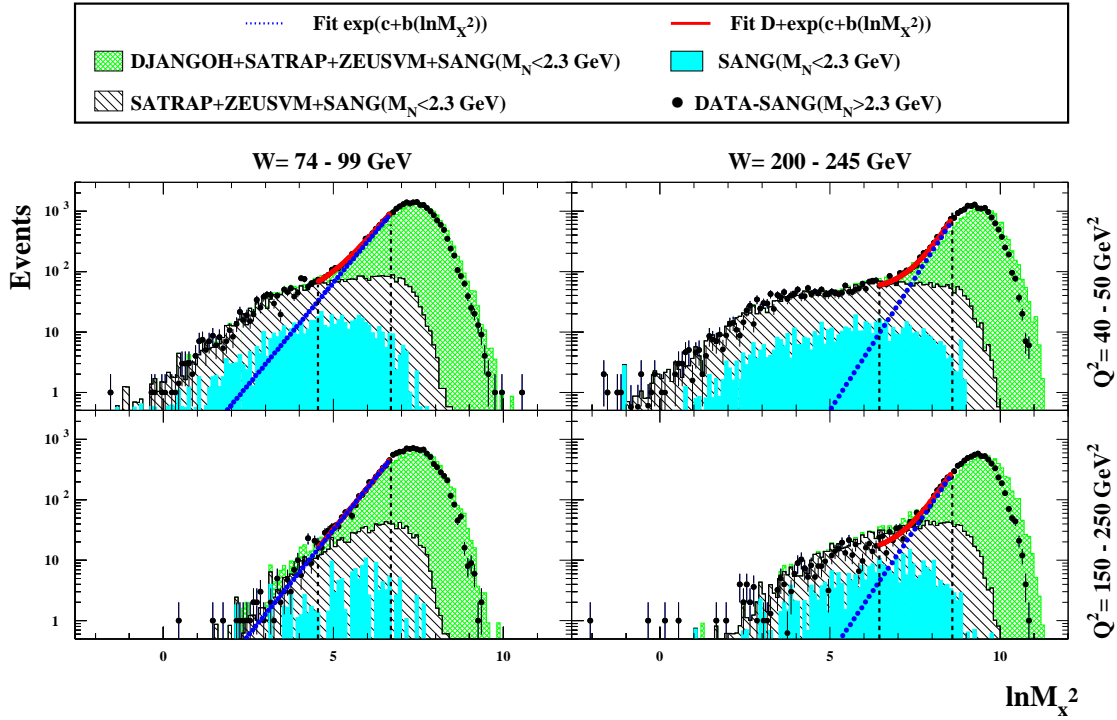,angle=270,width=15.2cm}}
\caption{Distributions of $\ln M^2_X$ ($M_X$ in units of \GeV) at the 
detector level for the $W = 74 - 99$\GeV, $W = 200 - 245$\GeV and $Q^2 = 40-50$\GeV$^2$, $Q^2 = 150-250$\GeV$^2$ bins. The points with error bars 
show the data, with the contribution from proton dissociation, as predicted by SANG, for $M_N > 2.3$ GeV subtracted. The light shaded areas show the non-peripheral contributions as 
predicted by DJANGOH plus the diffractive contributions from SATRAP+ZEUSVM+SANG($M_N < 2.3$ GeV). The diffractive contributions from $\gamma^* p \to XN$, $M_N < 2.3$\GeV, as predicted by SATRAP+ZEUSVM+SANG($M_N < 2.3$\GeV), are shown as hatched areas. The dark grey areas show the contribution of diffractive events in which the proton dissociates into a system $N$, with $M_N < 2.3$\GeV. The dash-dotted lines show the results for the non-diffractive contribution from fitting the data in the $\ln M^2 _X$ range delimited by the two vertical dashed lines.}
\label{f:lnmxsel}
\vfill
\end{figure}
\clearpage

\begin{figure}[p]
\vfill
\vspace*{-2.2cm}
\hspace*{0.7cm}
{\epsfig{file=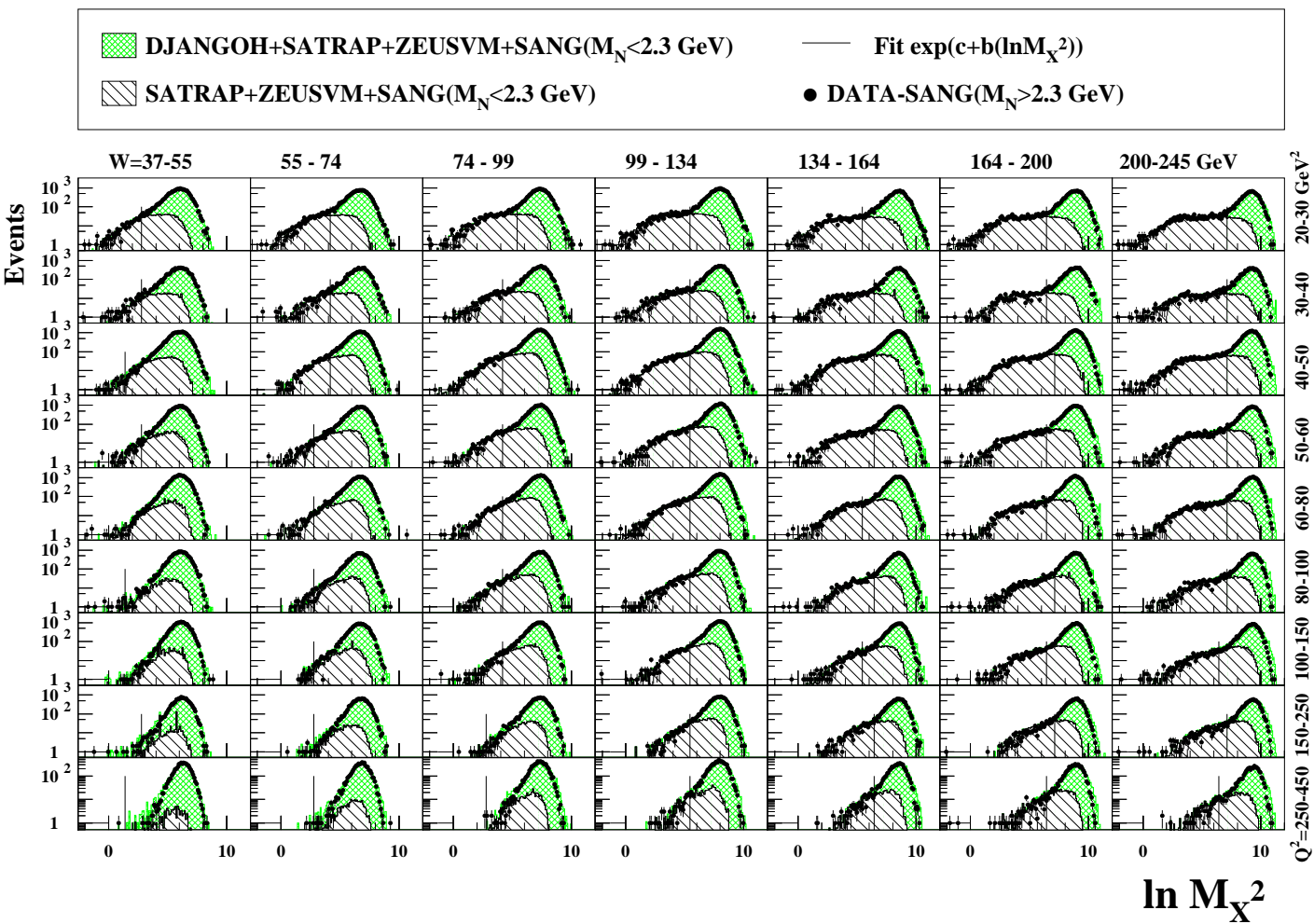,width=14.7cm,angle=0}}
\caption{Distributions of $\ln M^2_X$ ($M_X$ in units of\GeV) at the detector level for different ($W$, $Q^2$) bins. The points with error bars 
show the data, with the contribution from proton dissociation, as predicted by SANG, for $M_N > 2.3$ GeV subtracted. The diffractive contributions from $\gamma^* p \to XN$, $M_N < 2.3$\GeV, as predicted by SATRAP+ZEUSVM+SANG($M_N < 2.3$\GeV), are shown by the hatched histograms. The cross-hatched histograms show the non-peripheral contributions as predicted by DJANGOH plus the diffractive contributions. The slanted straight lines show the result for the non-diffractive contribution from fitting the data in the range $\ln W^2 - 4.4 < \ln M^2_X < \ln W^2 - 2.2$. The vertical lines indicate the maximum $M_X$ value up to which diffractive cross sections are determined.}
\label{f:lnmxall}
\vfill
\end{figure}
\clearpage

\begin{figure}[p]
\begin{center}
\includegraphics[angle=0,totalheight=20cm]{./{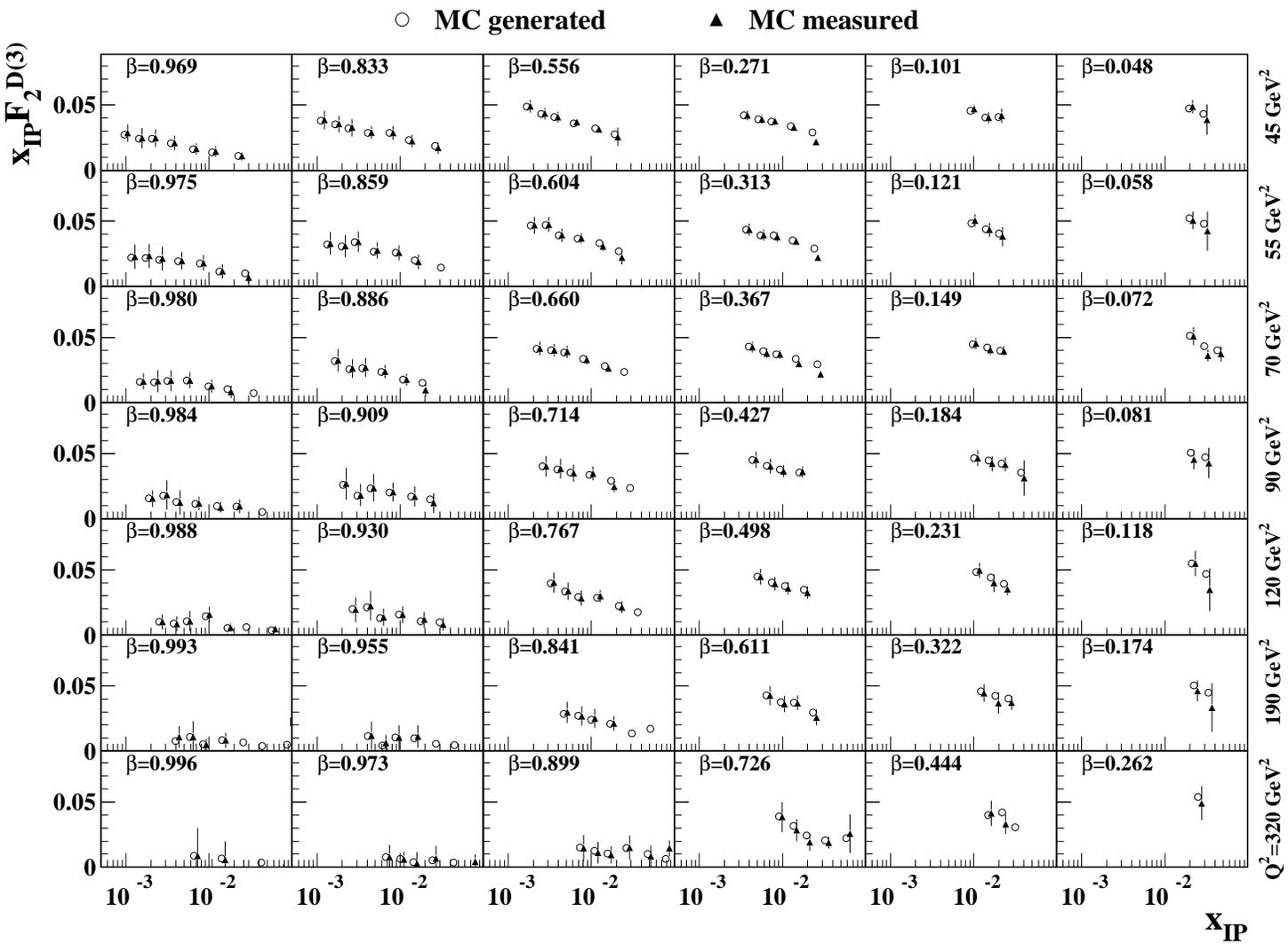}}
\caption{The diffractive structure function of the proton multiplied by $\xpom$, $\xpom F^{D(3)}_2$, as a function of $\xpom$ for different regions of $\beta$ and $Q^2$: comparison of the MC generated values (open points) with the MC measured values (solid triangles) as determined via the fit to the $\ln M^2_X$ distributions. The $\xpom F^{D(3)}_2$ values for MC measured are shown at values of $\xpom$ increased by 10\% .} 
\label{f:xpfd3vsxpgw}
\end{center}
\vfill
\end{figure}

\begin{figure}[p]
\vfill
{\epsfig{file=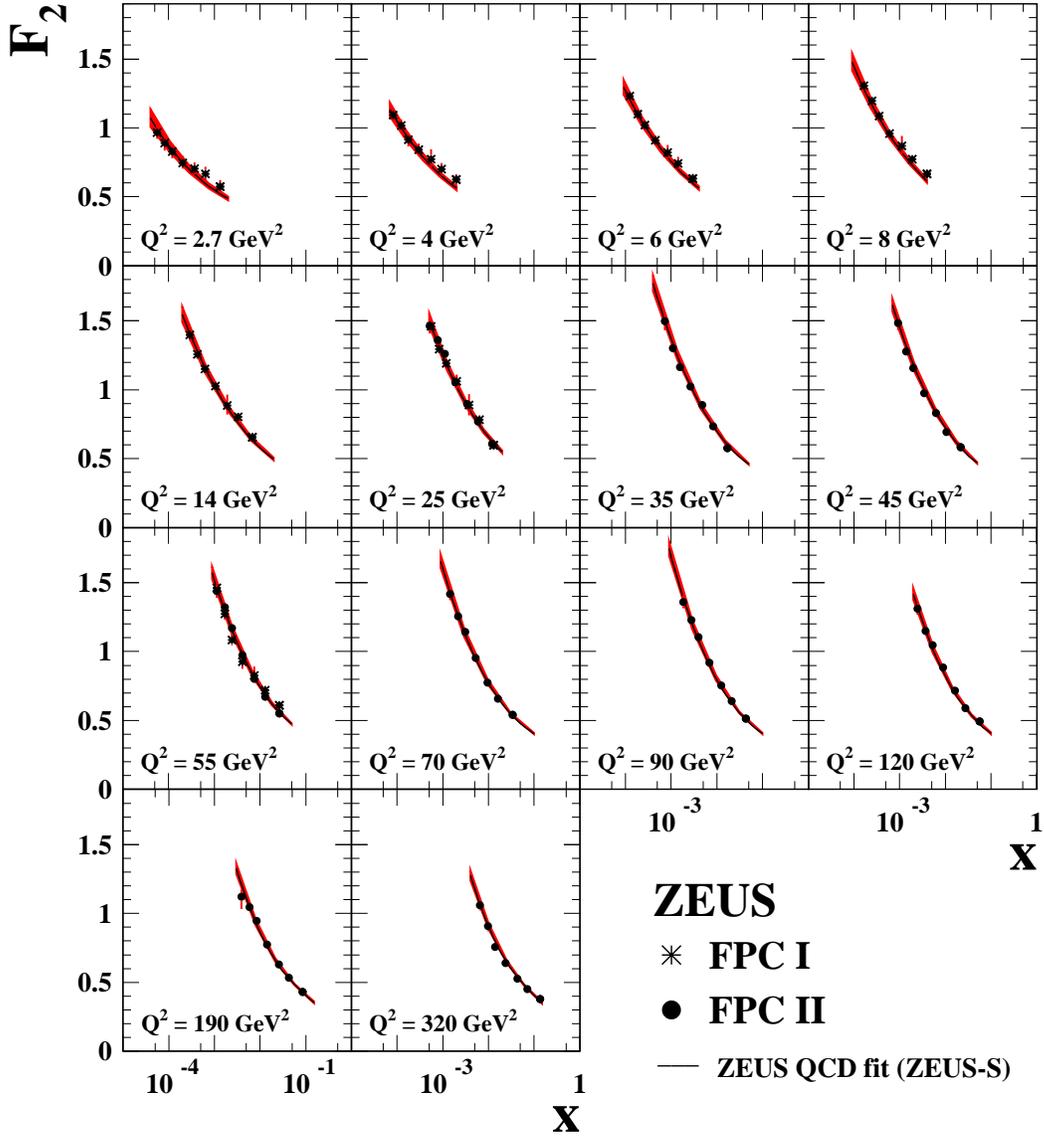,width=15cm}}
\caption{The proton structure function $F_2$ versus $x$ for the $Q^2$ values indicated. The results from this analysis, FPC~II, are shown together with those from the FPC~I analysis. The inner error bars show the statistical uncertainties and the full bars the statistical and systematic systematic uncertainties added in quadrature. The line shows the result of ZEUS-S NLO QCD fit with its uncertainty band.}
\label{f:fpcf2lh}
\vfill
\end{figure}
\clearpage

\begin{figure}
\begin{center}
\includegraphics[angle=90,totalheight=9.5cm]{./{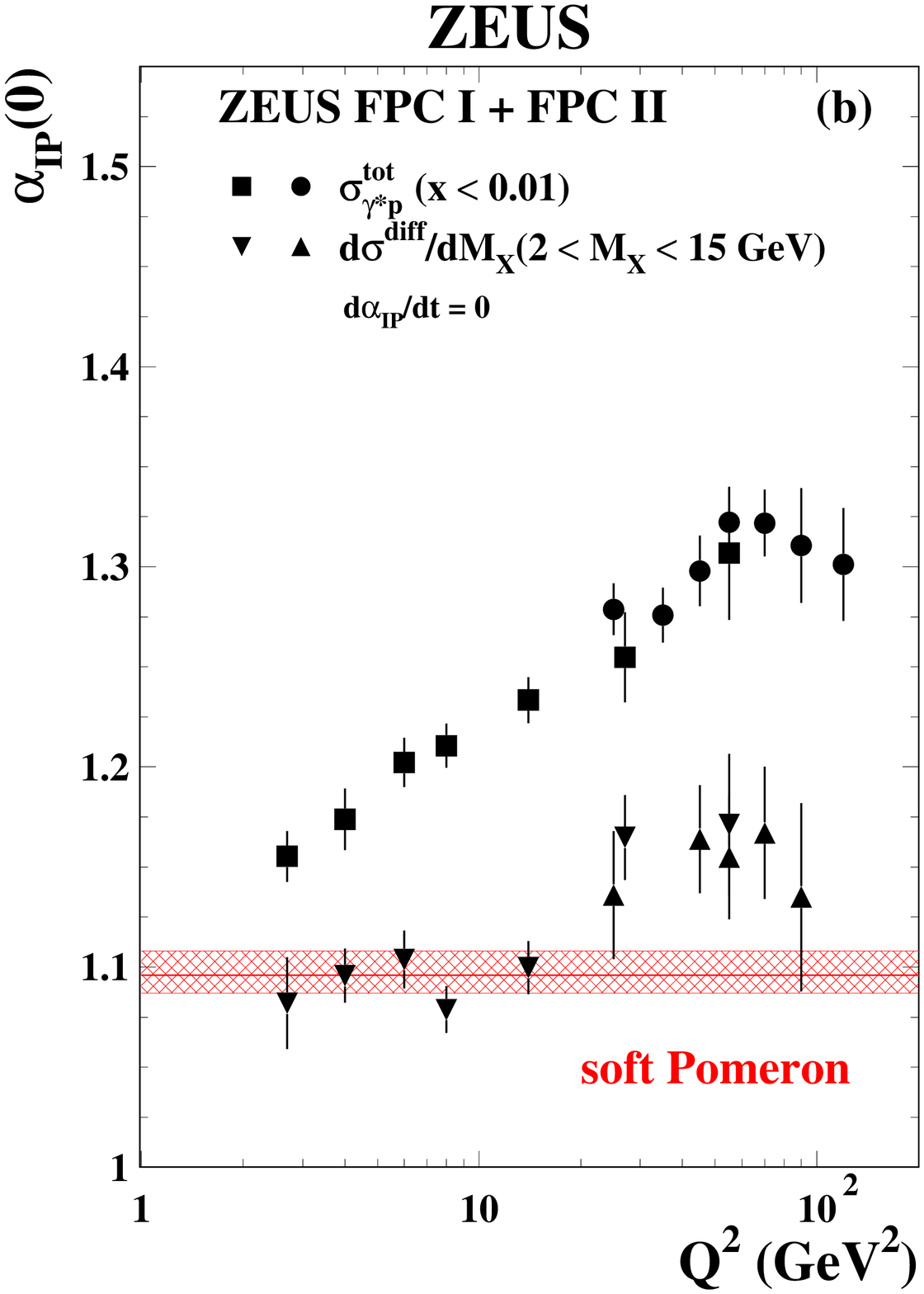}}
\includegraphics[angle=90,totalheight=9.5cm]{./{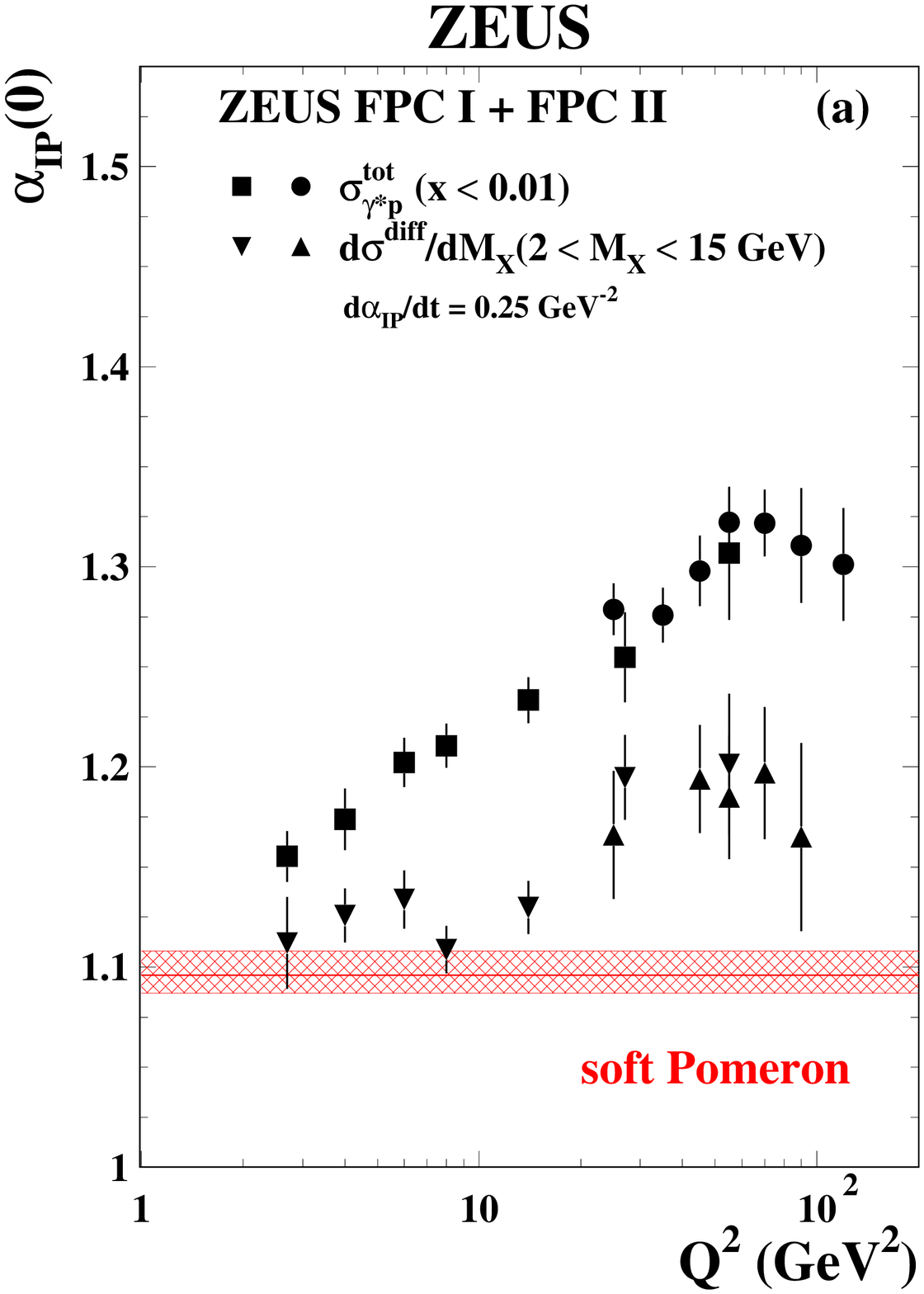}}
\caption{The intercepts of the Pomeron trajectory, $\alpha^{\rm tot}_{\pom}(0)$
and $\alpha^{\rm diff}_{\pom}(0)$, as a function of $Q^2$, obtained from the 
$W$ dependences of the total $\gamma^*p$ cross section and of the 
diffractive cross section, $d\sigma^{\rm diff}_{\gamma^* p \to XN}/dM_X$ for 
$2< M_X < 15$\GeV, from the FPC~I and FPC~II analyses, for (a) $\alpha_{\pom}^{\prime} = 0.25$\GeV$^{-2}$ and (b) $\alpha_{\pom}^{\prime} = 0$. The error bars show the sum of the statistical and systematic uncertainties added in quadrature. The shaded bands show the expectation for the soft Pomeron. }
\label{f:apomcomb}
\end{center}
\end{figure}

\begin{figure}[p]
\vspace*{-2cm}
\begin{center}
\includegraphics[angle=90,totalheight=10cm]{./{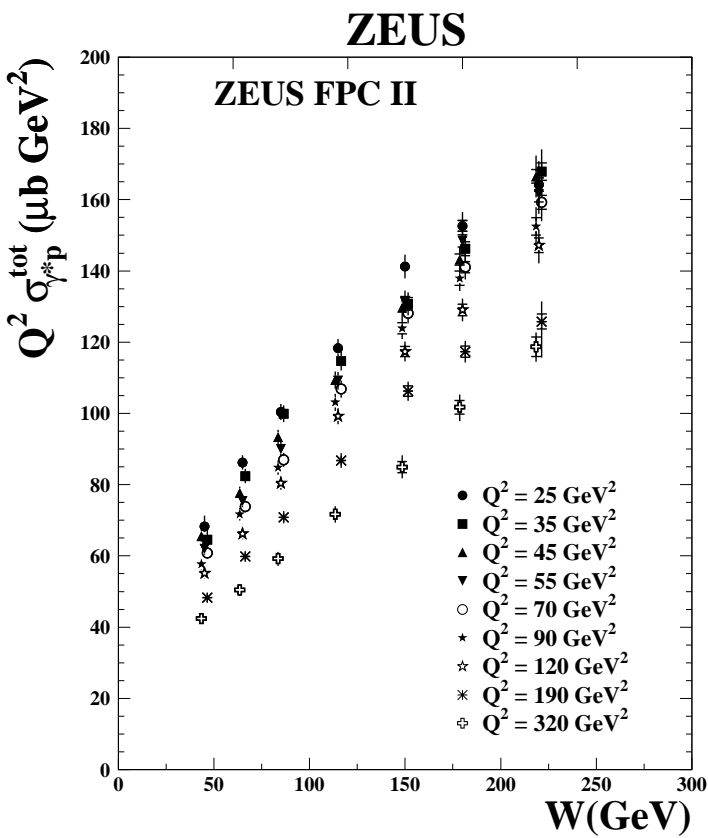}}
\includegraphics[angle=90,totalheight=10cm]{./{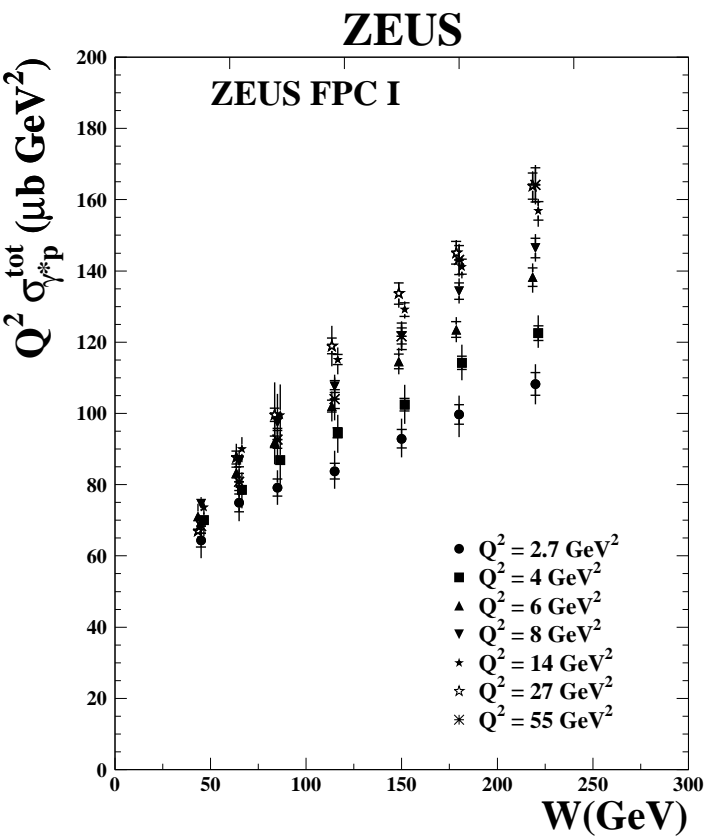}}
\caption{The total virtual photon-proton cross section, $\sigma^{\rm tot}_{\gamma^{\ast} p}$, multiplied by $Q^2$, as a function of $W$, for the $Q^2$ intervals indicated. The inner error bars show the statistical uncertainties and the full bars the statistical and systematic uncertainties added in quadrature. For better visibility, the points for adjacent values of $Q^2$ were shifted in $W$ by zero, +1.5\GeV or -1.5\GeV. Data are shown (a): from the FPC~I analysis; (b) from the FPC~II analysis.}
\label{f:sigtotlh}
\end{center}
\end{figure}

\begin{figure}
\begin{center}
\includegraphics[totalheight=12cm]{./{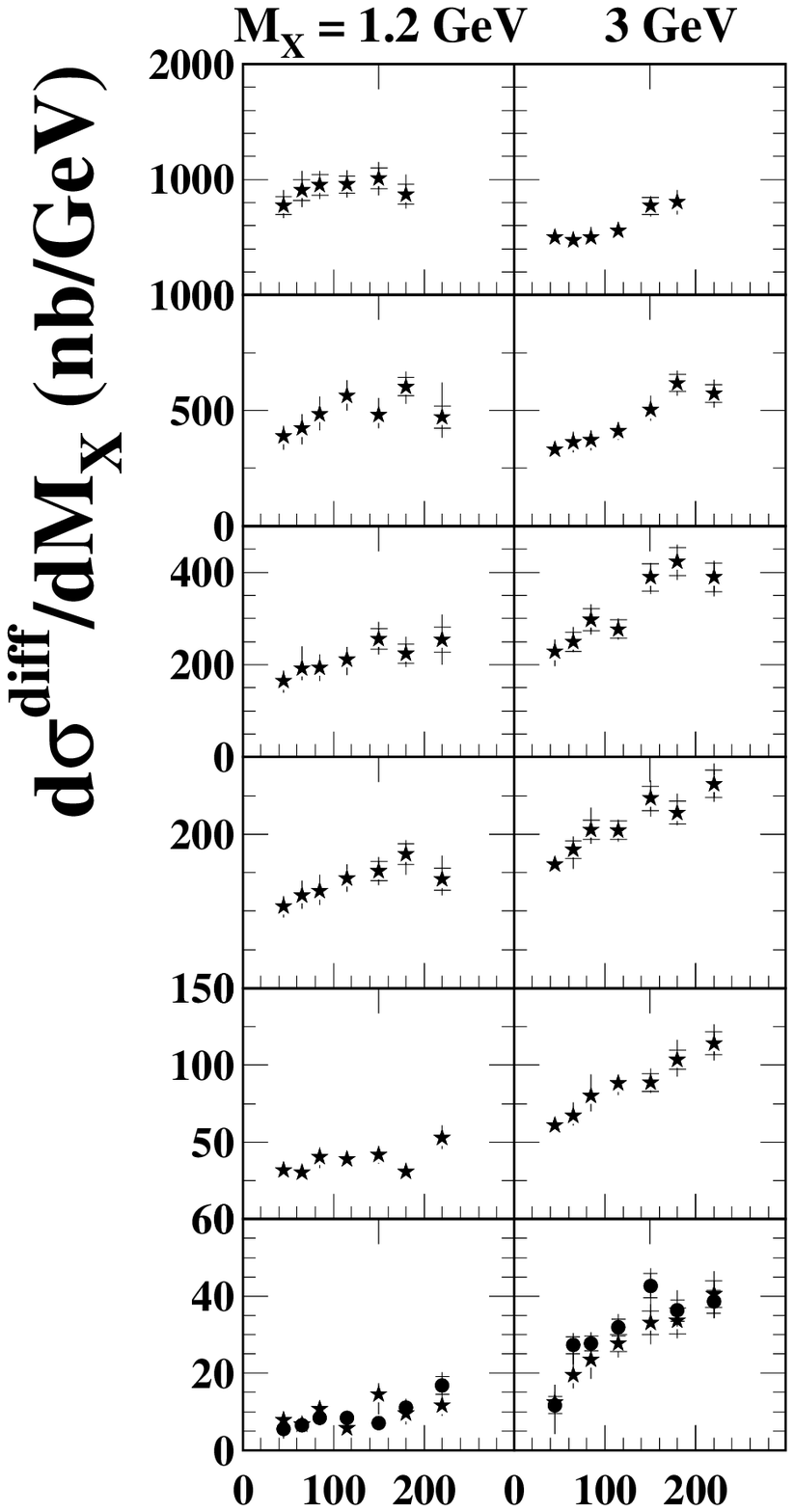}}
\includegraphics[totalheight=12cm]{./{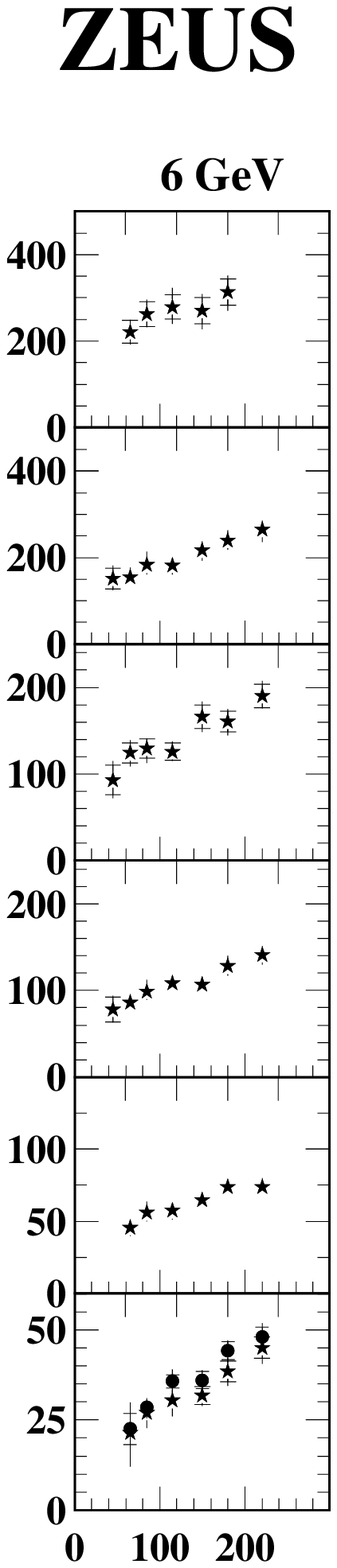}}
\includegraphics[totalheight=12cm]{./{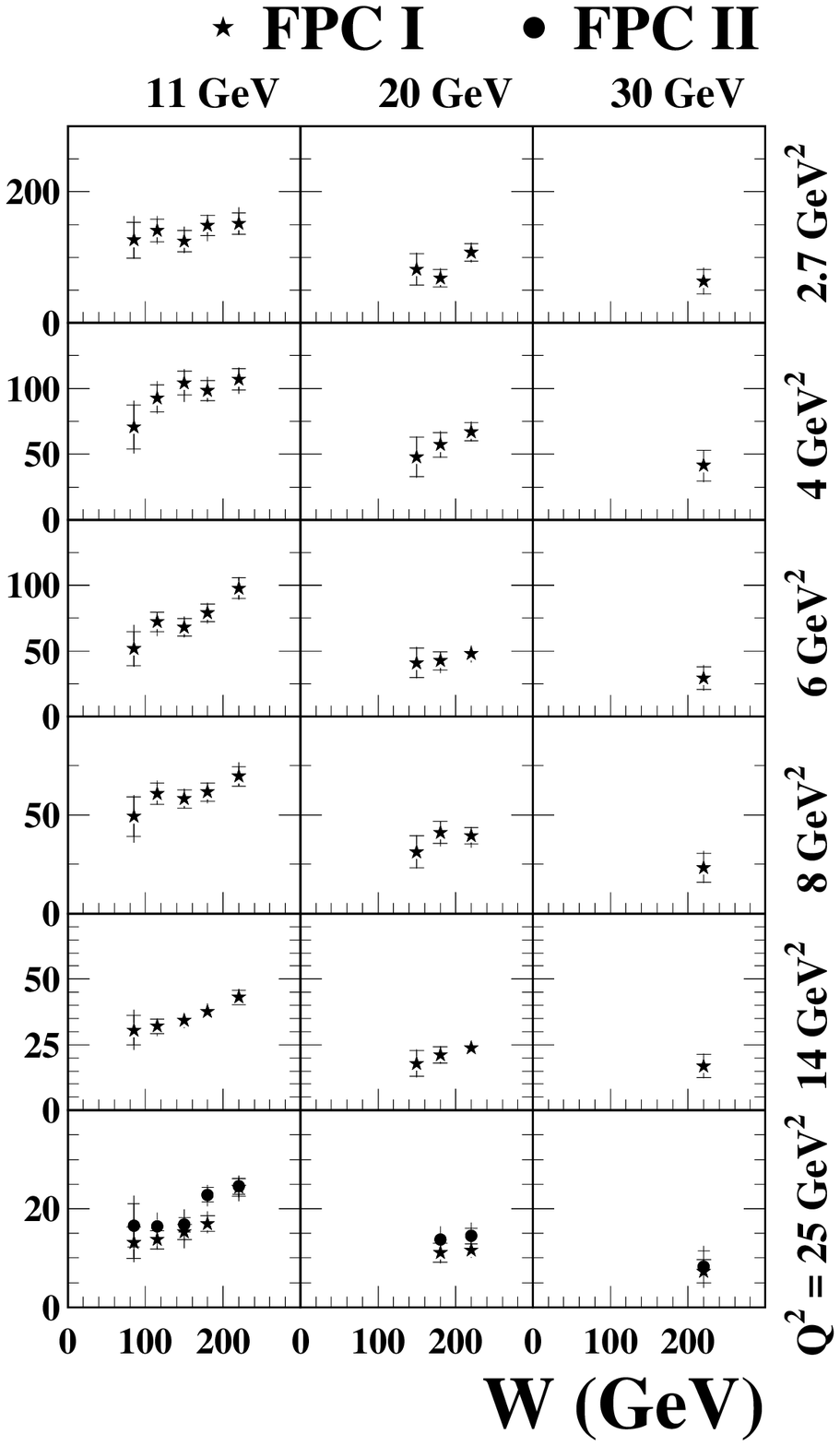}}
\caption{The differential cross sections, $d\sigma^{\rm diff}_{\gamma^*p \to XN}/dM_X$, $M_N < 2.3$\GeV, as a function of $W$ for bins of $Q^2$ and of $M_X$, for FPC~I data (stars) and FPC~II data (dots), for $Q^2$ between 2.7 and 25$\GeV^2$.  The inner error bars show the statistical uncertainties and the full bars the statistical and systematic uncertainties added in quadrature.}
\label{f:dsigdmxlh2.7.25}
\end{center}
\end{figure}

\begin{figure}
\begin{center}
\includegraphics[totalheight=15cm]{./{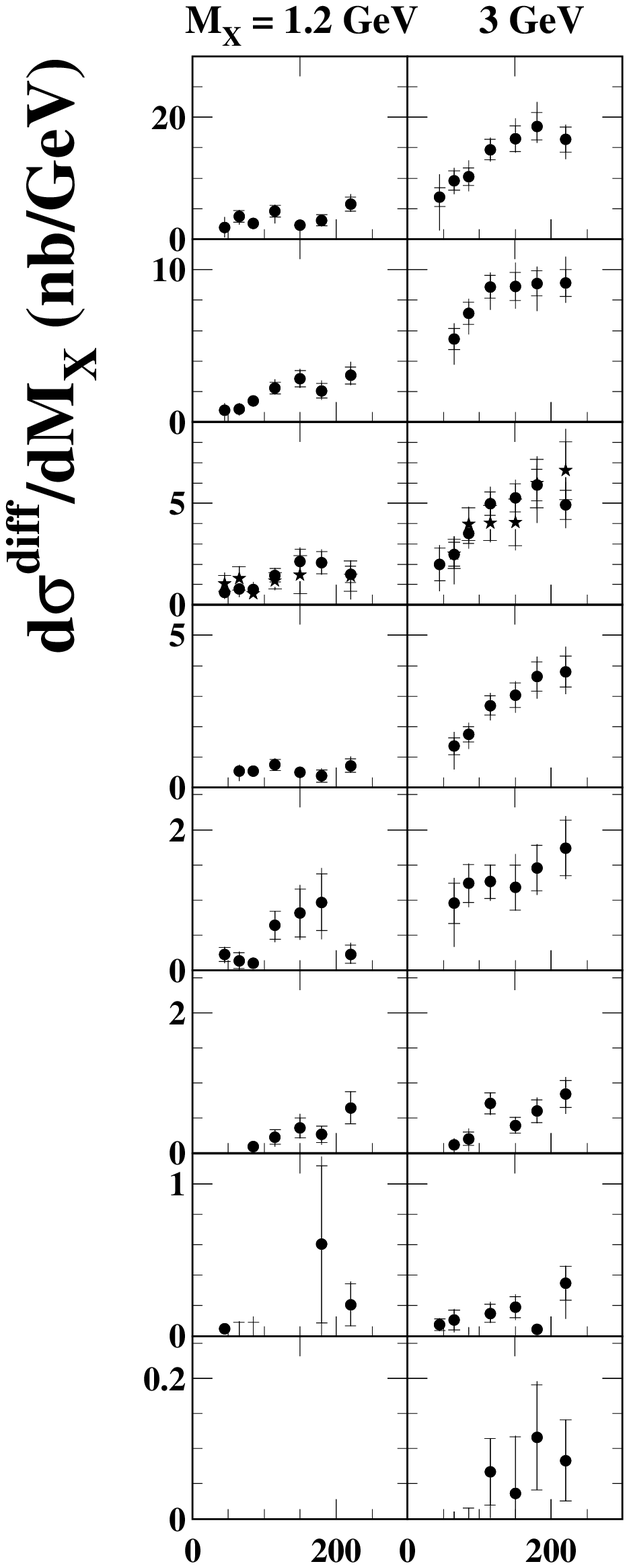}}
\includegraphics[totalheight=15cm]{./{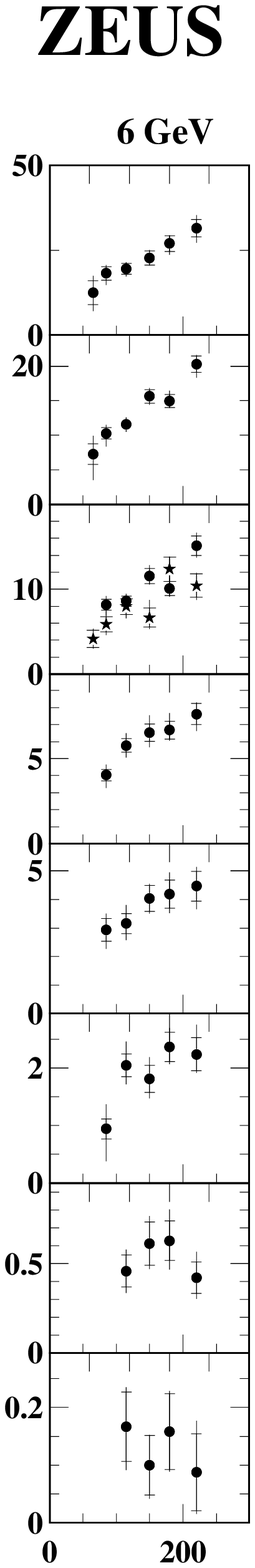}}
\includegraphics[totalheight=15cm]{./{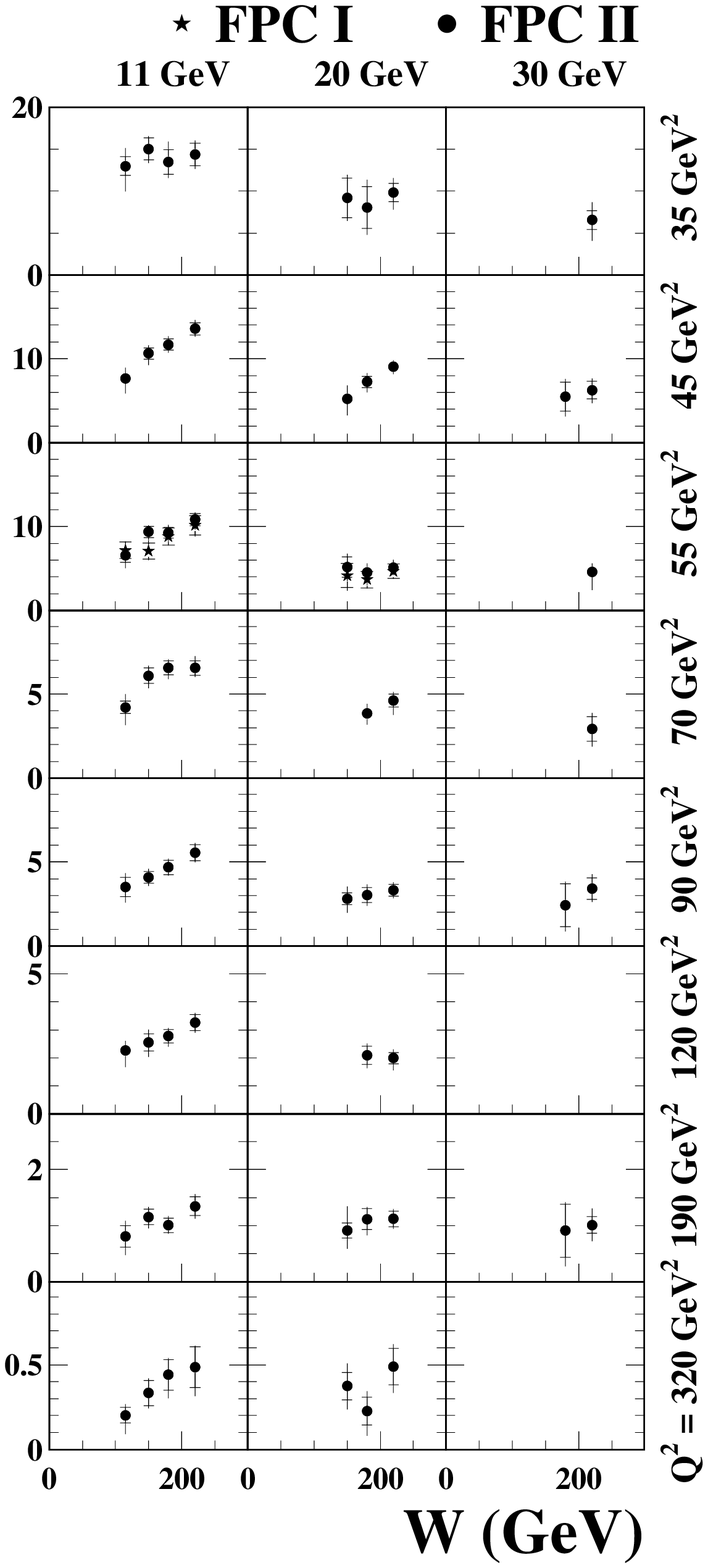}}
\caption{The differential cross sections, $d\sigma^{\rm diff}_{\gamma^*p \to XN}/dM_X$, $M_N < 2.3$\GeV, as a function of $W$ for bins of $Q^2$ 
and of $M_X$, for FPC~I data (stars) and FPC~II data (dots), for $Q^2$ between 35 and 320\GeV$^2$. The inner error bars show the statistical uncertainties and the full bars the statistical and systematic uncertainties added in quadrature. 
For display purposes, some of the cross section values at $Q^2 = 320$\GeV$^2$ are not shown but given in Tables~\ref{t:dsigdmx1} --~\ref{t:dsigdmx6}.}
\label{f:dsigdmxlh.35.320}
\end{center}
\end{figure}

\begin{figure}[p]
\vfill
{\epsfig{file=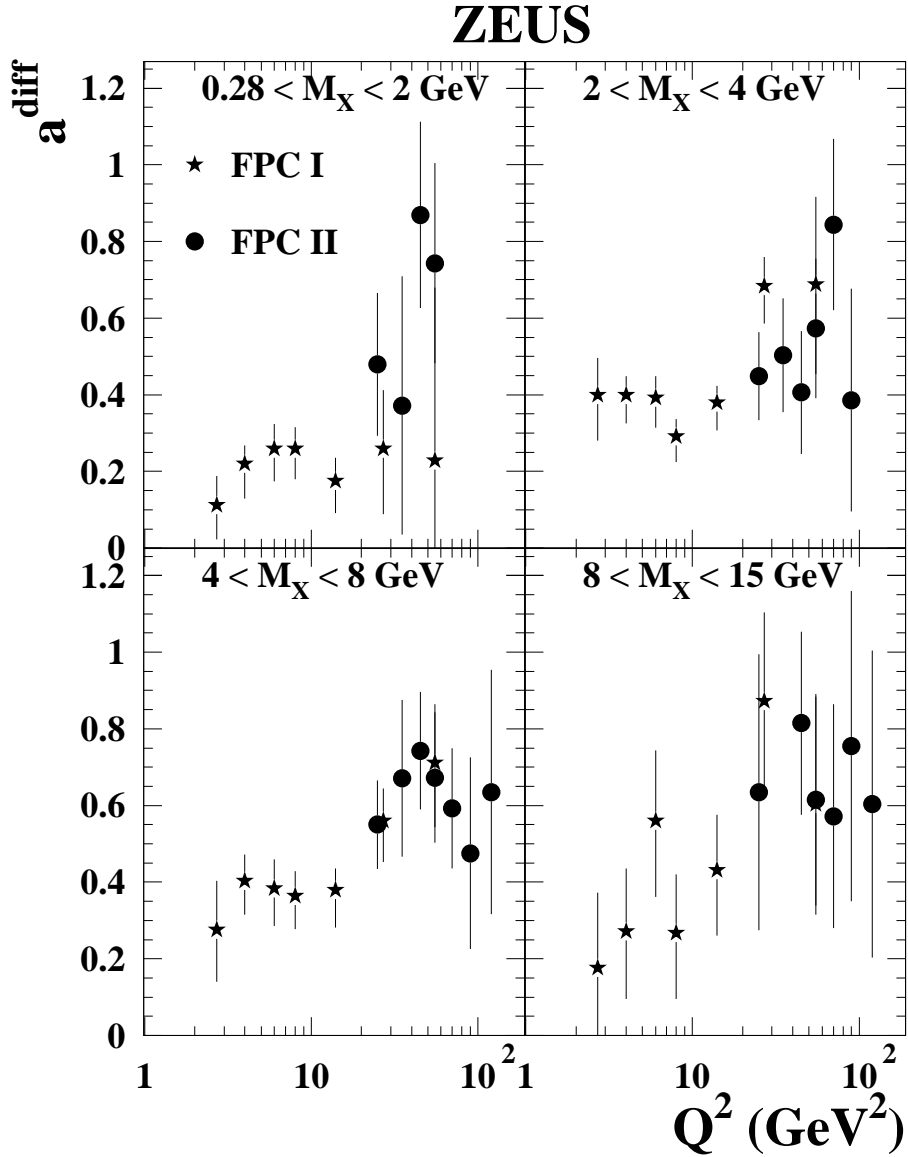,width=17cm}}
\caption{The power $a^{\rm diff}$, obtained from fitting the diffractive cross section to the form $d\sigma^{\rm diff}_{\gamma^* p \to XN}(M_X,W, Q^2)/dM_X \propto (W/W_0)^{a^{\rm diff}}$, $W_0 = 1$\GeV, as a function of $Q^2$ for the different $M_X$ bins indicated, for FPC~I data (stars) and FPC~II data (dots). The error bars show the statistical and systematic uncertainties added in quadrature.}
\label{f:difslope12}
\vfill
\end{figure}

\begin{figure}
\vspace*{-1cm}
\begin{center}
\includegraphics[totalheight=11cm]{./{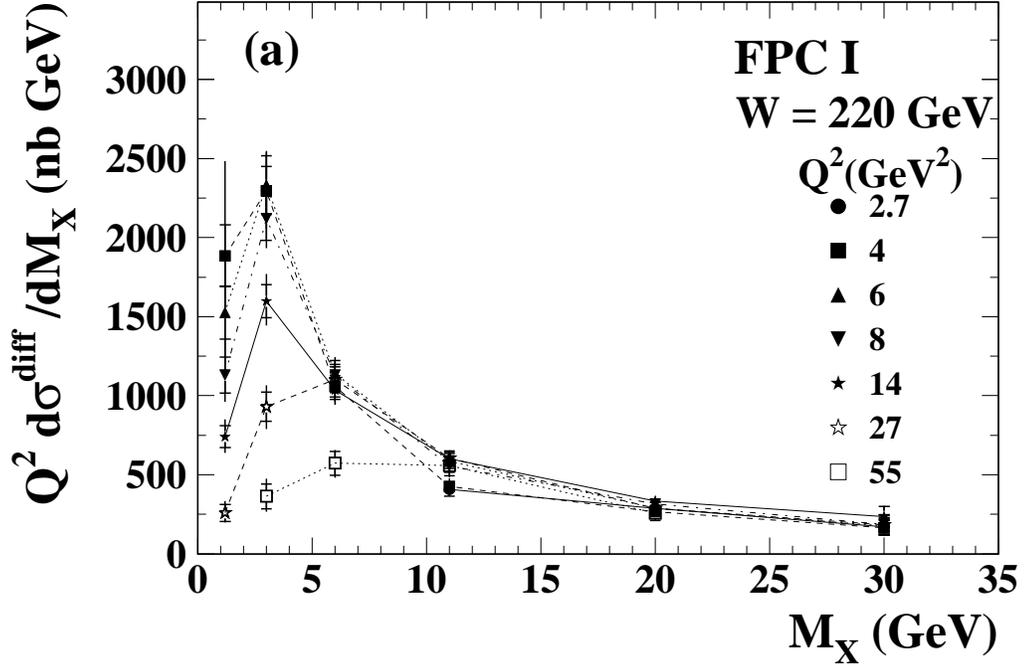}}
\end{center}
\vspace*{-1cm}
\begin{center}
\includegraphics[totalheight=11cm]{./{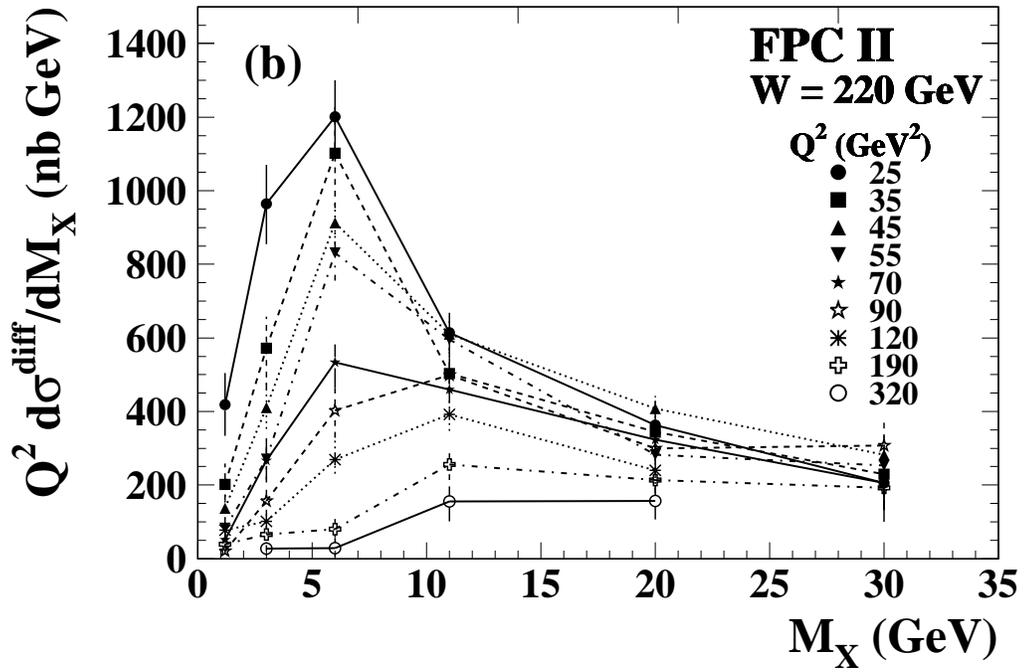}}
\caption{The diffractive cross section multiplied by $Q^2$,  $Q^2 d\sigma^{\rm diff}_{\gamma^*p \to XN}/dM_X$, $M_N < 2.3$\GeV, $W = 220$\GeV as a function of $M_X$ for the $Q^2$ values indicated, for (a) FPC~I data and (b) FPC~II data. The inner error bars show the statistical uncertainties and the full bars the statistical and systematic uncertainties added in quadrature.}
\label{f:dsdmmxlh}
\end{center}
\end{figure}

\begin{figure}
\vspace*{-1.5cm}
\begin{center}
\includegraphics[angle=90,totalheight=10cm]{./{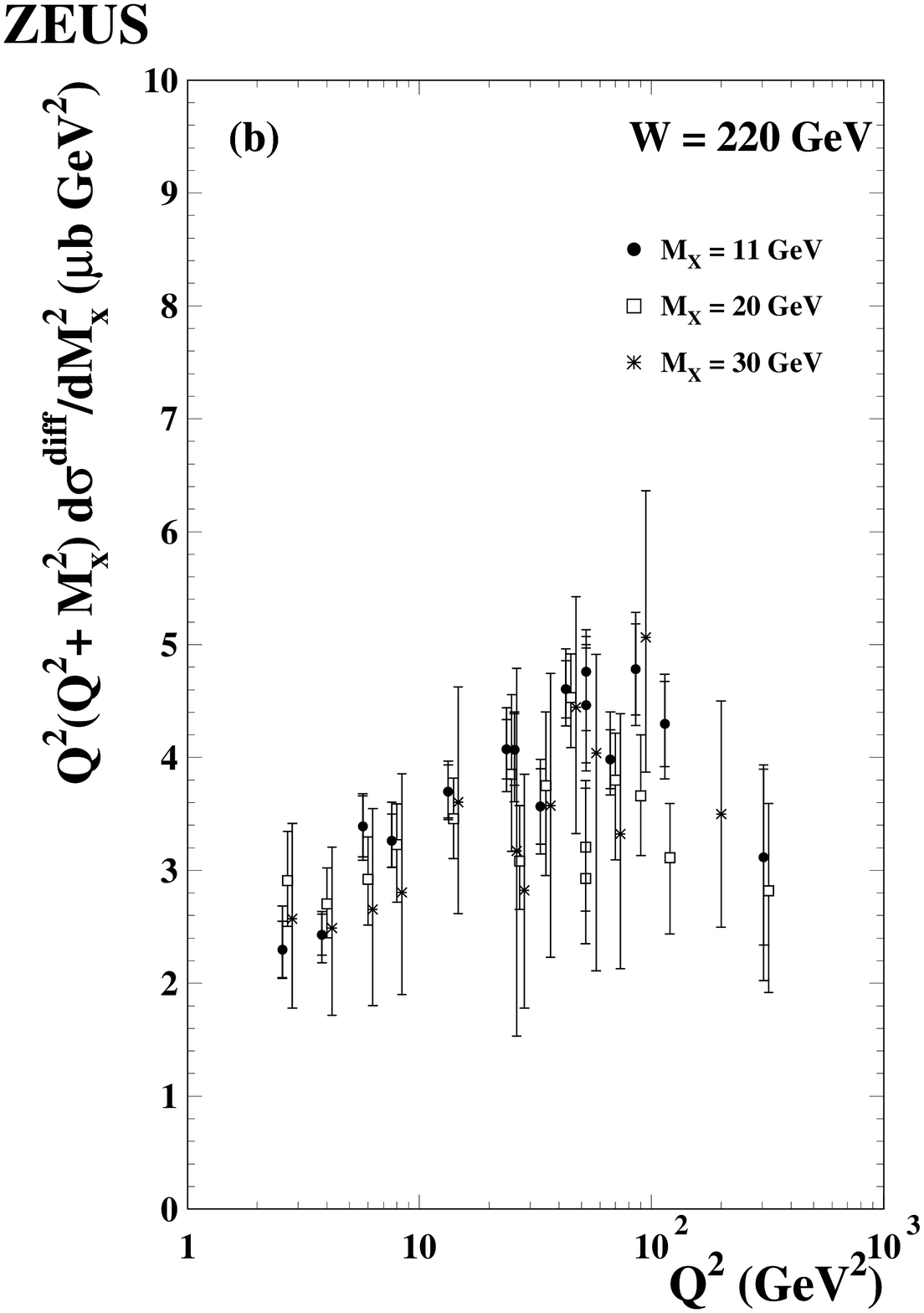}}
\includegraphics[angle=90,totalheight=10cm]{./{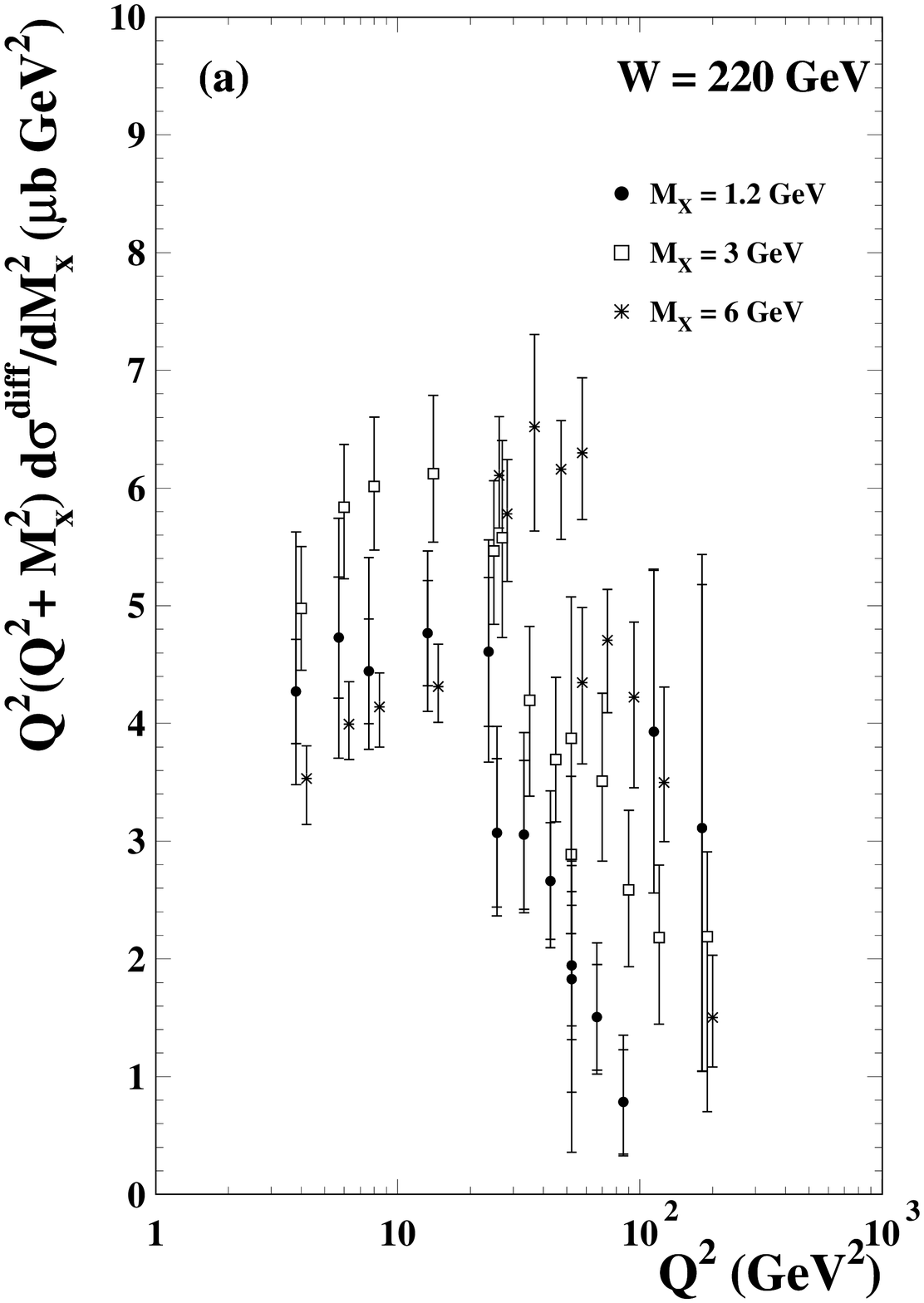}}
\caption{The diffractive cross section multiplied by $Q^2(Q^2+M^2_X)$,  $Q^2(Q^2+M^2_X)d\sigma^{\rm diff}_{\gamma^*p \to XN}/dM^2_X$, $M_N < 2.3$\GeV, for $W = 220$\GeV as a function of $Q^2$ for (a) $M_X = 1.2, 3, 6 $\GeV and (b) $M_X = 11, 20, 30$\GeV. Shown are the combined results from the FPC~I data and FPC~II data. The inner error bars show the statistical uncertainties and the full bars the statistical and systematic uncertainties added in quadrature. For better visibility, in each figure the $x$-axis values of the data points with the lowest (highest) value of $M_X$ have been decreased (increased) by a factor of $1.05$. }
\label{f:dsdmq2lh}
\end{center}
\end{figure}

\begin{figure}[p]
\vfill
{\epsfig{file=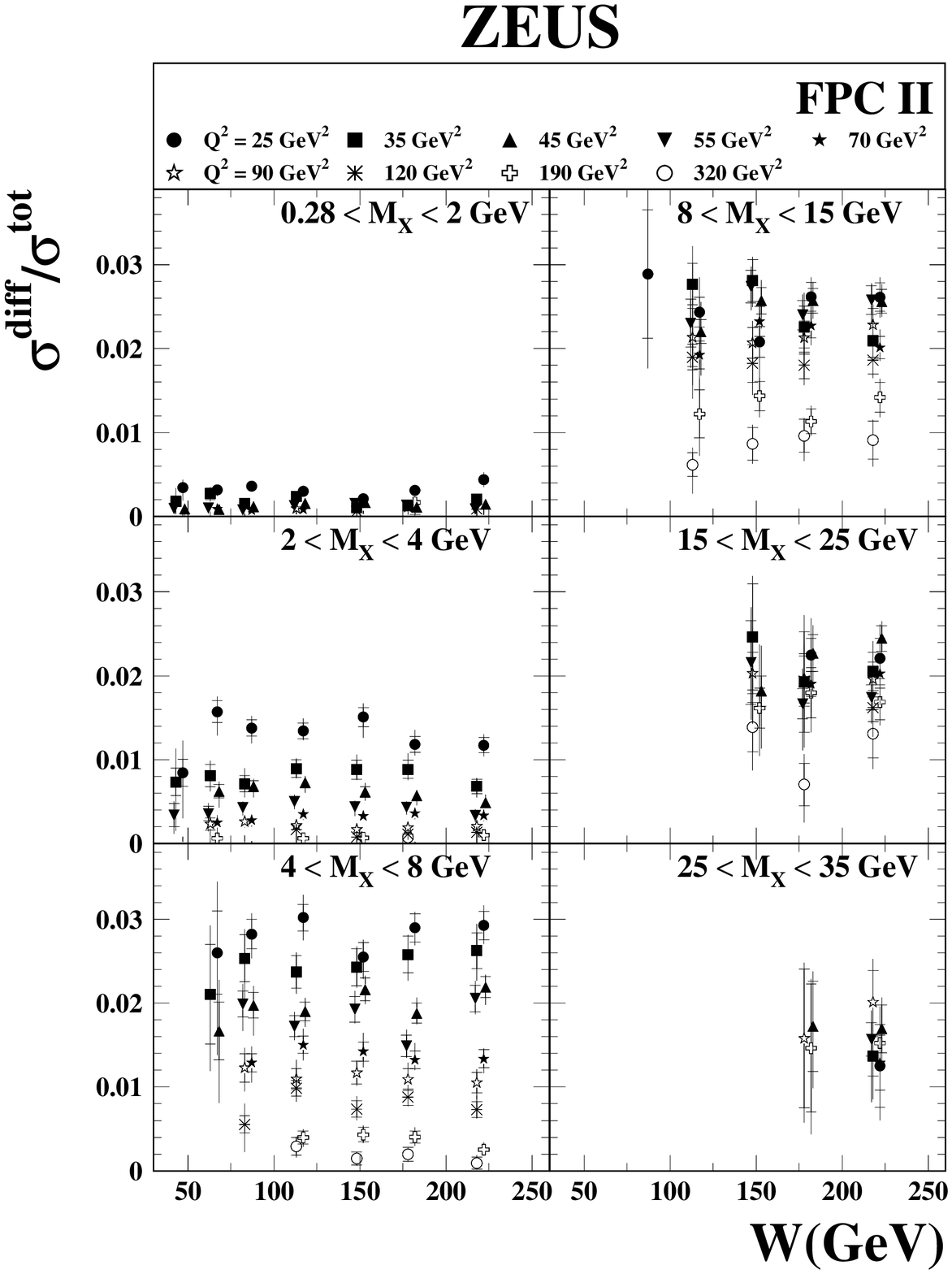,width=14.7cm,clip=}}
\caption{The ratio of the diffractive cross section, integrated over the $M_X$ intervals indicated, $\int^{M_b}_{M_a} dM_X d\sigma^{\rm diff}_{\gamma^* p \to XN, M_N < 2.3 {\rm GeV}}/dM_X$, to the total $\gamma ^{\ast}p$ cross section, $r^{\rm diff}_{\rm tot} = \sigma^{\rm diff}/\sigma^{\rm tot}_{\gamma^*p}$, as a function of $W$ for the $Q^2$ values indicated, for the FPC~II data. The inner error bars show the statistical uncertainties and the full bars the statistical and systematic uncertainties added in quadrature.}
\label{f:rdiftot}
\vfill
\end{figure}

\begin{figure}[p]
\vfill
{\epsfig{file=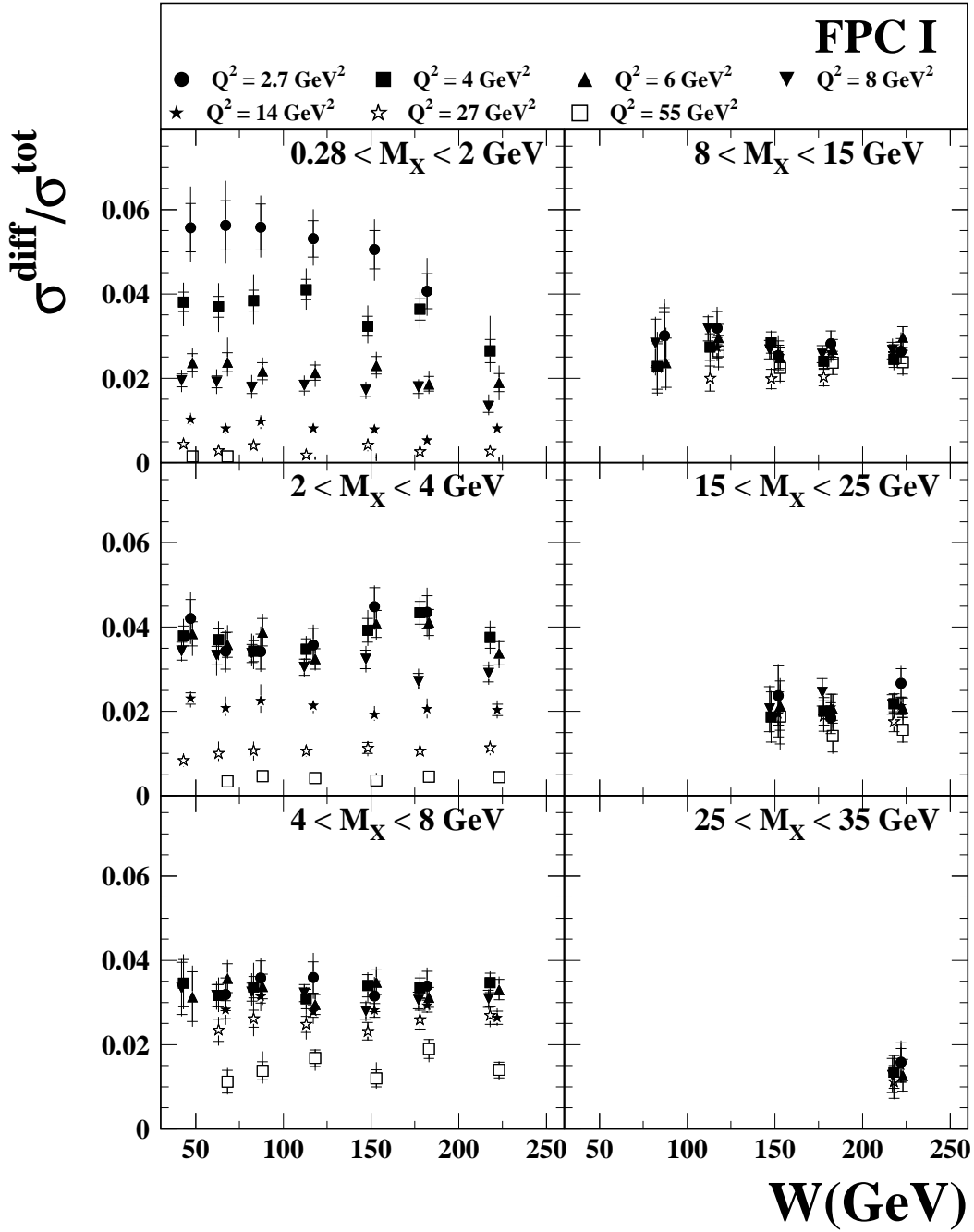,width=14.7cm,clip=}}
\caption{The ratio of the diffractive cross section, integrated over the $M_X$ intervals indicated, $\int^{M_b}_{M_a} dM_X d\sigma^{\rm diff}_{\gamma^* p \to XN, M_N < 2.3 {\rm GeV}}/dM_X$, to the total $\gamma ^{\ast}p$ cross section, $r^{\rm diff}_{\rm tot} = \sigma^{\rm diff}/\sigma^{\rm tot}_{\gamma^*p}$, as a function of $W$ for the $Q^2$ intervals indicated. The inner error bars show the statistical uncertainties and the full bars the statistical and systematic uncertainties added in quadrature. The plot shown is taken directly from the FPC~I paper.}
\label{f:rdiftotl}
\vfill
\end{figure}

\begin{figure}[p]
\vspace*{-1.5cm}
\vfill
{\epsfig{file=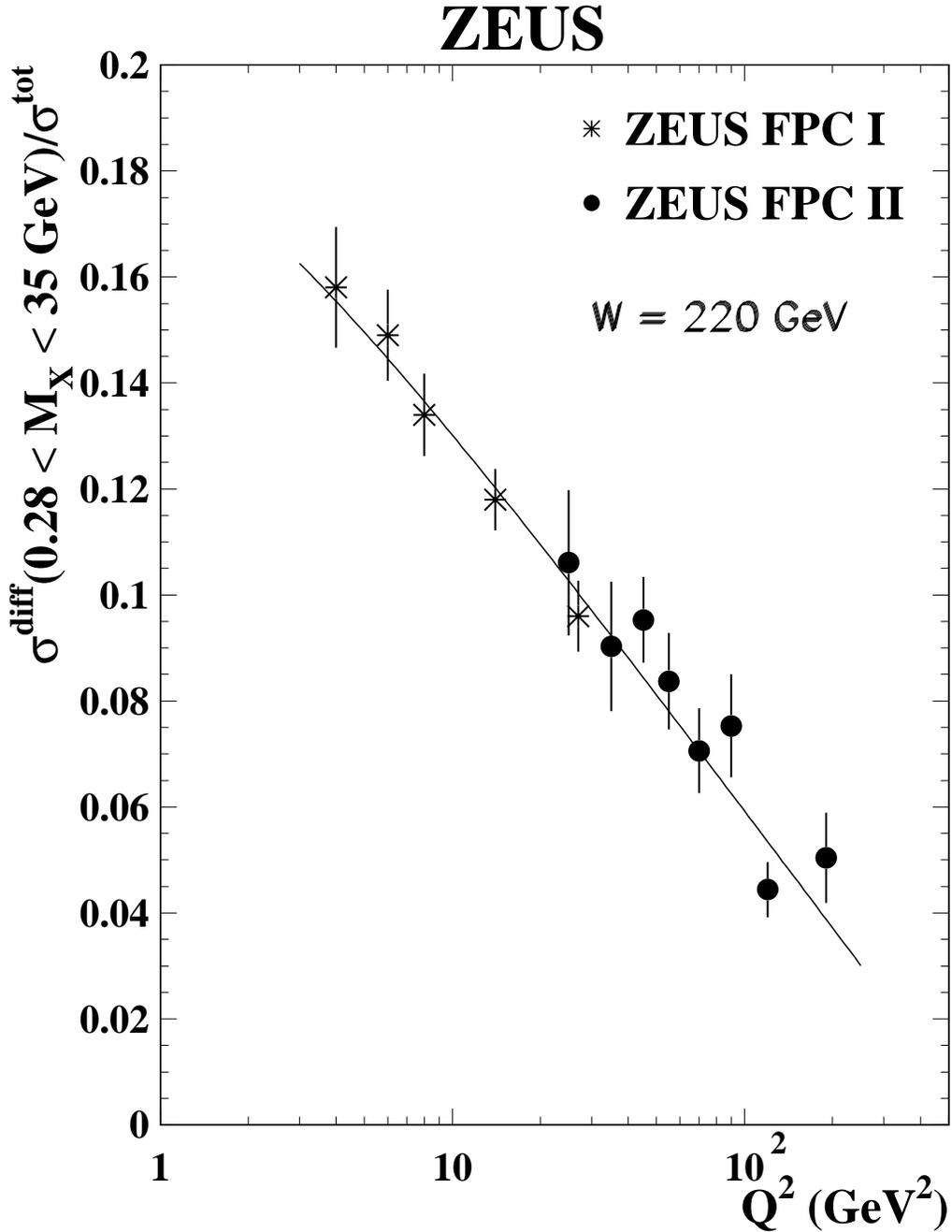,width=14.7cm,clip=}}
\caption{The ratio of the diffractive cross section, integrated over $0.28 < M_X < 35$\GeV, to the total $\gamma ^{\ast}p$ cross section, at $W = 220$\GeV as a function of $Q^2$. The error bars represent the statistical and systematic uncertainties added in quadrature. Shown are the combined data from FPC~I (stars) and FPC~II (dots).  The line shows the result of fitting the data to the form $r = a  - b \cdot \ln (1+Q^2)$, see text.}
\label{f:rdiftot220}
\vfill
\end{figure}

\begin{figure}
\vspace*{-1.5cm}
\begin{center}
{\epsfig{file=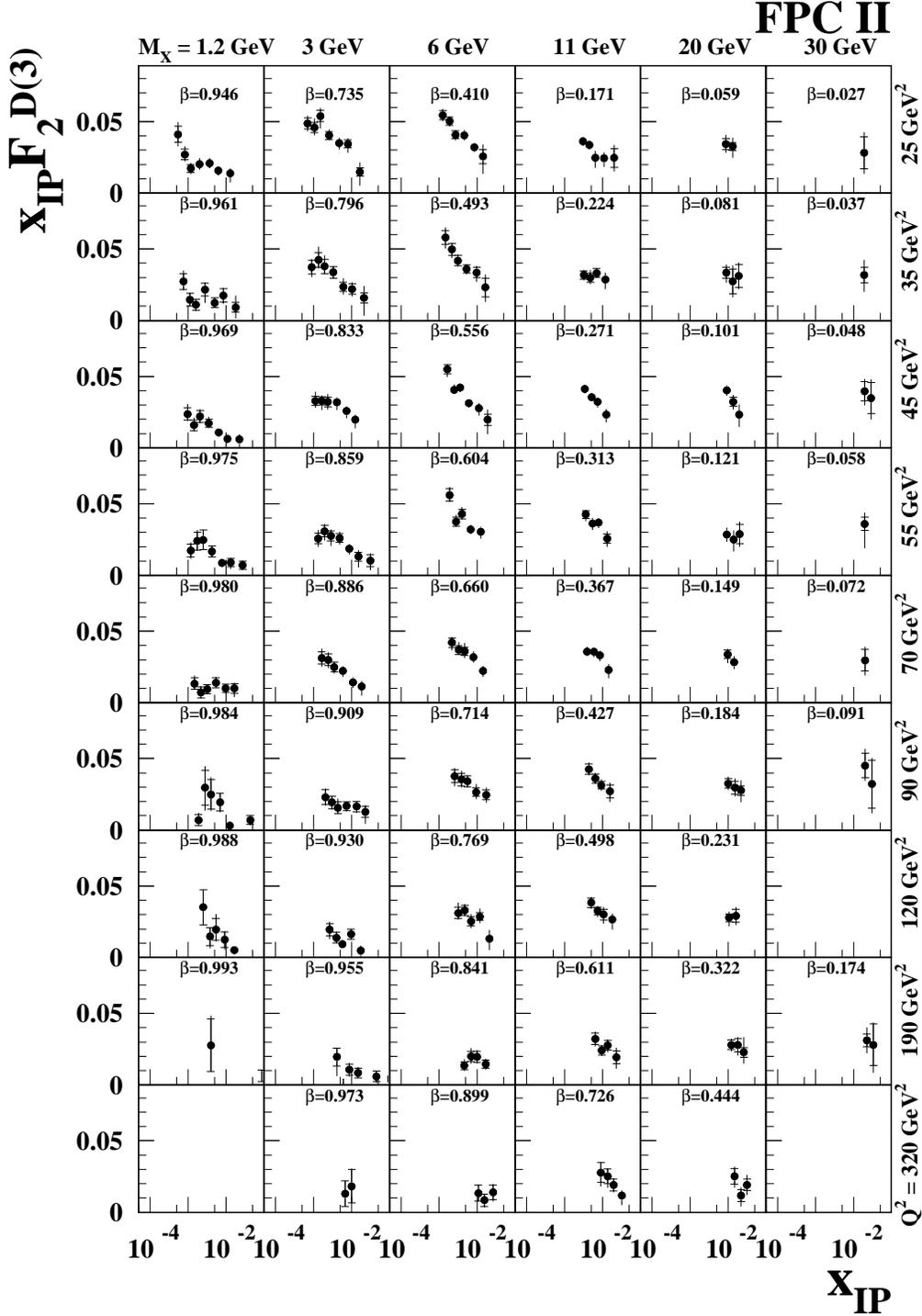,width=14.3cm}}
\caption{The diffractive structure function of the proton multiplied by $\xpom$, $\xpom F^{D(3)}_2$, for $\gamma^*p \to XN$, $M_N < 2.3$\GeV as a function of $\xpom$ for different regions of $\beta$ and $Q^2$, for the FPC~II data. The inner error bars show the statistical uncertainties and the full bars the statistical and systematic uncertainties added in quadrature. For display purposes, some of the $\xpom F^{D(3)}_2$ values at $Q^2 = 90, 120, 190$ and $320$\GeV$^2$ with large uncertainties are not shown but given in Tables~\ref{t:f2d301} --~\ref{t:f2d305}.}
\label{f:f2d3vsxph}
\end{center}
\end{figure}
\clearpage

\begin{figure}[p]
\centerline
{\epsfig{file=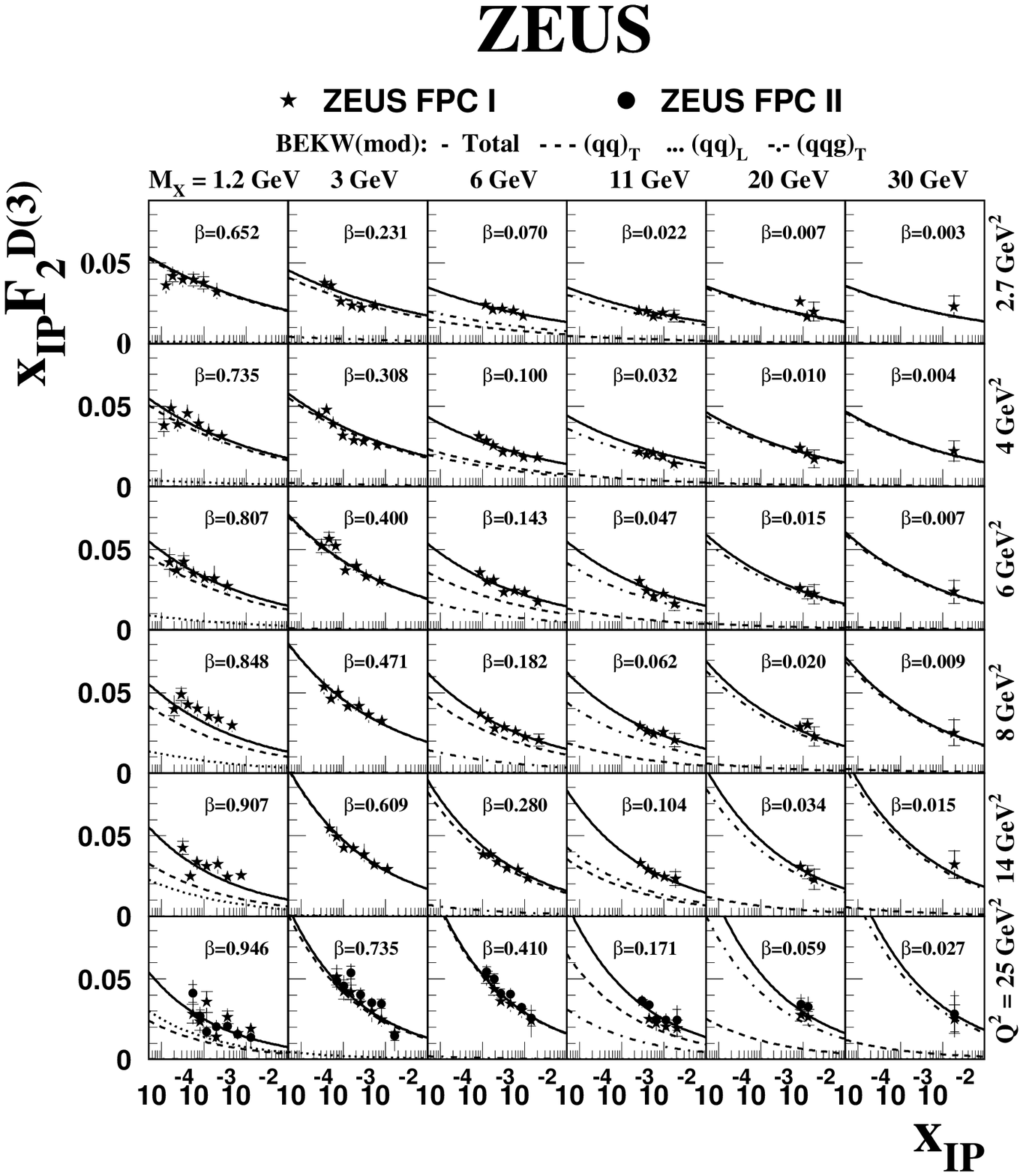,width=15cm}}
\caption{The diffractive structure function of the proton multiplied by $\xpom$, $\xpom F^{D(3)}_2$, for $\gamma^*p \to XN$, $M_N < 2.3$\GeV as a function of $\xpom$ for different regions of $\beta$ and $Q^2 \le 25$\GeV$^2$: both  FPC~I data (stars) and FPC~II data (dots) are shown. The inner error bars show the statistical uncertainties and the full bars the statistical and systematic uncertainties added in quadrature. The curves show the results of the BEKW(mod) fit for the contributions from $(q \overline{q})$ for transverse (dashed) and longitudinal photons (dotted) and for the $(q \overline{q}g)$ contribution for transverse photons (dashed-dotted) together with the sum of all contributions (solid).}
\label{f:f2d3vsxplh1}
\end{figure}
\clearpage

\begin{figure}[p]
\vspace*{-3.8cm}
\centerline
{\epsfig{file=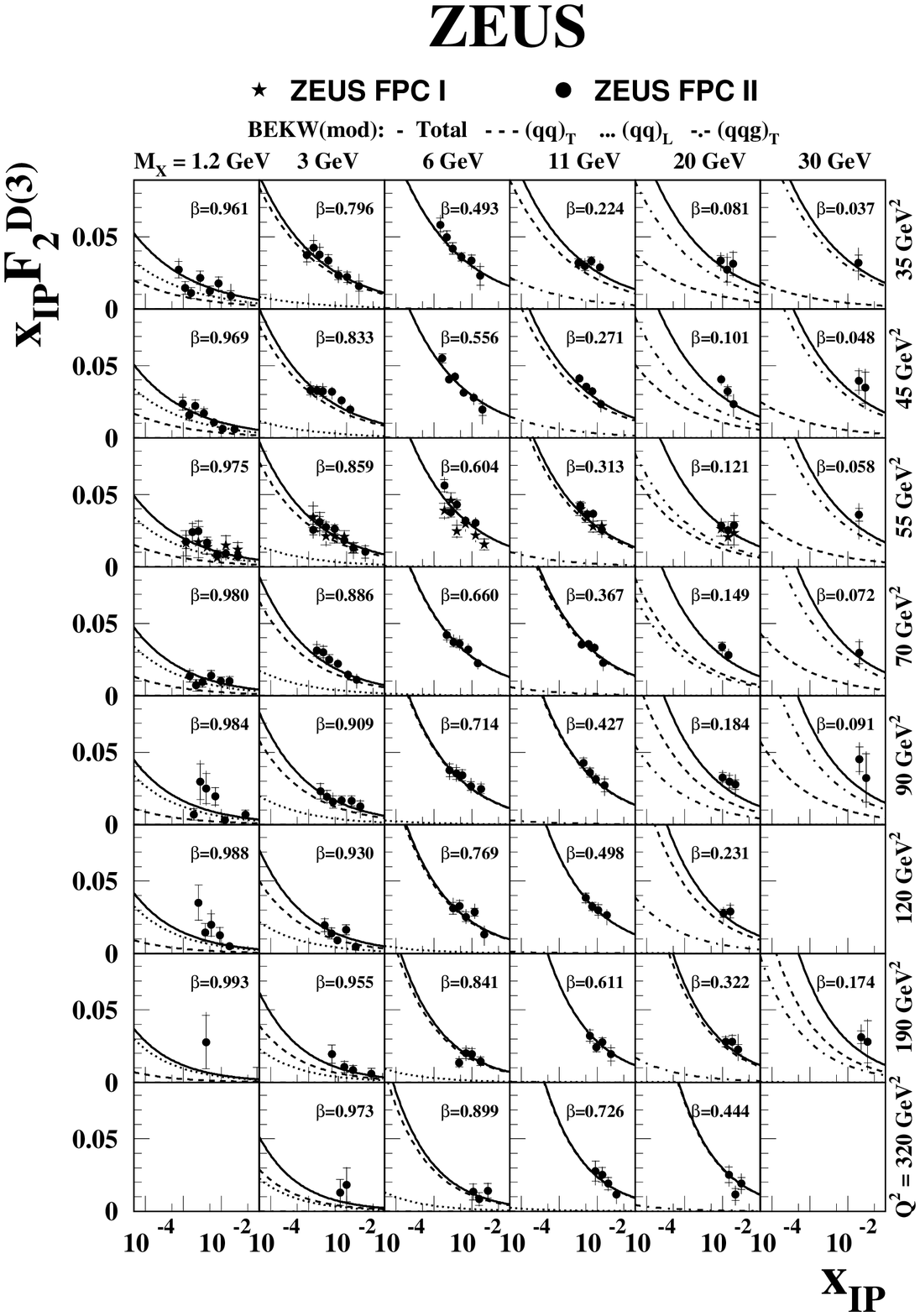,width=15cm}}
\caption{The diffractive structure function of the proton multiplied by $\xpom$, $\xpom F^{D(3)}_2$, for $\gamma^*p \to XN$, $M_N < 2.3$\GeV as a function of $\xpom$ for different regions of $\beta$ and $Q^2 \ge 35$\GeV$^2$: both FPC~I data (stars) and FPC~II data (dots) are shown. The inner error bars show the statistical uncertainties and the full bars the statistical and systematic uncertainties added in quadrature. The curves show the results of the BEKW(mod) fit for the contributions from $(q \overline{q})$ for transverse (dashed) and longitudinal photons (dotted) and for the $(q \overline{q}g)$ contribution for transverse photons (dashed-dotted) together with the sum of all contributions (solid). For display purposes, some of the $\xpom F^{D(3)}_2$ values at $Q^2 = 90, 120, 190$ and $320$\GeV$^2$ with large uncertainties are not shown but given in Tables~\ref{t:f2d303} --~\ref{t:f2d305}.} 
\label{f:f2d3vsxplh2}
\end{figure}
\clearpage

\begin{figure}[p]
\centerline
{\epsfig{file=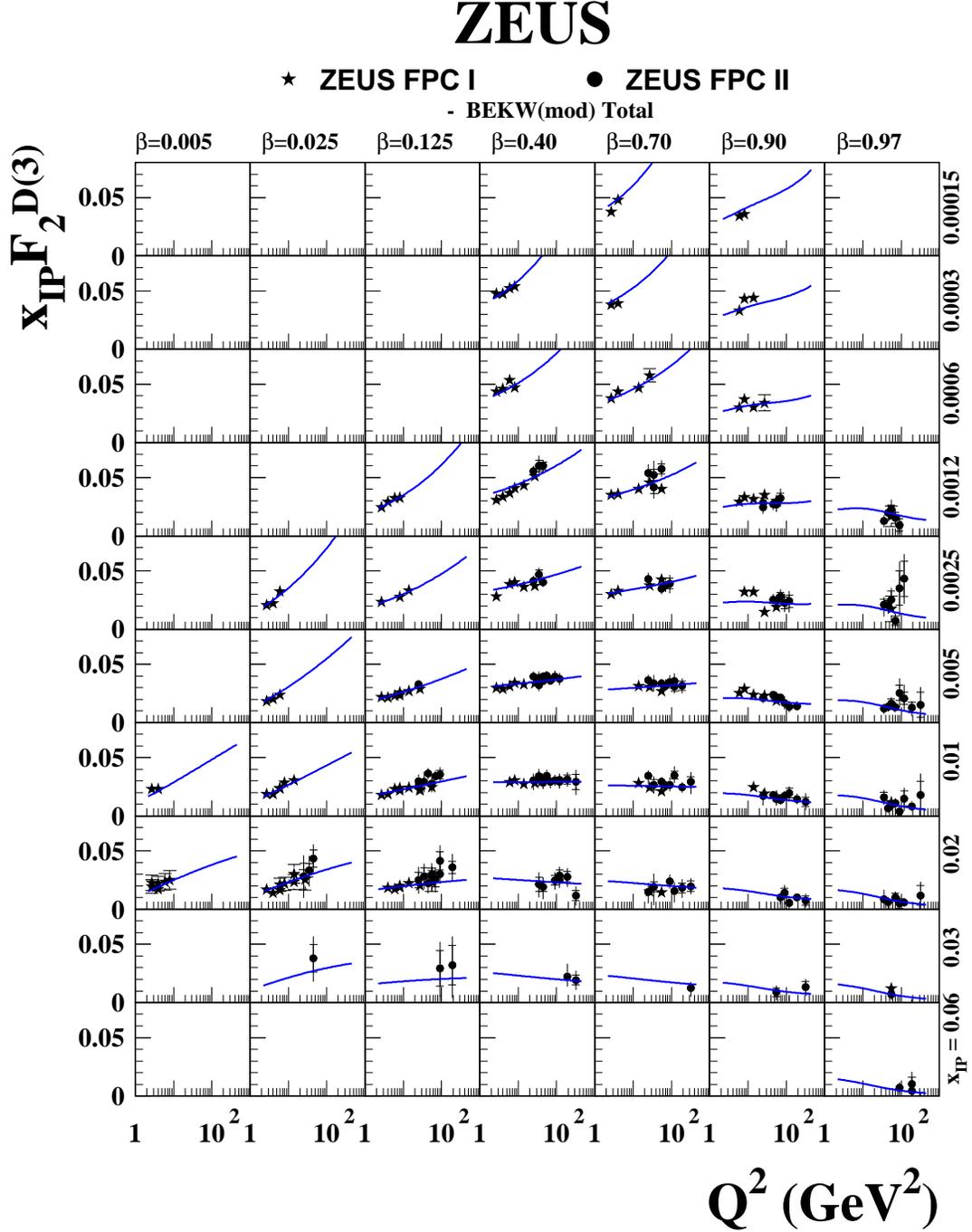,width=16.cm}}
\vspace*{-0.5cm}
\caption{The diffractive structure function of the proton multiplied by $\xpom$, $\xpom F^{D(3)}_2$, as a function of $Q^2$ for different regions of $\beta$ and $\xpom$ from the FPC~I data (stars) and FPC~II data (dots).  The inner error bars show the statistical uncertainties and the full bars the statistical and systematic uncertainties added in quadrature. The curves show the result of the BEKW(mod) fit to the data.}
\label{f:f2d3vsq2lh}
\end{figure}
\clearpage

\begin{figure}
\begin{center}
\includegraphics[angle=90,totalheight=9.5cm]{./{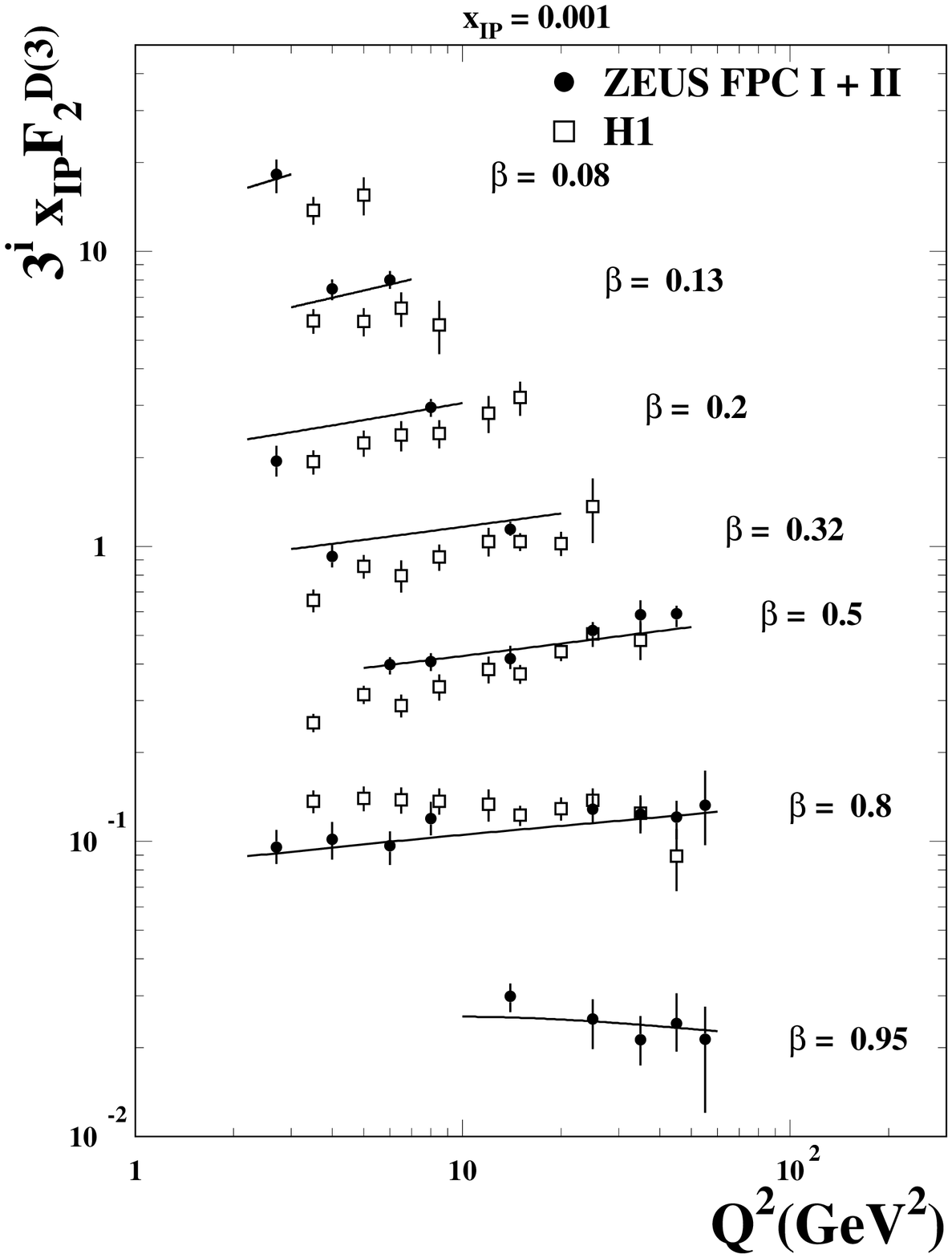}}
\includegraphics[angle=90,totalheight=9.5cm]{./{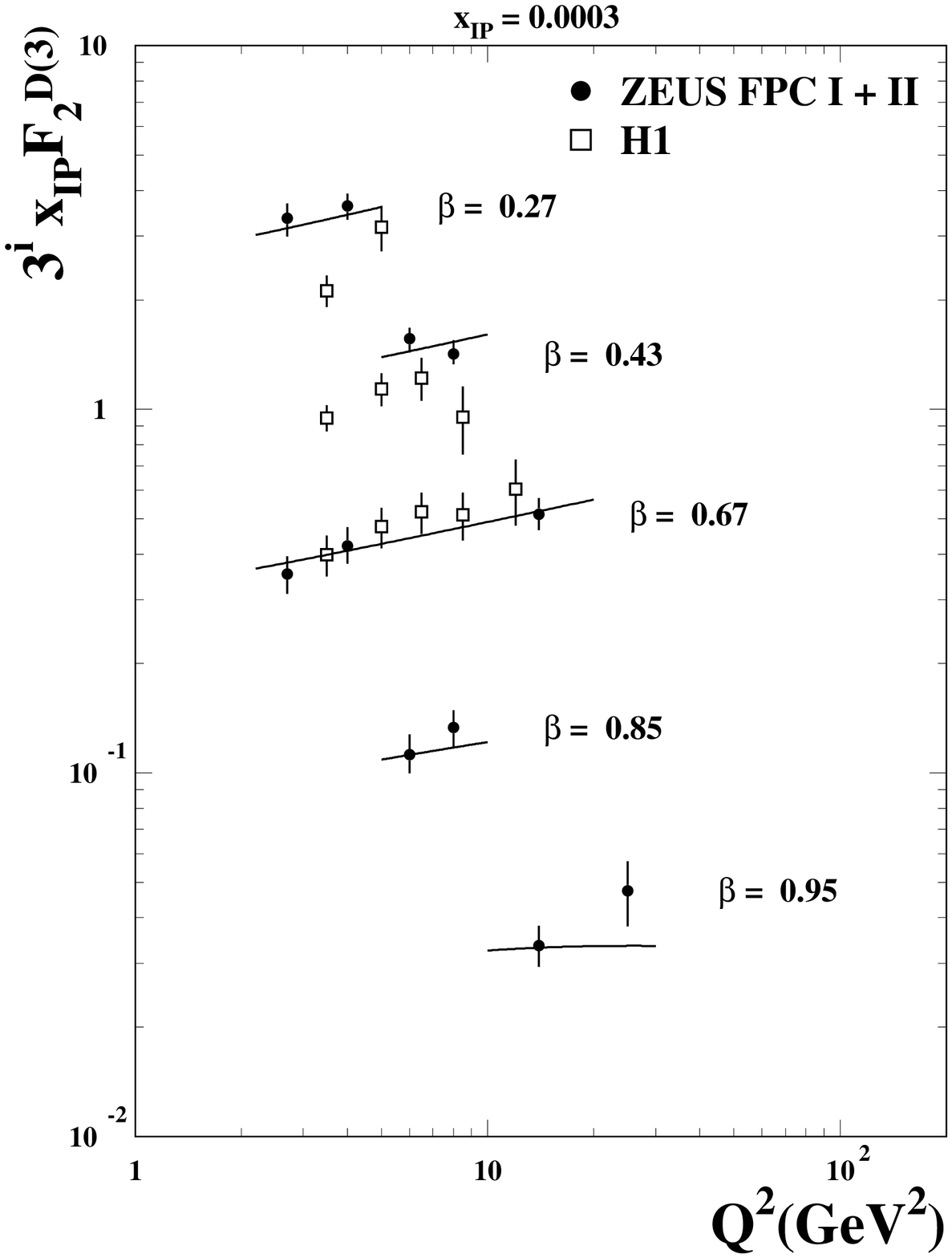}}
\caption{The diffractive structure function of the proton multiplied by
   $\xpom$, $\xpom F^{D(3)}_2$, as a function of $Q^2$ for fixed $\xpom = 0.0003$ and $\xpom = 0.001$ as indicated, for different values of $\beta$. The results of the FPC~I and FPC~II data are compared with those of H1. The data are multiplied by a factor of $3^i$ for better visibility with $i=0$ for the highest value of $\beta$, $i = 1$ for the next highest $\beta$, and so on. The curves show the result of the BEKW(mod) fit to the FPC~I and FPC~II data.}
\label{f:f2d3vsq2bxp00031zh}
\end{center}
\end{figure}

\begin{figure}
\begin{center}
\includegraphics[angle=90,totalheight=9.5cm]{./{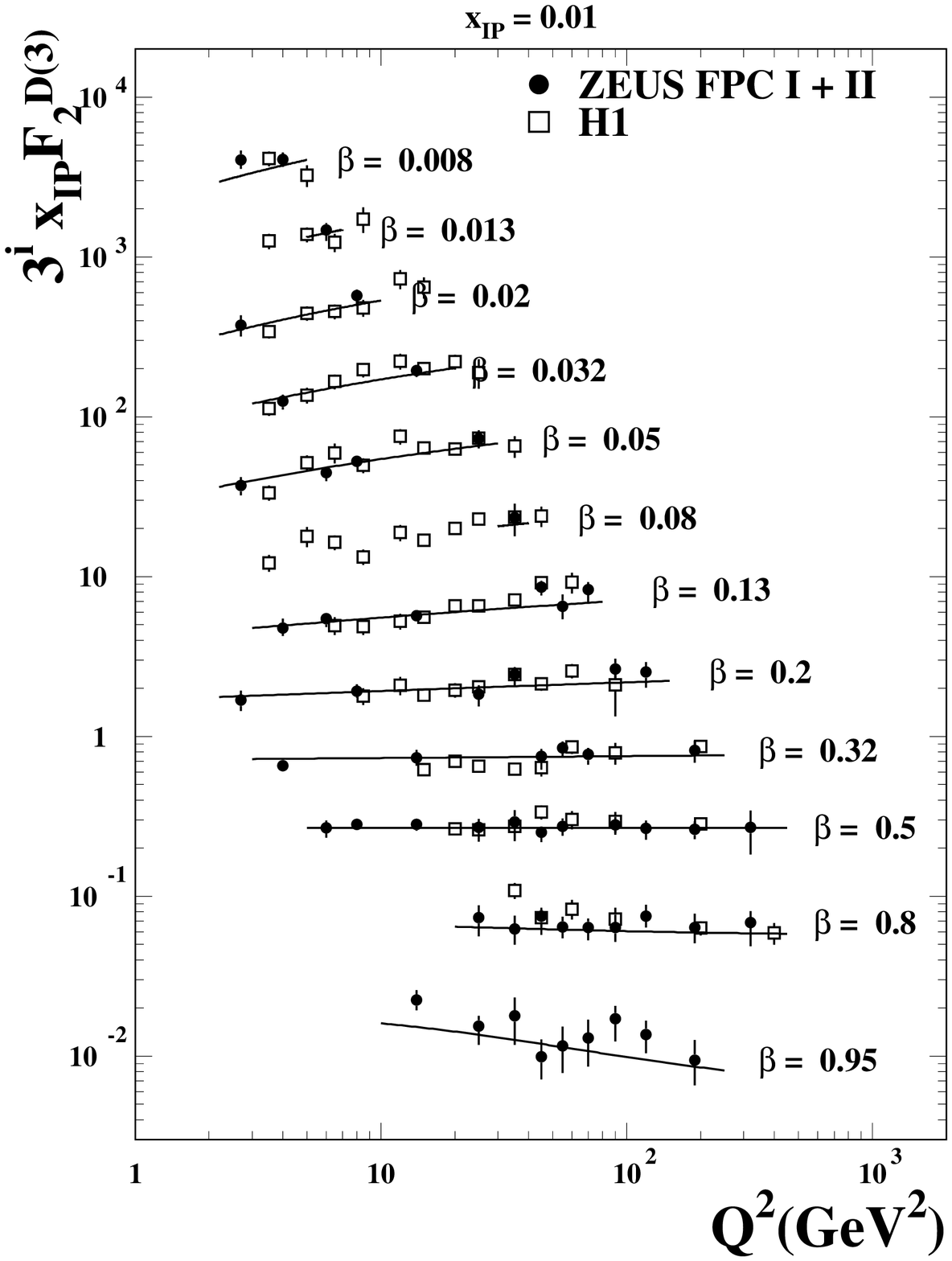}}
\includegraphics[angle=90,totalheight=9.5cm]{./{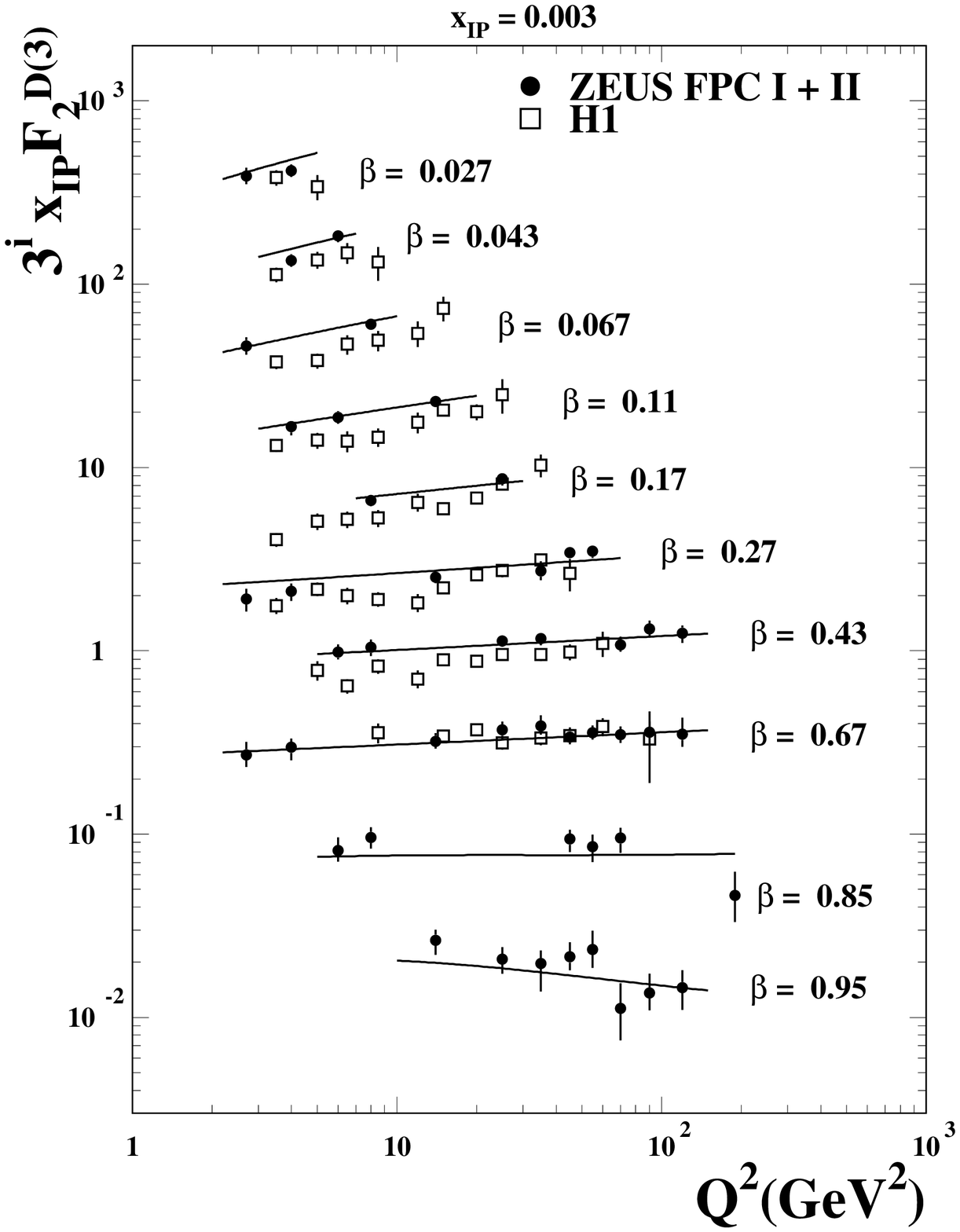}}
\caption{The diffractive structure function of the proton multiplied by
   $\xpom$, $\xpom F^{D(3)}_2$, as a function of $Q^2$ for fixed $\xpom = 0.003$  and $\xpom = 0.01$, as indicated, for different values of $\beta$. The results of the FPC~I data and FPC~II data are compared with those of H1. The data are multiplied by a factor of $3^i$ for better visibility with $i=0$ for the highest value of $\beta$, $i = 1$ for the next highest $\beta$, and so on. The curves show the result of the BEKW(mod) fit to the FPC~I and FPC~II data.}
\label{f:f2d3vsq2bxp0031zh}
\end{center}
\end{figure}

\begin{figure}
\begin{center}
\includegraphics[angle=90,totalheight=9.5cm]{./{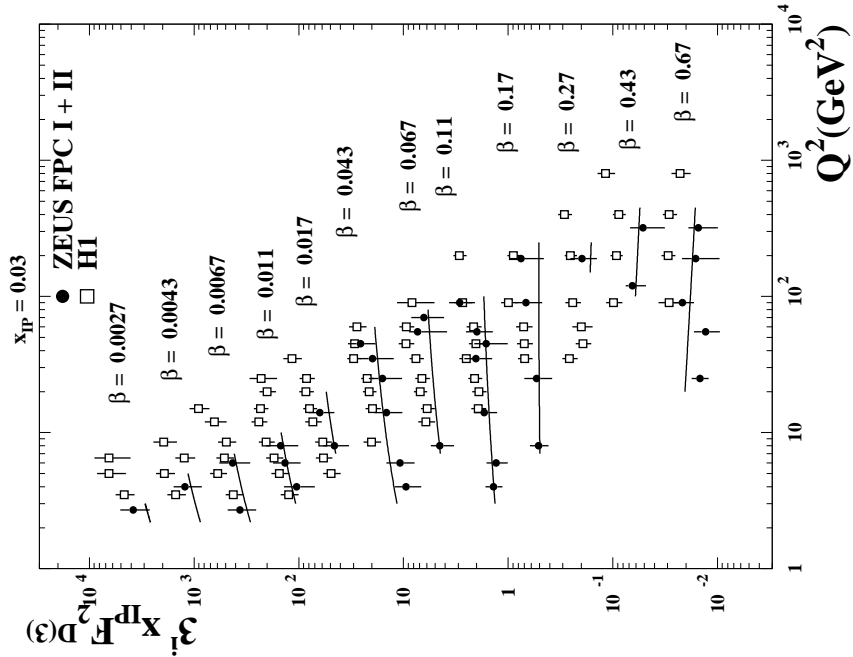}}
\caption{The diffractive structure function of the proton multiplied by
   $\xpom$, $\xpom F^{D(3)}_2$, as a function of $Q^2$ for fixed $\xpom = 0.03$, for different values of $\beta$. The results of the FPC~I and FPC~II data are compared with those of H1. The data are multiplied by a factor of $3^i$ for better visibility with $i=0$ for the highest value of $\beta$, $i = 1$ for the next highest $\beta$, and so on. The curves show the result of the BEKW(mod) fit to the FPC~I and FPC~II data.}
\label{f:f2d3vsq2bxp03zh}
\end{center}
\end{figure}

\begin{figure}
\begin{center}
\includegraphics[angle=90,totalheight=9.5cm]{./{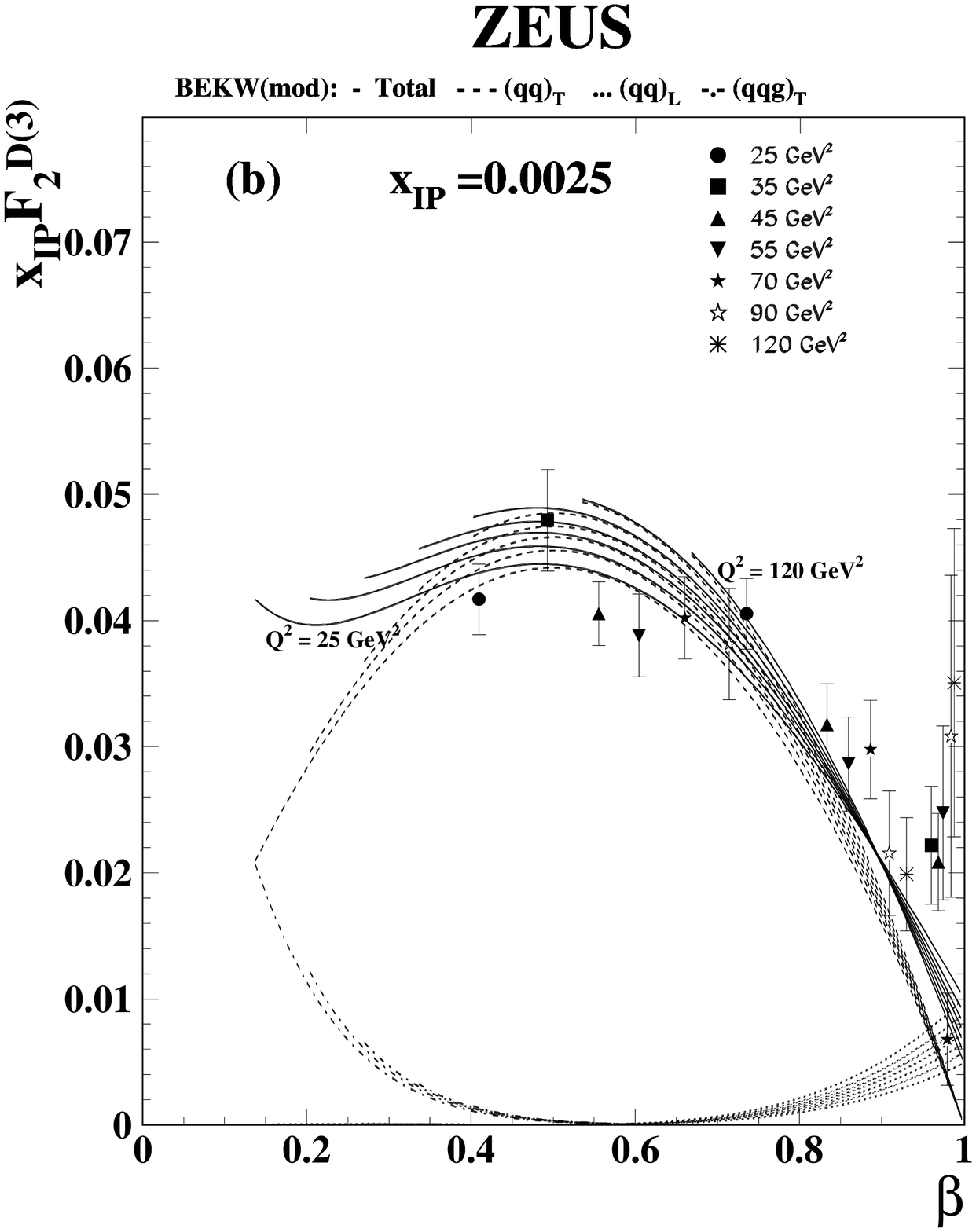}}
\includegraphics[angle=90,totalheight=9.5cm]{./{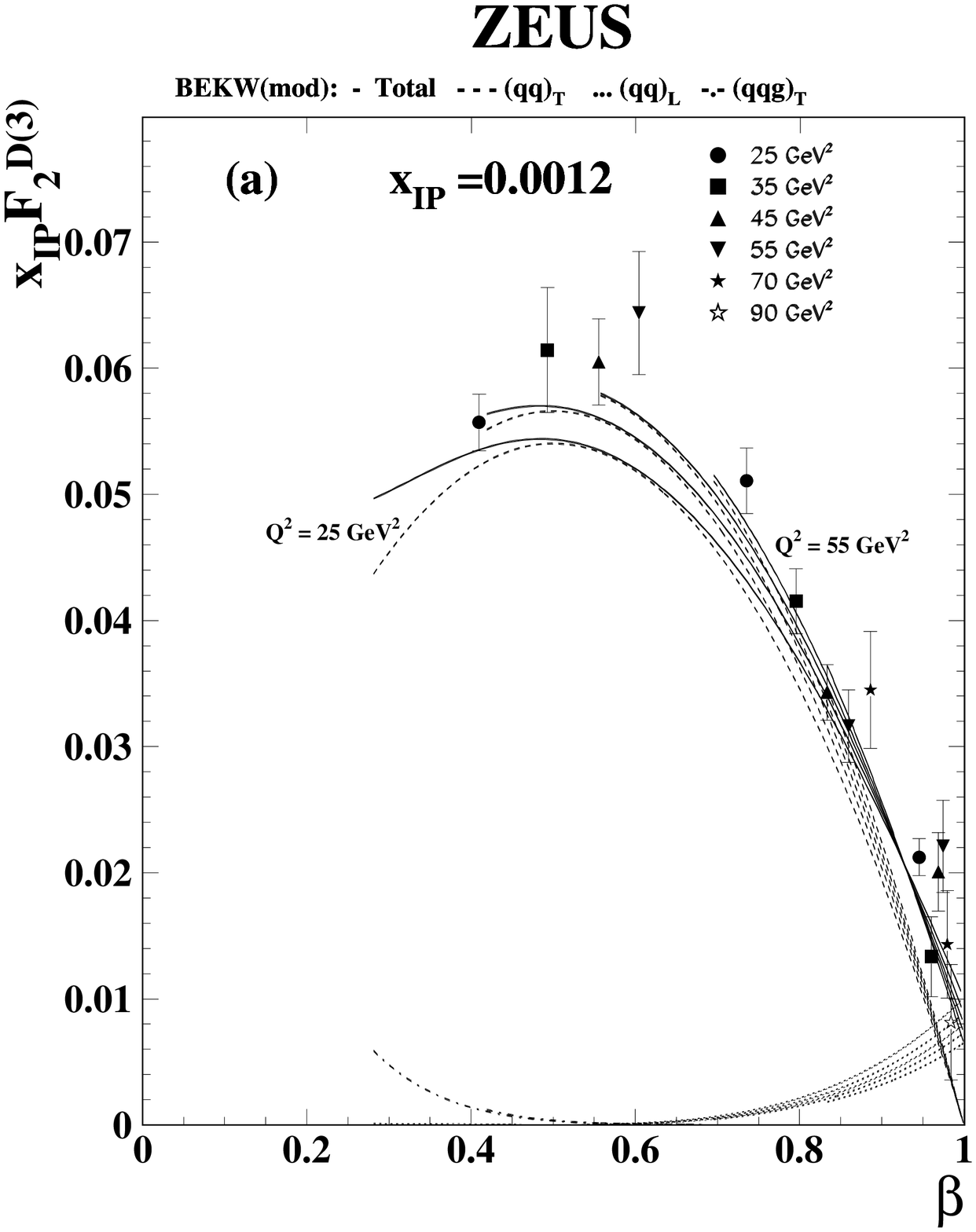}}
\caption{The diffractive structure function of the proton multiplied by
   $\xpom$, $\xpom F^{D(3)}_2$, as a function of $\beta$ for the $Q^2$ values indicated, at fixed (a) $\xpom = 0.0012$ and (b) $\xpom = 0.0025$, for the FPC~I and FPC~II data. The curves show the results of the BEKW(mod) fit for the contributions from $(q \overline{q})$ for transverse (dashed) and longitudinal photons (dotted) and for the $(q \overline{q}g)$ contribution for transverse photons (dashed-dotted) together with the sum of all contributions (solid), for the $\beta$-region studied for diffractive scattering.}
\label{f:f2d3vsbetah12}
\end{center}
\end{figure}

\begin{figure}
\begin{center}
\includegraphics[angle=90,totalheight=9.5cm]{./{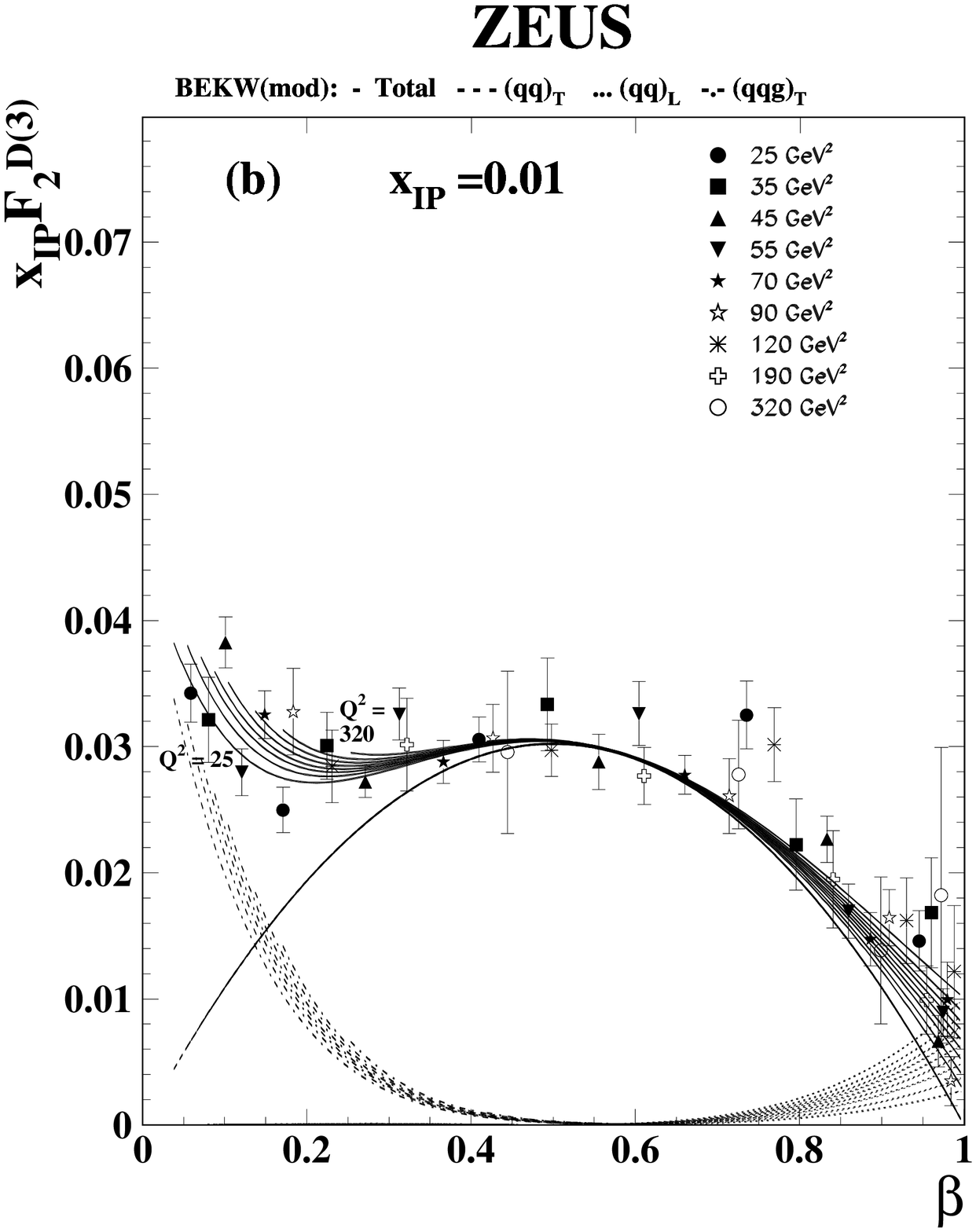}}
\includegraphics[angle=90,totalheight=9.5cm]{./{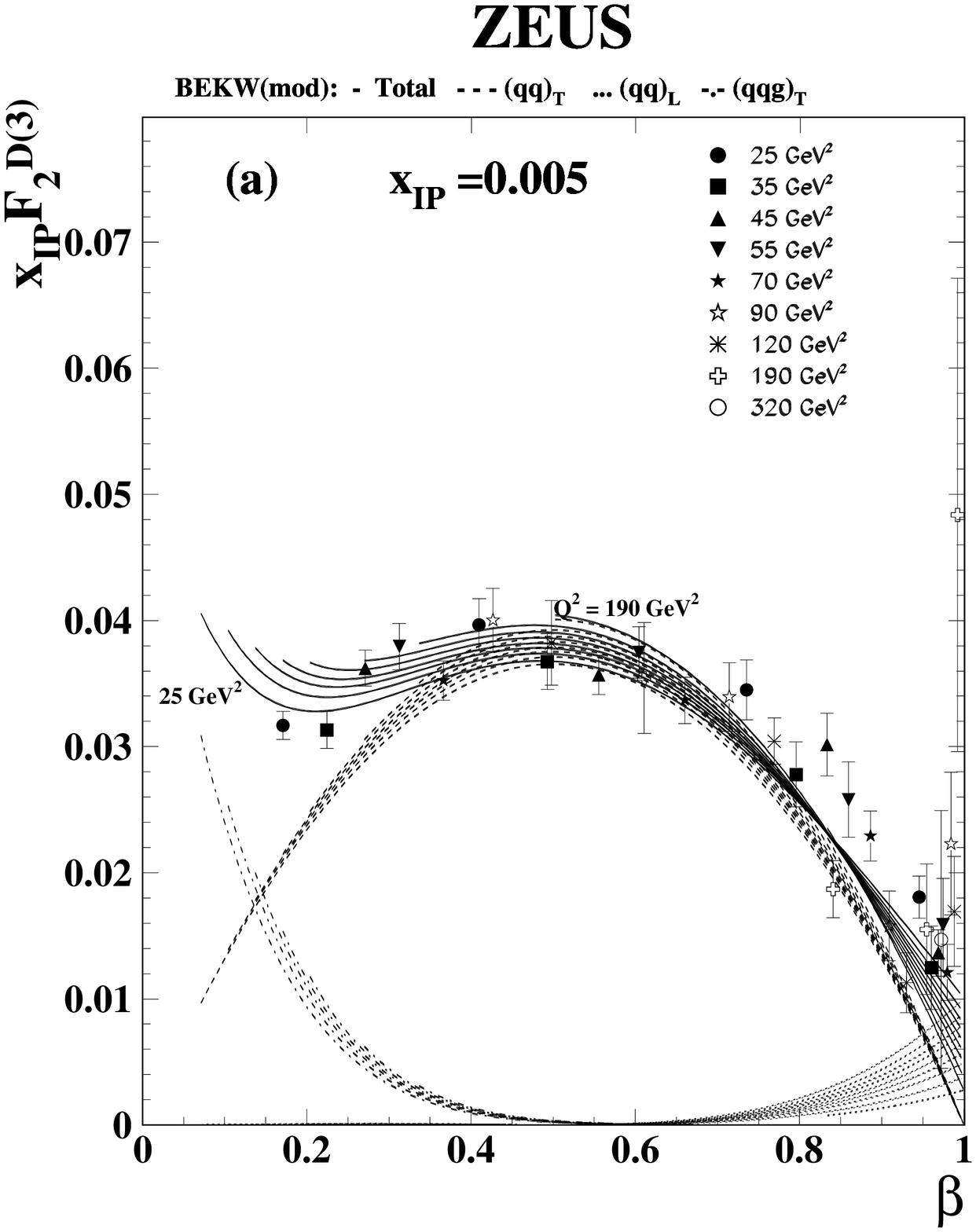}}
\caption{The diffractive structure function of the proton multiplied by
   $\xpom$, $\xpom F^{D(3)}_2$, as a function of $\beta$ for the $Q^2$ values indicated, at fixed (a) $\xpom = 0.005$ and (b) $\xpom = 0.01$, for the FPC~I and FPC~II data. The curves show the results of the BEKW(mod) fit for the contributions from $(q \overline{q})$ for transverse (dashed) and longitudinal photons (dotted) and for the $(q \overline{q}g)$ contribution for transverse photons (dashed-dotted) together with the sum of all contributions (solid), for the $\beta$-region studied for diffractive scattering.}
\label{f:f2d3vsbetah34}
\end{center}
\end{figure}

\begin{figure}
\begin{center}
\includegraphics[angle=90,totalheight=9.5cm]{./{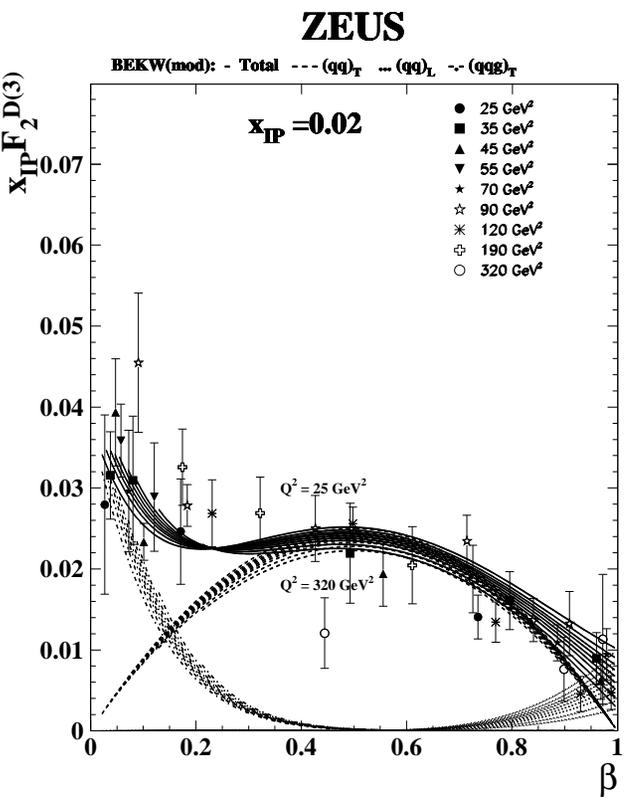}}
\caption{The diffractive structure function of the proton multiplied by
   $\xpom$, $\xpom F^{D(3)}_2$, as a function of $\beta$ for the $Q^2$ values indicated at fixed $\xpom = 0.02$, for the FPC~I and FPC~II data. The curves show the results of the BEKW(mod) fit for the contributions from $(q \overline{q})$ for transverse (dashed) and longitudinal photons (dotted) and for the $(q \overline{q}g)$ contribution for transverse photons (dashed-dotted) together with the sum of all contributions (solid), for the $\beta$-region studied for diffractive scattering.}
\label{f:f2d3vsbetah5}
\end{center}
\end{figure}


%
%
\end{document}